\documentclass[10pt,3p,final,twocolumn]{elsarticle}

\usepackage{graphicx}

\usepackage{amssymb}
\usepackage{amsthm}
\usepackage{upgreek}
\usepackage{float}
\usepackage{amsmath}
\usepackage{gensymb}
\usepackage{subfigure}
\usepackage{array}
\usepackage{url}
\usepackage{multirow}
\usepackage[para,online,flushleft]{threeparttable}

\usepackage{lineno}

\biboptions{sort&compress}

\journal{Nuclear Instruments and Methods A}

\begin{document}

\begin{frontmatter}

\title{The GRIFFIN Facility for Decay-Spectroscopy Studies at TRIUMF-ISAC}

\author[labelTRI]{A.B.~Garnsworthy\corref{cor1}}
 \ead{garns@triumf.ca}
 \cortext[cor1]{Corresponding author: 4004 Wesbrook Mall, Vancouver, BC, V6T 2A3, Canada}
\author[labelGuelph]{C.E.~Svensson}%
 
\author[labelTRI]{M.~Bowry}
\author[labelGuelph]{R.~Dunlop}
\author[labelGuelph]{A.D.~MacLean}%
\author[labelTRI]{B.~Olaizola} 
\author[labelReed]{J.K.~Smith}%
 
\author[labelGuelph,labelSulaimani]{F.A.~Ali}%
\author[labelSFU]{C.~Andreoiu}%
\author[labelTennessee]{J.E.~Ash} 
\author[labelReed]{W.H.~Ashfield}%
% ...
%

\address[labelTRI]{TRIUMF, 4004 Wesbrook Mall, Vancouver, BC, V6T 2A3, Canada}
\address[labelGuelph]{Department of Physics, University of Guelph, Guelph, ON, Canada, N1G 2W1}
\address[labelReed]{Department of Physics, Reed College, Portland, OR 97202, USA}
\address[labelSulaimani]{Department of Physics, College of Education, University of Sulaimani, P.O. Box 334, Sulaimani, Kurdistan Region, Iraq}
\address[labelSFU]{Department of Chemistry, Simon Fraser University, Burnaby, BC, V5A 1S6, Canada}
\address[labelSFU2]{Department of Physics, Simon Fraser University, Burnaby, BC, V5A 1S6, Canada}
\address[labelTennessee]{Department of Physics, Tennessee Technological University, Cookeville, TN 38505, USA}
\address[labelUBC]{Department of Physics and Astronomy, University of British Columbia, Vancouver, BC, V6T 1Z4, Canada}
\address[labelMines]{Department of Physics, Colorado School of Mines, Golden, CO 80401, USA}
\address[labelUVic]{Department of Physics and Astronomy, University of Victoria, Victoria, British Columbia V8P 5C2, Canada}
\address[labelSurrey]{Department of Physics, University of Surrey, Guildford, Surrey, GU2 7XH, UK}
\address[labelQueens]{Department of Physics, Queen's University, Kingston, ON,  K7L 3N6, Canada}
\address[labelMexico]{Universidad Nacional Aut\'{o}noma de M\'{e}xico, Instituto de Ciencias Nucleares, AP 70-543, M\'{e}xico City 04510, DF, M\'{e}xico}
\address[labelUK]{Department of Chemistry, University of Kentucky, Lexington, Kentucky 40506-0055, USA}
\address[labelUK2]{Department of Physics \& Astronomy, University of Kentucky, Lexington, Kentucky 40506-0055, USA}
\address[labelLSU]{Department of Physics and Astronomy, Louisiana State University, Baton Rouge, LA 70803, USA}

\fntext[fn1]{Present Address: Department of Physics and Applied Physics, University of Massachusetts Lowell, Lowell, MA 01854, USA}\fntext[fn1a]{Present Address: TRIUMF, 4004 Wesbrook Mall, Vancouver, BC, V6T 2A3, Canada}
\fntext[fn1b]{Present Address: Nuclear Futures Institute, Bangor University, Dean Street, Bangor, Gwynedd, LL57 1UT, UK}
\fntext[fn3]{Present Address: Department of Physics, University of Surrey, Guildford, Surrey, GU2 7XH, UK}
\fntext[fn3A]{Present Address: Lawrence Livermore National Laboratory, 7000 East Ave, Livermore, CA 94550, USA}
\fntext[fn3AB]{Present Address: Nuclear Physics Institute of ASCR,
250 68 \u{R}e\u{z}. Prague, Czech Republic}
\fntext[fn3B]{Present Address: Department of Physics, University of Toronto, Toronto, ON M5S 1A7, Canada}
\fntext[fn2]{Present Address: Idaho National Laboratory, Idaho Falls, Idaho 83415, USA.}
\fntext[fn10]{Present Address: Department of Physics, Lund University, 22100 Lund, Sweden}
\fntext[fn8]{Present Address: Lawrence Berkeley National Laboratory, Berkeley, CA 94720, USA}
\fntext[fn6]{Present Address: University of Jyv\"{a}skyl\"{a}, Department of Physics, P.O. Box 35, FI-40014 Jyv\"{a}skyl\"{a}, Finland}
\fntext[fn5]{Present Address: Oliver Lodge Laboratory, The University of Liverpool, Liverpool, L69 7ZE, UK}
\fntext[fn9]{Present Address: Department of Physics, University of Chicago, Chicago, Illinois 60637, USA}

\begin{abstract}
Gamma-Ray Infrastructure For Fundamental Investigations of Nuclei, GRIFFIN, is a new high-efficiency $\gamma$-ray spectrometer designed for use in decay spectroscopy experiments with low-energy radioactive ion beams provided by TRIUMF's Isotope Separator and Accelerator (ISAC-I) facility. GRIFFIN is composed of sixteen Compton-suppressed large-volume clover-type
high-purity germanium (HPGe) $\gamma$-ray detectors combined with a suite of ancillary detection systems and coupled to a custom digital data acquisition system. The infrastructure and detectors of the spectrometer as well as the performance characteristics and the analysis techniques applied to the experimental data are described.
\end{abstract}


\end{frontmatter}


\section{Introduction}
\label{sec:intro}
Large arrays of detectors for $\gamma$-ray measurements coupled with auxiliary particle detection systems provide a powerful and versatile tool for studying exotic nuclei through nuclear spectroscopy at radioactive ion beam facilities.
Decay spectroscopy experiments are performed at a number of radioactive beam facilities world-wide, such as with the EURICA setup at RIKEN \cite{EURIKA}, the beta counting station at NSCL \cite{NSCL-BCS}, the X-Array and SATURN at ANL \cite{XARRAY}, the ISOLDE decay station \cite{ISOLDE-IDS}, the implantation and decay station at Lanzhou \cite{Lanzhou-DecayStation}, the identification station at SPIRAL \cite{GANIL-ID-Station} and in the past the RISING setup at GSI \cite{RISING-Isomer,RISING-ActiveStopper} which is transitioning to DESPEC at FAIR \cite{DESPEC}.
Gamma-Ray Infrastructure For Fundamental Investigations of Nuclei, GRIFFIN \cite{Svensson14}, is a new experimental facility for radioactive decay studies at the TRIUMF-ISAC laboratory \cite{Dilling14} located in Vancouver, Canada.

\begin{figure*}
\centering
    \subfigure{
        {\includegraphics[width=1.0\linewidth]{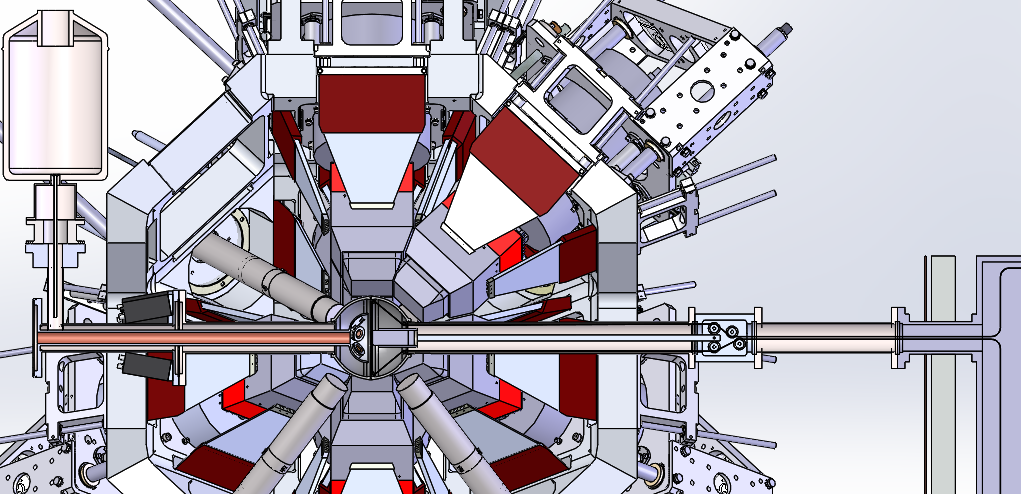}
        \label{fig:GRIFFIN_overview1}}
    }\\
    \subfigure{
        {\includegraphics[width=1.0\linewidth]{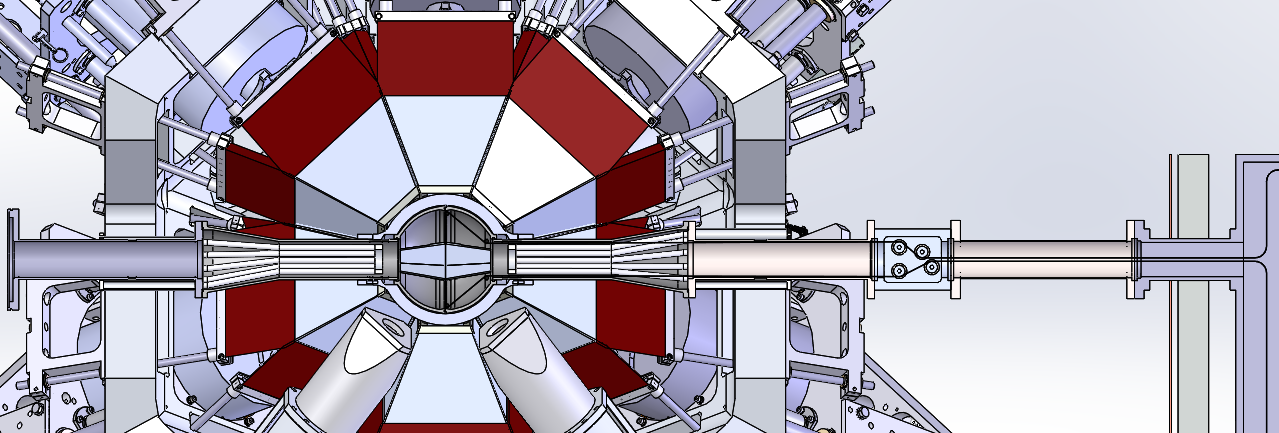}
        \label{fig:GRIFFIN_overview2}}
    }\\
\caption{Two possible configurations of the GRIFFIN spectrometer. The beam is delivered from the left. The Pb wall shielding the tape box can be seen on the right. {\bf Upper panel} showing PACES in the upstream (left) half of the chamber and the fast $\beta$ scintillator in the downstream (right) chamber. In the `High-efficiency' mode, the HPGe detectors are at 11\,cm and the LaBr$_3$(Ce) detectors at 12.5\,cm from the implantation point on the tape. {\bf Lower panel} with SCEPTAR in the upstream and downstream chambers. In the `Optimized peak-to-total' mode, the HPGe and LaBr$_3$(Ce) detectors are retracted to 14.5\,cm and 13.5\,cm, respectively, in order that they can be fully Compton and background suppressed with BGO shields. The 20\,mm Delrin absorber is also installed in the lower panel.}
\label{fig:GRIFFIN_overview}
\end{figure*}

GRIFFIN is used for decay spectroscopy research with stopped low-energy radioactive ion beams. High $\gamma$-ray detection efficiency provided by an array of sixteen Compton-suppressed high-purity germanium (HPGe) clover detectors \cite{Rizwan2016} is combined with a suite of ancillary detection systems.
Many of the ancillary detector systems, infrastructure and techniques used with GRIFFIN were originally developed for use with the 8$\pi$ spectrometer during its 11 years of operation at ISAC \cite{Garnsworthy2015,Garrett2015,Garnsworthy2014,Ball2012,Garrett2007,Zganyar2007,Ball2005,Svensson2003}.
These include an array of plastic scintillators for $\beta$ particle detection (SCEPTAR), a set of five in-vacuum LN$_2$-cooled lithium-drifted silicon detectors for conversion electron measurements (PACES), and an array of eight LaBr$_3$(Ce) scintillators and a fast $\beta$ scintillator for fast-timing measurements. Two possible configurations of the GRIFFIN array and the ancillary detectors are shown in Figure \ref{fig:GRIFFIN_overview}.
In addition, GRIFFIN can couple to the DESCANT array \cite{Garrett2014,Bildstein2015} of neutron detectors for $\beta$-delayed neutron emission studies with exotic neutron-rich beams.
This powerful combination of detectors to be used with the radioactive-ion beams from ISAC and in the future the Advanced Rare-IsotopE Laboratory (ARIEL) \cite{Dilling14}, supports a broad program of research in the areas of nuclear structure, nuclear astrophysics, and fundamental interactions.

The data acquisition (DAQ) system for GRIFFIN takes advantage of custom-designed digital electronics \cite{Garnsworthy2017}. Unique features of the DAQ system enable half-life and branching ratio measurements with levels of precision better than $\pm$0.05\%. The system is also capable of effectively collecting signals from HPGe crystals at counting rates up to 50\,kHz while maintaining good energy resolution, detection efficiency and spectral quality \cite{Garnsworthy2017}. 

This article provides a full description of the infrastructure and detectors of the GRIFFIN spectrometer as well as the performance characteristics and includes a discussion of the analysis techniques applied to the experimental data. 
Section \ref{sec:infrastructure} details the infrastructure of the facility and Section \ref{sec:hardware} describes the detector hardware. Typical data analysis techniques are discussed in Section \ref{sec:analysis-performance} together with the performance achieved with the spectrometer. Finally, Section \ref{sec:Future} presents a look towards future development plans, and a summary is given in Section \ref{sec:summary}.

\section{Infrastructure}
\label{sec:infrastructure}

\subsection{Beamline and Tape Box}
The GRIFFIN spectrometer is used to study the decay of stopped radioactive ion beams produced by the ISOL (isotope separation on-line) method. The beam is delivered at a typical energy of 20 to 40\,keV. The low energy beam transport (LEBT) beamlines \cite{Baartman2014} use exclusively electrostatic bending and focusing elements. Pressures of $<$3$\times10^{-7}$\,Torr are typical for the beamlines leading to GRIFFIN. The final beam focus at GRIFFIN is typically $<$3\,mm in diameter with a transverse, 2$\sigma$ emittance of 10\,$\mu$m.

Vacuum in the GRIFFIN chamber is provided by an Agilent Varian Turbo-V 1000HT turbo-molecular pump located 95\,cm from the array centre on the upstream side, and an Oerlikon Leybold TurboVac 361 turbo-molecular pump located 160\,cm downstream of the array centre in the tape box. The turbo pumps are backed by an Agilent Varian TriScroll vacuum pump.

The radioactive ion beam delivered to the spectrometer is implanted at the mutual centers of the detector arrays onto the tape of the moving tape collector (MTC). The tape itself is standard 12.7\,mm wide computer tape constructed of Mylar with a thin layer of iron oxide on one side. In experiments involving gaseous species, Mylar tape aluminized on one side is used to prevent diffusion after implantation. 
In this case, the beam is implanted into the 6.4\,mg/cm$^2$ aluminized layer of the tape. A continuous $\sim$135\,m loop of tape is loaded into the system in the case of the standard Mylar tape. Simple splices connected with Kapton adhesive tape join the loop during the loading process. The aluminized tape is heavier so a shorter continuous loop ($\sim$36\,m) is used. Once loaded, the tape is entirely inside the vacuum system.

\begin{figure}
\centering
\includegraphics[width=0.9\linewidth]{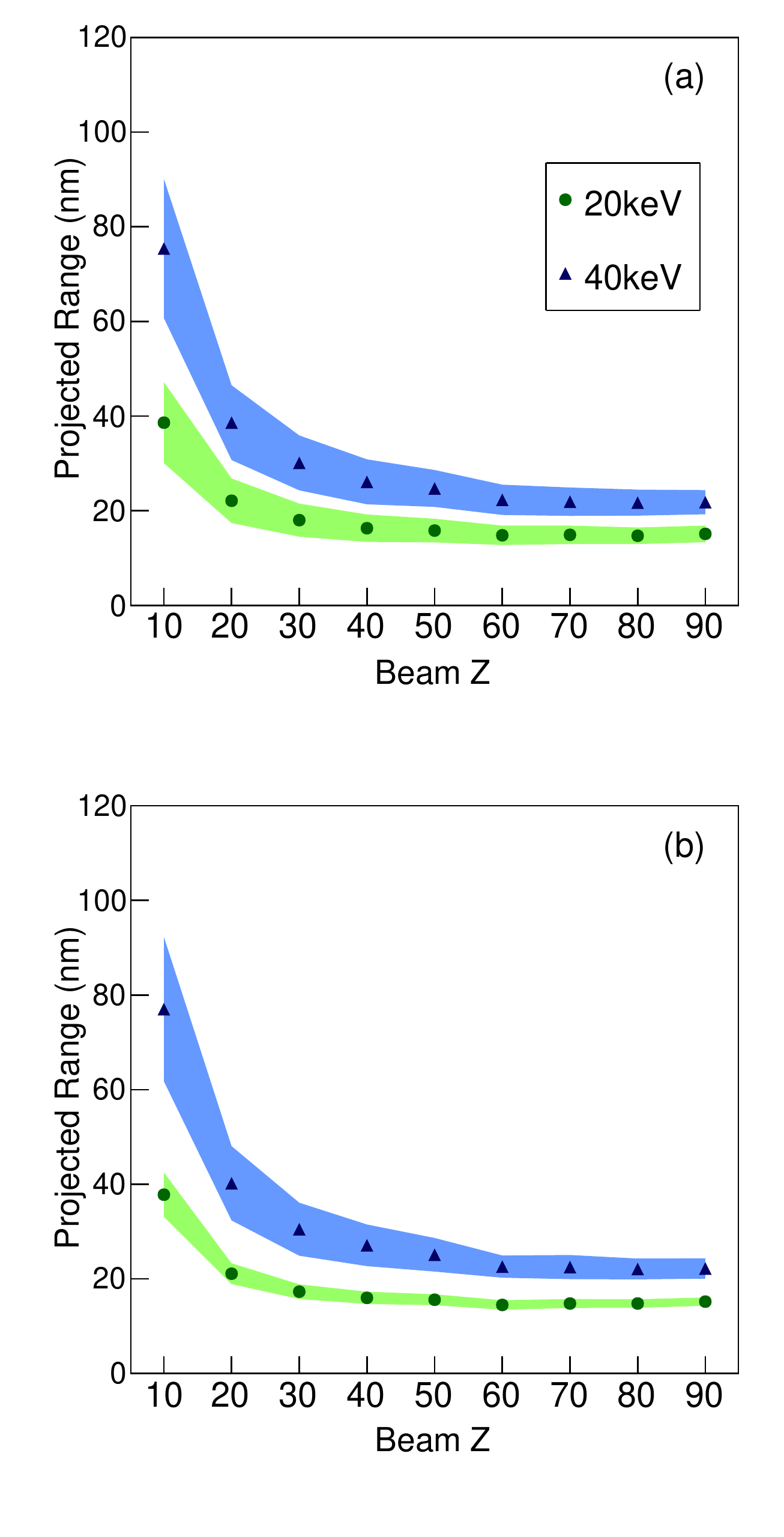}
\caption{Projected range of beam ions delivered to the GRIFFIN spectrometer tape system at energies of 20 and 40\,keV and implanted into the (a) thin iron oxide layer of the standard Mylar tape and (b) aluminized Mylar tape calculated using SRIM-2013 \cite{SRIM-2013}. The bands indicate the calculated longitudinal straggling of the projected range.}
\label{fig:Beam_Projected_Range}
\end{figure}

\begin{figure}
\centering
\includegraphics[width=0.9\linewidth]{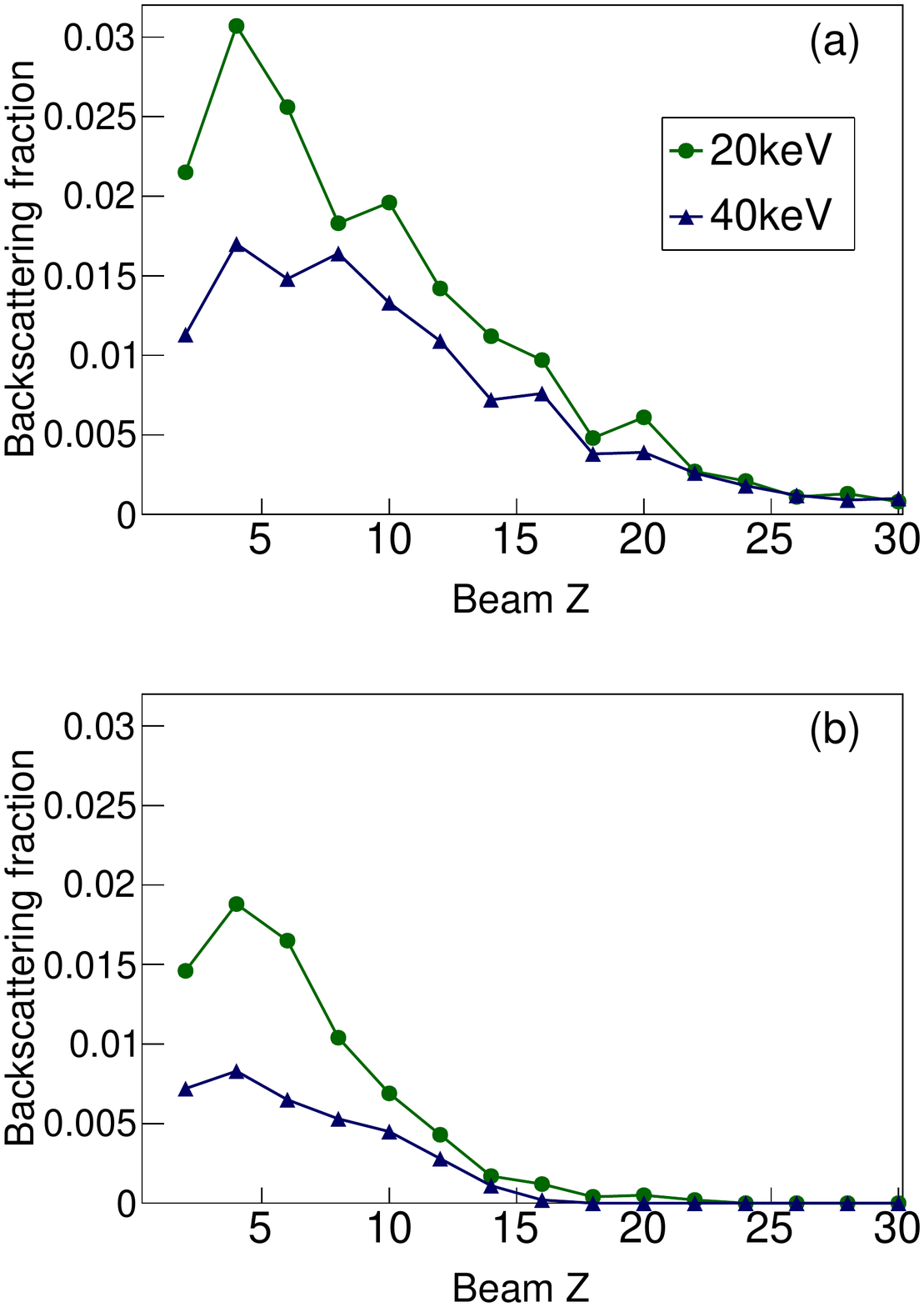}
\caption{Backscattering fraction for incident beam ions delivered to the GRIFFIN spectrometer tape system at energies of 20 and 40\,keV and implanted into the (a) thin iron oxide layer of standard Mylar tape and (b) aluminized Mylar tape calculated using SRIM-2013 \cite{SRIM-2013}.}
\label{fig:Beam_BackScattering}
\end{figure}

The distance into the tape at which the delivered beam ions are implanted (projected range) is shown in Figure \ref{fig:Beam_Projected_Range}. These projected ranges are calculated using SRIM-2013 \cite{SRIM-2013} and the error bars indicate the longitudinal straggling calculated with the same program. Here the iron oxide layer is modeled as a mixture of 40\% iron oxide ($Fe_2O_3$) and 60\% polyurethane ($C_{27}H_{36}N_{2}O_{10}$) by volume with a density of 2.896\,mg/cm$^2$. In the worst case of $Z>70$ ions delivered at 20\,keV into the aluminized Mylar tape, the projected range is 15$\pm$2\,nm. Figure \ref{fig:Beam_BackScattering} shows the backscattering fraction calculated using SRIM-2013 \cite{SRIM-2013} and demonstrates that this is only relevant for beams of the lightest elements. However, even when a good ion-optics tune is established, beam ions can miss the tape as a result of other processes. The primary process for this to occur is charge exchange with residual gas molecules in the straight section of beamline upstream of the last electrostatic element. This process neutralizes the beam particles which then drift to the experimental chamber instead of being focused by the final electrostatic elements.
The fraction of beam ions that are not implanted into the tape is pressure dependent but is less than 1\% for pressures $<$1$\times10^{-6}$\,Torr.

The tape is pulled through a pathway of rollers by rubber-surfaced pulleys moved by an Allen-Bradley (Model Number: 193507) servomotor. The pulleys are spring-mounted so that tension on the tape can be kept to a minimum. A set of brushes provides electrical grounding to the surface of the tape to avoid the build up of static charge from the implanted ion beam. The continuous loop of tape is allowed to cascade freely in the tape box region.

The servomotor is controlled by a custom built controller box. Although the controller box can be programmed with any control parameters, there are usually three speed settings at which the tape can be moved. A slow continuous movement of 13.2(1)\,cm/s is used only during setup for loading a new tape into the system. The other modes of operation are described in Section \ref{sec:cycles}.

Two sets of sensors located at different positions along the tape path ensure that the tape has not broken. A light-emitting diode and light sensor combination rely on the fact that the tape is opaque. The tape passes between these devices, blocking the light from the sensor, and a fault alert is raised if the sensor detects light.

A wall of 5\,cm thick lead bricks blocks the detectors from direct sight of the tape box. A 3.2\,mm thick copper sheet sits between the lead wall and the detectors to attenuate the flux of Pb X rays that is produced in the lead wall by incident photons and neutrons.

This tape system has proven to be an extremely reliable design and has very rarely experienced an issue during standard operation.

\subsection{Mechanical Support Structure}
\begin{figure}
\centering
    \subfigure{
        {\includegraphics[width=1.0\linewidth]{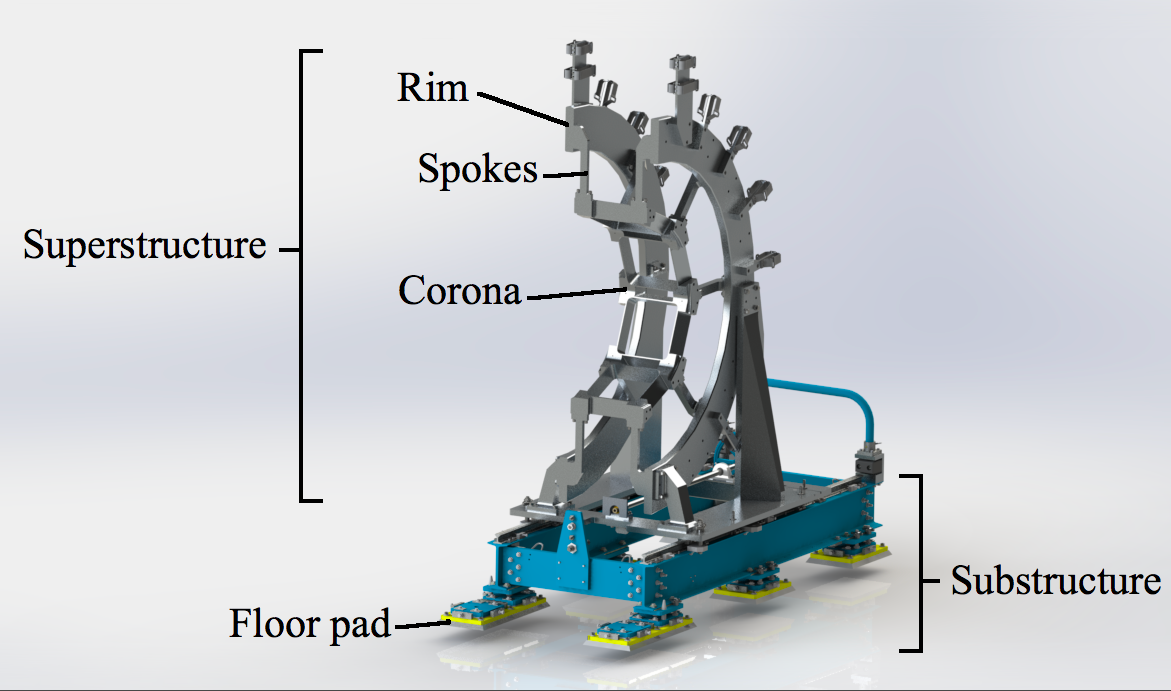}
        \label{fig:Structure1}}
    }\\
    \subfigure{
        {\includegraphics[width=1.0\linewidth]{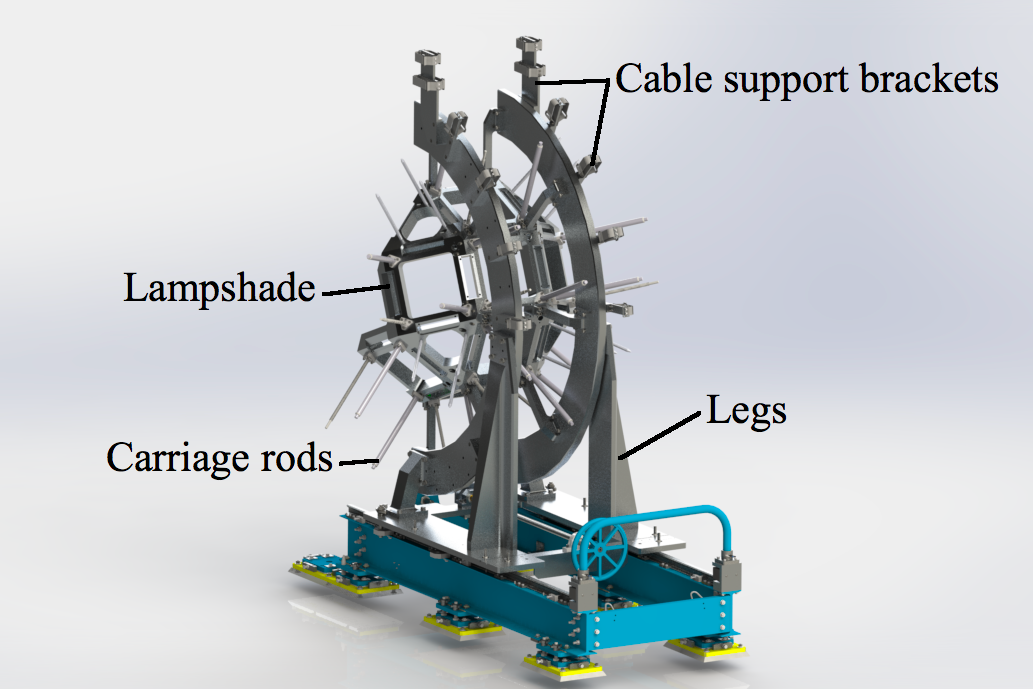}
        \label{fig:Structure2}}
    }\\
\caption{Main components of the GRIFFIN mechanical support structure.}
\label{fig:GRIFFIN_Structure}
\end{figure}

The GRIFFIN mechanical support structure, shown in Figure \ref{fig:GRIFFIN_Structure}, is composed of three sub-systems: a superstructure, a substructure and a set of detector carriages.

The superstructure supports the detector units as two halves of the overall rhombicuboctahedral geometry and itself consists of two inner ``coronas'' that each hold four GRIFFIN detector units at 90\,$^{\circ}$ relative to the beamline, and four ``lampshades'' that each hold two GRIFFIN detector units at either 45 or 135\,$^{\circ}$ relative to the ISAC-I beamline. Four C-shaped ``rims'' provide the overall spherical backbone of the superstructure. A set of ``spokes'' mount the coronas and lampshades to the rims, and ``legs'' position the superstructure at the correct beamline height of 1.676\,m above the floor of the experimental hall. The corona and lampshade pieces were machined from a single piece of M-1 aluminium. The rest of the superstructure and detector carriages are machined from 6061-T6 aluminium.

The substructure is a combination of steel I-beams and rails that couple the superstructure to the floor through a set of electrically-isolating floor pads, and allows linear motion of the two hemispheres of the array for access to the vacuum chambers. Each hemisphere opens to a distance of 2\,m from the beamline. The rigid substructure is adjustable at each floor pad in order to achieve alignment and positioning with respect to the beamline. Electrical isolation is implemented in several places. Each detector unit is electrically isolated from its support carriage. The substructure is electrically isolated from the floor at the floor pads.

An individual detector carriage supports each detector component and is mounted on a set of guide rods. This arrangement allows both precise detector alignment and radial motion capabilities for each detector position. Cables and liquid nitrogen supply lines are routed around the rims to each detector position by a set of supporting brackets and straps.

\subsection{Liquid Nitrogen Filling System}
The HPGe detectors require a regular supply of liquid nitrogen to maintain them at their $\approx$100\,K operational temperature. 
Detectors are filled from a local Cryofab CLPB-240-PS2 240\,litre phase separator through a manifold employing 24\,VDC solenoid valves. The liquid nitrogen is supplied from this manifold by short lengths of Tygothane tubing wrapped with Armaflex thermal pipe insulation.
 
The autofill system that controls the manifolds includes a PLC-based logic component that executes the fill processes as well as provides a graphical-user interface for control and monitoring. The custom-built PLC-based autofill control system was designed and assembled by the TRIUMF controls group. Fill processes are executed at regular 8-hour intervals. During a fill process, following a pre-cooling of the manifold, the liquid inlet valve to each enabled HPGe clover detector position or PACES dewar is opened and remains open until a sensor in the exhaust outlet for that detector reaches liquid nitrogen temperature.
The success or failure of a fill is reported by email and phone message following each fill cycle.

The local tank is used as a phase separator and is maintained at 15\,psig. The liquid level is monitored by a Custom Biogenics liquid level controller, model LLCHP-1B, which also allows the setting of low- and high-level limits. 
The liquid level in the dewar is derived from a sensor that measures the difference in hydrostatic pressure between the bottom and top of the tank. When the measured liquid level drops below the low limit, the monitor enters `fill' mode, and remains in that mode until the sensor reports a value exceeding the upper limit. In fill mode, the control box turns on a Type B 120\,VAC outlet that energizes two Redhat 8222 cyrogenic solenoid valves, one each at the liquid-in and vent ports of the phase separator. Liquid nitrogen is supplied through a permanent, rigid, vacuum-jacketed line that connects to the ISAC-I building LN$_2$ supply tank. 
The flexible vacuum jacketed hose has joints every 21\,m. Each segment of hose has a 15\,psig pressure relief valve, so that boiling liquid that may be trapped within the hose escapes rather than causes the hose to rupture.

\subsection{Electronics Enclosure}
A dedicated climate-controlled electronics enclosure houses the GRIFFIN data acquisition system and associated infrastructure. The structure is constructed of Bosch 45$\times$90 and 45$\times$45 profile extrusions \cite{BOSCH} with Plexiglass panels enclosing the space. The inside space is 3.1\,m by 2\,m with an overall height of 3.2\,m. Sliding doors on the front and rear sides allow access to the front and rear sides of the electronics racks. Penetrations in the ceiling and on one side allow for cable access.

The six electronics racks with a height of 2.64\,m are electrically isolated from the structure. 
Electrical power is provided to the electronics through a single-point-ground Hammond Power solutions Inc. 75\,kVA super isolation transformer.
The racks are securely mounted on a false floor formed of perforated tiles and raised by 30.5\,cm from the concrete floor of the experimental hall to allow air circulation. A dedicated Liebert Deluxe System 3 (Model number: DE138W-CAEI9517), 10-tonne air-conditioning unit maintains a stable temperature and humidity level within the electronics enclosure.

\subsection{Low- and High-Voltage Supplies}
The low-voltage power for the preamplifiers of each HPGe clover is provided from a dedicated custom Nuclear Instrumentation Module (NIM) module located in a high-power NIM crate \cite{NIM}. There are three NIM crates in the GRIFFIN electronics enclosure to accommodate the 16 preamplifier power modules, and the custom preamplifier supply modules for the GRIFFIN Compton and background suppression shields.
Mesytec MHV-4 4-channel NIM voltage supply modules are used to provide low-voltage power to the preamplifiers of the PACES and LaBr$_3$(Ce) detectors. These MHV-4 modules are all located in a single NIM crate.

High voltage is supplied from modules housed in CAEN high-voltage mainframes of the SY1527LC and SY4527 type \cite{CAEN}. The A1832PE 12-channel, 6\,kV/1\,mA CAEN module with individual channel shutdown is used to supply the high voltage to each HPGe crystal. The bias shutdown signal (BSD) is provided from the custom GRIFFIN preamplifier power supply module and is derived from a BSD signal provided from the detector alarm card. The operating voltages of the HPGe crystals are between 3.5 and 4\,kV with leakage currents less than 1\,nA.

Sixteen A7030P 48-channel, 3\,kV/1\,mA CAEN modules with a single Radiall 52-pin connector supply high voltage independently to each of the 40 PMTs in each of the sixteen sets of GRIFFIN Compton and background suppression shields. The high voltage for the PMTs of the ancillary detector suppression shields is supplied from four 12-channel A1733P CAEN modules with SHV (safe high voltage) connectors. Typical operating voltages for each of these PMTs are between 750 and 950\,V with leakage currents of $\sim$500\,$\mu$A.

Two 12-channel A1733N CAEN modules each with up to 3\,mA at 3\,kV are used to provide the negative high voltage to the SCEPTAR PMTs and the PMT of the fast $\beta$ scintillator. Typical voltages supplied are between -1100 and -1300\,V with leakage currents of $\sim$500\,$\mu$A. The same type of module is used to supply the -500\,V bias to the five PACES Si(Li) detectors.

The high voltages for the LaBr$_3$(Ce) detectors are supplied by dual-channel iseg NHQ204M NIM modules \cite{iseg} capable of supplying 3\,mA at 4\,kV. The bias voltage is adjusted to produce an output signal amplitude of 662\,mV for the 662\,keV $\gamma$ ray from a $^{137}$Cs source. The voltages supplied are approximately -1400\,V with leakage currents of $\sim$400\,$\mu$A.

\subsection{Cables}
Two cable trays support the low-voltage, high-voltage and signal cables between the GRIFFIN electronics enclosure and the spectrometer. The upper cable tray is used for all the HPGe and suppression shield cabling which enter the electronics shack through a penetration in the ceiling. The length of this cable run is 21\,meters. The lower cable tray is used for all ancillary detector cabling and has a shorter cable length of 19\,meters as it enters the electronic enclosure through the wall closest to the spectrometer.
 
There is a separate cable bundle for each HPGe position of the array. Each bundle consists of low-voltage, high-voltage, and signal cables packaged in an electrically grounded flexible sheath formed with a knitted wire mesh of tin-plated-copper-clad steel. For convenience, a panel of bulkhead connectors is located on each HPGe carriage which connects the long cables (21\,m) with short cables (0.75\,m) within the carriage. 
Low voltage is supplied to each clover detector by a dedicated low-voltage cable with CPC power connectors at the supply module and bulkhead. The short low-voltage cable connects to the detector with a 9-pin D-sub connector. This cable also carries the bias shutdown signal from the clover to the GRIFFIN preamplifier power supply module.
High voltage to the HPGe crystals is supplied by dedicated RG59 cables with SHV connectors.
Each of the 40 PMTs in each of the GRIFFIN Compton and background suppression shields receives high voltage through a single 48-conductor cable with Radiall 52-pin connectors on each end. This long cable is broken out into 48 LEMO connectors at the bulkhead which the short RG174 cables from the detectors can plug into directly.
The signals of the HPGe and suppression shield are read out with two 16-conductor RG174 cable bundles with CPC connectors at the bulkhead. The long signal cables are broken out at both ends into 16 short individual RG174 cables with SMA plugs for connection to the clover or GRIF-16 digitizer module.

Each of the 20 PMTs of the SCEPTAR array of plastic scintillators is provided with high voltage by a separate RG59 cable with SHV connector. The signal from each PMT is read out on a 0.5\,m RG174 cable to a patch panel close to the spectrometer. The signal is amplified and shaped by a custom preamplifier and then connected to the GRIF-16 digitizer by a long RG58 cable.

Low voltage is supplied to the preamplifier of each PACES detector by a multi-conductor cable with 9-pin D-sub connectors. High voltage for the silicon diode is provided to the detector through the preamplifier and is supplied by a RG59 cable with a SHV connector. The output signal of each preamplifer is connected by a BNC connector to a RG58 cable and a SMA connector to the GRIF-16 digitizer.

High voltage is supplied to the PMT of each LaBr$_3$(Ce) detector and ancillary detector shield BGO crystal by a RG59 cables with a SHV connector. Low voltage is supplied to the LaBr$_3$(Ce) PMT base by a multi-conductor cable with 9-pin D-sub connectors, and to a preamplifier box for each ancillary detector Compton and background suppression shield in the same way. The signal used for fast timing is taken from the anode of the PMT connected to the NIM electronics with a LMR-400 low-loss cable. This same type of low-loss cable is also used for the signal from the fast $\beta$ scintillator. The dynode signal is connected by a BNC connector to a RG58 cable and a SMA connector to the GRIF-16 digitizer. The output signals from the ancillary detector Compton and background suppression shield preamplifier boxes are connected by RG174 cables with and SMA connectors to the GRIF-16 digitizer.

\subsection{Computing}
A series of rack-mounted computers for data collection and storage are located in the GRIFFIN electronics enclosure. One computer serves as the frontend machine and interface for MIDAS (Maximum Integration Data Acquisition System) \cite{MIDAS}. There is a disk server with capacity for several hundred terabytes of storage. This space is sufficient for the temporary storage of data collected in several experimental runs and there is significant room for future expansion. Typical experimental runs performed so far have been 2-10 days in length and collected between 1 and 15~TB of data each. Following an experiment, the data are copied to other locations for analysis and long-term storage.

\section{Description of Detector Hardware and Read-out}
\label{sec:hardware}
In this section the various detector systems of the GRIFFIN facility are described in detail. The in-vacuum ancillary detector systems are constructed as individual modular pieces of the vacuum chamber with common vacuum couplings. This approach allows for simple and rapid reconfiguration of the spectrometer to meet the requirements of the particular experimental study. The in-vacuum ancillary detector systems include an array of plastic scintillators for $\beta$ particle detection (SCEPTAR, Section \ref{sec:SCEPTAR}), a set of five in-vacuum LN$_2$-cooled lithium-drifted silicon detectors for conversion electron measurements (PACES, Section \ref{sec:PACES}), and a single fast-$\beta$ scintillator. The fast $\beta$ scintillator, also referred to as the zero-degree scintillator (ZDS) is used in conjunction with an array of eight LaBr$_3$(Ce) scintillators for fast-timing measurements (Section \ref{sec:LaBr3}). The DESCANT neutron array (Section \ref{sec:DESCANT}) can also be coupled with the GRIFFIN array with the removal of four HPGe clovers.
The combinations of ancillary detector systems presently available with the GRIFFIN spectrometer are listed in Table \ref{tab:configurations}.

\begin{table}[!htpb]
\centering
\caption{Combinations of ancillary detector systems that can be operated with the GRIFFIN spectrometer.}
\label{tab:configurations}
\small
\begin{tabular}{cccc}
\hline
HPGe & Ancillary & Upstream & Downstream \\
Clovers & positions & chamber & chamber \\
\hline
16 & 8 LaBr$_3$(Ce) & SCEPTAR & SCEPTAR \\ 
16 & 8 LaBr$_3$(Ce) & SCEPTAR & ZDS \\ 
15 & 8 LaBr$_3$(Ce) & PACES & SCEPTAR \\ 
15 & 8 LaBr$_3$(Ce) & PACES & ZDS \\ 
\\
\multicolumn{4}{l}{With the DESCANT array:}\\
12 & 4 LaBr$_3$(Ce) & SCEPTAR & SCEPTAR \\ 
12 & 4 LaBr$_3$(Ce) & SCEPTAR & ZDS \\ 
11 & 4 LaBr$_3$(Ce) & PACES & SCEPTAR \\ 
11 & 4 LaBr$_3$(Ce) & PACES & ZDS \\ 
\hline
\end{tabular}
\end{table}

\subsection{Experiment Cycling Mode}
\label{sec:cycles}
The radioactive beam from TRIUMF-ISAC-I is implanted into the tape of the GRIFFIN moving tape collector system at the central focus of the detector arrays. During data collection, the tape is operated either in a continuous tape moving mode where the tape moves at a constant speed of 26.3(1)\,cm/s, or in a series of cycles that involve a fast tape movement followed by a period where the tape is stationary. The fast movements are either a short move of 94\,cm in 996\,ms, or a long move of 165\,cm in 1320\,ms. Each of these tape movements is dominated by ramping-up and ramping-down of the tape-moving speed in a controlled manner so as to minimize stress on the tape and the chance of breakage. The short move brings the implantation point to the centre of a thick steel beam pipe just in front of the Pb wall (visible in Figure \ref{fig:GRIFFIN_overview}). The long move brings the implantation point to about 10\,cm behind the Pb wall.

A typical cycle involves a tape movement, background data collection, data collection while the radioactive beam is being delivered, and finally a period of data collection with the radioactive beam blocked. Blocking of the beam is achieved by an electrostatic kicker located in the ISAC basement level well away from the experimental hall. The relative duration of each stage within the cycle can vary between tens of milliseconds and tens of minutes in different experiments, depending on the half-lives and relative beam intensities of the isotopes involved, as well as the goal of the measurement. Control of this cycling mode is coordinated by the GRIFFIN digital data acquisition system \cite{Garnsworthy2017} and can be easily modified by the user for cycle optimization.

\subsection{Detector Coordinate System}
\label{sec:coordinate}
The GRIFFIN beamline is installed in the ISAC-I experimental hall in a north-south orientation and delivers the radioactive ion beam to GRIFFIN towards the south. The two hemispheres of the mechanical support structure open and close in the east-west direction.

The beam implantation location on the tape of the moving tape collector is at the origin of the GRIFFIN coordinate system.
Downstream refers to a vector for which the component along the beam direction is positive. For example, the tape box is located downstream of the implantation position. Upstream has a negative beam-direction component.

The GRIFFIN coordinate system is defined by the polar vector $\vec{r}$=($r$,$\theta$,$\phi$):\\

\noindent
$r$ is the distance from the origin to the front face of the detector element,\\
$\theta$=0$^{\circ}$ is along the beamline in the downstream (south) direction, \\
$\theta$=$\pi$/2=90$^{\circ}$, $\phi$=0 points horizontally east,\\
$\theta$=$\pi$/2=90$^{\circ}$, $\phi$=$\pi$/2=90$^{\circ}$ points straight up,\\
$\theta$=$\pi$/2=90$^{\circ}$, $\phi$=$\pi$=180$^{\circ}$ points horizontally west,\\
$\theta$=$\pi$/2=90$^{\circ}$, $\phi$=3$\pi$/2=270$^{\circ}$ points straight down,\\
$\theta$=$\pi$=180$^{\circ}$ points upstream (north).\\

Detector position numbering is a function of physical location in the coordinate system and for a detector or detector element represents the spatial centre of the front face of the sensitive volume of the element. The detector positions of a specific type, such as the HPGe clover detectors, are numbered sequentially by increasing $\phi$ angle within each ring of increasing $\theta$ angle.

In the discussion of angular correlations of particles, the relative angular difference between two different detectors in the spectrometer will be examined and is defined as $\Theta$.

\subsection{HPGe Clovers}
\label{sec:HPGe}
\begin{figure}
\centering
\includegraphics[width=1.0\linewidth]{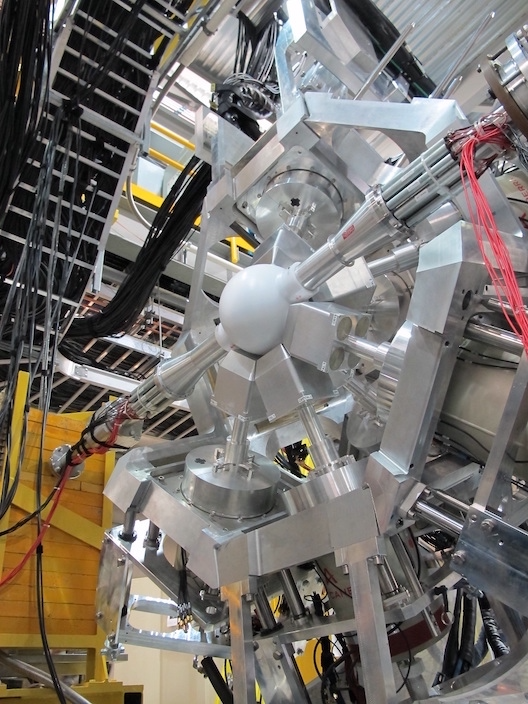}
\caption{The west hemisphere of the GRIFFIN array of HPGe clover detectors. The 20\,mm thick white Delrin absorber shell can be seen around the vacuum chamber and the Pb wall shielding the tape box is on the left.}
\label{fig:GRIFFIN}
\end{figure}

GRIFFIN is composed of sixteen high-purity germanium (HPGe) clover detectors arranged into a close-packed array. The detectors themselves are described in detail in Reference \cite{Rizwan2016}. Sixteen of the eighteen square faces of a rhombicuboctahedron are covered by HPGe at a source-to-detector distance of 11\,cm. One of the remaining square faces is used for the delivery of the low-energy radioactive ion beam from TRIUMF-ISAC-I and the final square is used for the in-vacuum moving tape collector system to remove long-lived activity from the chamber at the end of a measurement cycle. A photograph of the west hemisphere of the GRIFFIN spectrometer is shown in Figure \ref{fig:GRIFFIN}.
This detector arrangement is in a similar geometry to the TIGRESS spectrometer \cite{Svensson2005,Hackman2014} operating in the ISAC-II experimental hall for studies with accelerated radioactive beams. Table \ref{tab:HPGe_coordinates} lists the coordinate of the centre of each HPGe crystal in the GRIFFIN array with respect to the beam axis.

\begin{table}[!htpb]
\centering
\caption{Coordinates of the centre of each HPGe crystal in the GRIFFIN array at source-to-detector distances of 11 and 14.5\,cm. The coordinate system is defined in Section \ref{sec:coordinate}.}
\label{tab:HPGe_coordinates}
\scriptsize
\begin{tabular}{cccccc}
\hline
Array & Crystal & \multicolumn{2}{c}{11\,cm} & \multicolumn{2}{c}{14.5\,cm} \\
Position & Position & $\theta$ & $\phi$ & $\theta$ & $\phi$\\
& & (deg) & (deg) & (deg) & (deg) \\
\hline
1 & 1 & 36.5 & 83.4 & 37.9 & 80.1 \\
& 2&	55.1 & 79.0 & 53.2 & 77.2 \\
& 3&	55.1 & 56.0 & 53.2 & 57.8 \\
& 4&	36.5 & 51.6 & 37.9 & 54.9 \\
\hline
2 & 1 & 36.5 & 173.4 & 37.9 & 170.1\\
& 2 & 55.1 & 169.0 & 53.2 & 167.2\\
& 3 & 55.1 & 146.0 & 53.2 & 147.8\\
& 4 & 36.5 & 141.6 & 37.9 & 144.9\\
\hline
3& 1 & 36.5 & 263.4 & 37.9 & 260.1\\
& 2 & 55.1 & 259.0 & 53.2 & 257.2\\
& 3 & 55.1 & 236.0 & 53.2 & 237.8\\
& 4 & 36.5 & 231.6 & 37.9 & 234.9\\
\hline
4 & 1 & 36.5 & 353.4 & 37.9 & 350.1\\
& 2 & 55.1	 & 349.0 & 53.2 & 347.2\\
& 3 & 55.1	 & 326.0 & 53.2 & 327.8\\
& 4 & 36.5	 & 321.6 & 37.9 & 324.9\\
\hline
5 & 1 & 80.6 & 32.0 & 82.3 & 30.3\\
& 2 & 99.4 & 32.0 & 97.7 & 30.3\\
& 3 & 99.4 & 13.0 & 97.7 & 14.7\\
& 4 & 80.6 & 13.0 & 82.3 & 14.7\\
\hline
6 & 1 & 80.6 & 77.0 & 82.3 & 75.3\\
& 2 & 99.4 & 77.0 & 97.7 & 75.3\\
& 3 & 99.4 & 58.0 & 97.7 & 59.7\\
& 4 & 80.6 & 58.0 & 82.3 & 59.7\\
\hline
7 & 1 & 80.6 & 122.0 & 82.3 & 120.3\\
& 2 & 99.4 & 122.0 & 97.7 & 120.3\\
& 3 & 99.4 & 103.0 & 97.7 & 104.7\\
& 4 & 80.6 & 103.0 & 82.3 & 104.7\\
\hline
8 & 1 & 80.6 & 167.0 & 82.3 & 165.3\\
& 2 & 99.4 & 167.0 & 97.7 & 165.3\\
& 3 & 99.4 & 148.0 & 97.7 & 149.7\\
& 4 & 80.6 & 148.0 & 82.3 & 149.7\\
\hline
9 & 1 & 80.6 & 212.0 & 82.3 & 210.3\\
& 2 & 99.4 & 212.0 & 97.7 & 210.3\\
& 3 & 99.4 & 193.0 & 97.7 & 194.7\\
& 4 & 80.6 & 193.0 & 82.3 & 194.7\\
\hline
10 & 1 & 80.6 & 257.0 & 82.3 & 255.3\\
& 2 & 99.4 & 257.0 & 97.7 & 255.3\\
& 3 & 99.4 & 238.0 & 97.7 & 239.7\\
& 4 & 80.6 & 238.0 & 82.3 & 239.7\\
\hline
11 & 1 & 80.6 & 302.0 & 82.3 & 300.3\\
& 2 & 99.4 & 302.0 & 97.7 & 300.3\\
& 3 & 99.4 & 283.0 & 97.7 & 284.7\\
& 4 & 80.6 & 283.0 & 82.3 & 284.7\\
\hline
12 & 1 & 80.6 & 347.0 & 82.3 & 345.3\\
& 2 & 99.4 & 347.0 & 97.7 & 345.3\\
& 3 & 99.4 & 328.0 & 97.7 & 329.7\\
& 4 & 80.6 & 328.0 & 82.3 & 329.7\\
\hline
13 & 1 & 124.9 & 79.0 & 126.8 & 77.2\\
& 2 & 143.5 & 83.4 & 142.1 & 80.1\\
& 3 & 143.5 & 51.6 & 142.1 & 54.9\\
& 4 & 124.9 & 56.0 & 126.8 & 57.8\\
\hline
14 & 1 & 124.9 & 169.0 & 126.8 & 167.2\\
& 2 & 143.5 & 173.4 & 142.1 & 170.1\\
& 3 & 143.5 & 141.6 & 142.1 & 144.9\\
& 4 & 124.9 & 146.0 & 126.8 & 147.8\\
\hline
15 & 1 & 124.9 & 259.0 & 126.8 & 257.2\\
& 2 & 143.5 & 263.4 & 142.1 & 260.1\\
& 3 & 143.5 & 231.6 & 142.1 & 234.9\\
& 4 & 124.9 & 236.0 & 126.8 & 237.8\\
\hline
16 & 1 & 124.9 & 349.0 & 126.8 & 347.2\\
& 2 & 143.5 & 353.4 & 142.1 & 350.1\\
& 3 & 143.5 & 321.6 & 142.1 & 324.9\\
& 4 & 124.9 & 326.0 & 126.8 & 327.8\\
\hline
\end{tabular}
\end{table}

The gap between the vacuum chamber with an outer radius of 89\,mm and the GRIFFIN HPGe detectors can optionally be filled by a spherical shell of Delrin plastic with a thickness of 10 or 20\,mm. The purpose of this low-$Z$ absorber is to stop energetic $\beta$ particles from reaching the HPGe detectors while minimizing the production of bremsstrahlung background photons. The use of this absorber reduces the $\gamma$-ray detection efficiency at low energies (see Section \ref{sec:HPGeAbsEff} and Figure \ref{fig:lowEEffDelrin}), but the reduction in background is often of more benefit in the study of nuclei populated in a $\beta$ decay with a large $Q$ value.

\subsection{SCEPTAR}
\label{sec:SCEPTAR}
The principal ancillary detector system used in the GRIFFIN facility is the Scintillating Electron Positron Tagging ARray (SCEPTAR) consisting of 20 BC404 plastic scintillators located inside the vacuum chamber \cite{Garnsworthy2015}. These detectors subtend roughly 80\% of the solid angle and are used to detect $\beta$ particles emitted in the decay of a parent nucleus. 

\begin{figure}
\centering
\includegraphics[width=1.0\linewidth]{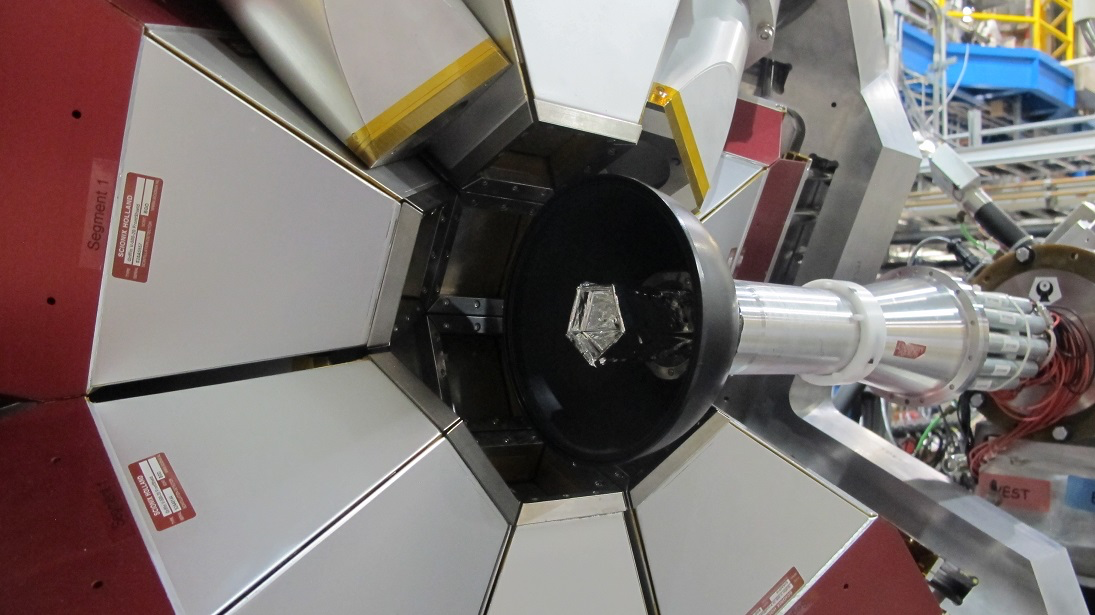}
\caption{The upstream hemisphere of the SCEPTAR array of plastic scintillators for $\beta$ tagging and the west hemisphere of the Compton-suppressed GRIFFIN spectrometer.}
\label{fig:SCEPTAR}
\end{figure}

\begin{table}[!htpb]
\centering
\caption{Coordinates of the centre of each SCEPTAR paddle which are at a source-to-detector distance of 3\,cm. The coordinate system is defined in Section \ref{sec:coordinate}.}
\label{tab:SCEPTAR_coordinates}
\small
\begin{tabular}{ccc}
\hline
SCEPTAR   &	$\theta$  & $\phi$  \\
Paddle  &	(deg)   & (deg)  \\
\hline
1& 37.4 & 18.0 \\
2& 37.4 & 90.0 \\
3& 37.4 & 162.0 \\
4& 37.4 & 234.0 \\
5& 37.4 & 306.0 \\
6& 79.2 & 18.0 \\
7& 79.2 & 90.0 \\
8& 79.2 & 162.0 \\
9& 79.2 & 234.0 \\
10& 79.2 & 306.0 \\
\\
11& 100.8 & 18.0 \\
12& 100.8 & 90.0 \\
13& 100.8 & 162.0 \\
14& 100.8 & 234.0 \\
15& 100.8 & 306.0 \\
16& 142.6 & 18.0 \\
17& 142.6 & 90.0 \\
18& 142.6 & 162.0 \\
19& 142.6 & 234.0 \\
20& 142.6 & 306.0 \\
\hline
\end{tabular}
\end{table}

The twenty trapezoidal paddles are arranged in four pentagonal rings concentric with the beam axis. Table \ref{tab:SCEPTAR_coordinates} lists the coordinates of the centre of each SCEPTAR paddle with respect to the beam axis. This geometry matches that of the 8$\pi$ spectrometer \cite{Garnsworthy2015} and originally there was a one-to-one correspondence between scintillator and HPGe crystal for the purpose of a bremsstrahlung veto with the associated HPGe crystal. 
The two upstream rings are shown in Figure \ref{fig:SCEPTAR}.
Light produced in the 1.5\,mm thick, trapezoidal-shaped scintillator is collected by a 1.5\,mm thick ultra-violet transmitting (UVT) acrylic light guide that is contoured and glued to a 1\,cm diameter UVT acrylic light guide rod. The light guide rod is integrated with the vacuum chamber as an optical vacuum feedthrough.
A ring of ten 14\,mm diameter Hamamatsu H3165-10 photomultiplier tubes (PMTs) assemblies sit around the outside of the beam pipe, coupled to the light-guide rods on the atmospheric side of the optical feedthrough flange located 40\,cm from the centre of the array.

High voltage is provided separately to each PMT from a CAEN A1733N module. The output signal from each PMT is amplified and shaped by a custom preamplifier so the signal is of a suitable form to be processed by the 14-bit, 100\,Msamples/s GRIF-16 digitizer \cite{Garnsworthy2017}. In the future, the signal directly from the PMT could be processed using a digitizer with a higher sampling rate.

\subsection{PACES}
\label{sec:PACES}
\begin{figure}
\centering
\includegraphics[width=1.0\linewidth]{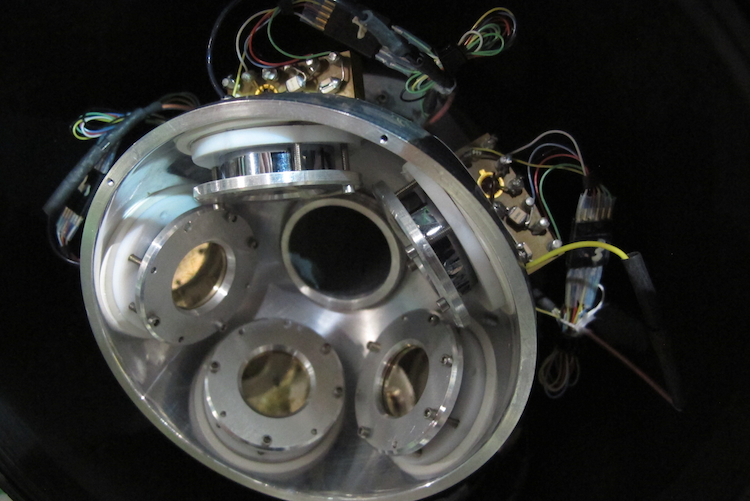}
\caption{The five Si(Li) detectors of the PACES array for internal conversion electron spectroscopy viewed from downstream. Components of the first stage of the preamplifier for each detector can be seen on the rear side of the cold finger.}
\label{fig:PACES}
\end{figure}

The Pentagonal Array of Conversion Electron Spectrometers (PACES) is a set of five cryogenically-cooled lithium-drifted silicon [Si(Li)] detectors located inside the vacuum chamber for internal conversion electron spectroscopy. A photograph of the detectors is shown in Figure \ref{fig:PACES}.
The field-effect transistor (FET) for each counter is also located inside the vacuum chamber on the same cold finger as the detectors.
The remainder of the preamplifier is mounted on the outside of the chamber on a vacuum feedthrough flange located 43\,cm from the array centre. It is a modified Canberra 2002C type with BNC and SHV connectors for signals and high voltage. Low-voltage preamplifier power is provided on a 9-pin D-sub connector to each module from a dedicated channel of a Mesytec MHV-4 4-channel NIM voltage supply. The output signals from the PACES preamplifiers are processed by a GRIF-16 digitizer within the GRIFFIN DAQ system \cite{Garnsworthy2017}.

Each cylindrical silicon diode is 200\,mm$^2$ and of 5\,mm thickness. There is a thin coating of gold over the entire crystal surface. Each crystal is mounted in a custom mount integrated with the cold finger. The detectors are arranged at approximately 72-degree intervals around the beam axis on the upstream hemisphere. Table \ref{tab:PACES_coordinates} lists the coordinates of the centre of each Si(Li) crystal in the PACES array with respect to the beam axis. The bias voltage of -500\,V supplied from a CAEN A1733N module is provided separately to each diode through an aluminum ring in contact with the outer radius of the front surface and the diode is grounded through a spring-mounted contact pin located at the centre of the rear surface. The front biasing ring provides a small amount of collimation to the detector so that the region visible to incident electrons is 198\,mm$^2$. 
The detector is mounted in such a way that it is tilted by 24\,$^{\circ}$ with respect to a direct observation of the implantation point on the tape. As a result, the solid angle is not defined by a circle, but by an ellipse with axis diameters of 15.9\,mm and 13.4\,mm corresponding to an area of 167\,mm$^2$. The implantation point to centre-of-detector surface distance is 31.5\,mm, so the solid angle coverage is 1.48\% of 4$\pi$ for each detector and 7.4\% of 4$\pi$ in total.

\begin{table}[!htpb]
\centering
\caption{Coordinates of the centre of each crystal in the PACES array which are at a source-to-detector distance of 3.15\,cm. The coordinate system is defined in Section \ref{sec:coordinate}.}
\label{tab:PACES_coordinates}
\small
\begin{tabular}{ccc}
\hline
PACES   &	$\theta$  & $\phi$  \\
Detector  &	(deg)   & (deg)  \\
\hline
1& 120.0 & 21.0 \\
2& 120.0 & 94.0 \\
3& 120.0 & 166.0 \\
4& 120.0 & 237.0 \\
5& 120.0 & 313.0 \\
\hline
\end{tabular}
\end{table}

Cooling of the detectors and FETs is achieved by a cold finger that is in direct contact with liquid nitrogen provided by a drip-feed dewar. The HPGe detector of position 13 must be removed to accommodate the PACES dewar.
Cooling of the PACES detectors from room temperature takes 8 hours to reach an equilibrium situation in which the crystals are held at a temperature of $\sim$148\,K.
The cryogenic pumping provided by the cold finger reduces the pressure in the chamber by almost one order of magnitude.

\subsection{LaBr$_3$(Ce) and Fast $\beta$ Scintillator}
\label{sec:LaBr3}
\begin{figure}
\centering
\includegraphics[width=1.0\linewidth]{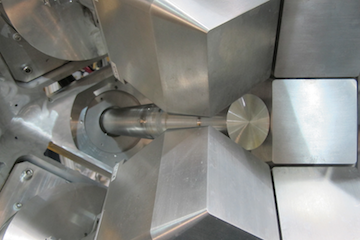}
\caption{A LaBr$_3$(Ce) detector in one of the ancillary detector positions of the west hemisphere of GRIFFIN.}
\label{fig:Labr3}
\end{figure}

In addition to the 18 square faces for HPGe and beam entry/tape removal, the rhombicuboctahedral geometry of GRIFFIN provides 8 triangular faces which can be used for ancillary detectors. The principal ancillary detector employed in these positions is a set of cerium-doped lanthanum bromide (LaBr$_3$(Ce)) scintillators for fast-coincidence-timing measurements of $\gamma$ rays. A photograph showing one of these detectors mounted in the west hemisphere of the array can be seen in Figure \ref{fig:Labr3}. Table \ref{tab:LaBr_coordinates} lists the coordinates of the centre of each LaBr$_3$(Ce) crystal in the array with respect to the beam axis.

The detectors are manufactured by St.~Gobain Crystals (BrilLanCe(380), Model number: 51-S-21/2) providing a single closed aluminum unit containing the crystal and PMT (Model number: R2083). The cylindrical crystal has a diameter and length of 5.1\,cm and has a Ce doping level of $\sim 5\%$. A removable preamplifier base (Model number: AS2612) is also used providing two output signals as a BNC connector from the dynode and anode. 
A SHV connector is used for high-voltage supply to the PMT and a 9-pin D-sub connector for low-voltage supply to the base. The anode output signal is used for fast-timing purposes. The dynode signal is used for an energy measurement. A custom modification has been added to the dynode signal by the addition of a second fast amplification stage in order to stabilize the signal at high counting rates before it is presented to the GRIF-16 digitizer \cite{Garnsworthy2017}.

A fast $\beta$ scintillator, also referred to as the zero-degree scintillator (ZDS), can be mounted in the vacuum chamber at 0$^\circ$ to the beam axis. This detector is a circular BC422Q plastic scintillator of 25\,mm diameter and 1\,mm thickness. It is coupled directly to a Hamamatsu H6533 PMT located entirely in vacuum. The whole assembly is mounted on a movable rod that allows the distance from the source location to be adjusted without the need for a vacuum cycle. The closest position is within a few millimeters from the tape subtending a solid angle of approximately 25\% of 4$\pi$. The scintillator can be retracted to a distance of around 5\,cm downstream of the tape so that the count rate can be lowered to an acceptable level without decreasing the amount of delivered beam activity. The motion, high-voltage and signal feedthroughs are located in a small box, the centre of which is located 75\,cm downstream of the array centre. The signal from the PMT is split using a Texas Instruments 8\,GHz Ultra Wideband Fully Differential Amplifier (Model number: LMH5401EVM) to provide separate signals for timing and energy evaluation without a reduction of signal quality.

A set of NIM-based electronics is used for fast timing within the GRIFFIN setup. The anode signal from each LaBr$_3$(Ce) detector as well as the signal from the fast $\beta$ scintillator are used as input to an Ortec 935 Constant Fraction Discriminator (CFD). The output logic signal of the CFD is used in a series of Lecroy 429A fan-in/fan-out logic modules to derive coincident timing logic between specific pairs of detectors. These logic modules ultimately provide the start and stop input signals to an Ortec 566 Time-to-Amplitude Converter (TAC) module. Eight such modules are used for the timing between the various pairs of the eight LaBr$_3$(Ce) detectors and the single fast $\beta$ scintillator. The TAC outputs can be up to 10\,V in amplitude and are thus attenuated in a special rear-transition card before being digitized in a GRIF-16 digitizer \cite{Garnsworthy2017}.

The dynode signal from each LaBr$_3$(Ce) detector as well as the signal from the fast $\beta$ scintillator is input directly to a GRIF-16 digitizer for energy evaluation. Although there is a roughly 1.5\,$\mu$s latency between the digitized LaBr$_3$(Ce) energy and TAC signals they can be easily correlated in the off-line data analysis using timestamp differences.

The signals from an Ortec 462 Time Calibrator module operated at a low rate are also simultaneously input directly to the start and stop of each of the TAC modules. These signals are easily distinguished in the data analysis from the real detector coincidence signals due to a lack of correlation with energy events. These time calibrator signals can be used to monitor and correct for any system timing drifts due to effects such as temperature fluctuations.

\begin{table}[!htpb]
\centering
\caption{Coordinates of the centre of each LaBr$_3$(Ce) crystal in the GRIFFIN array which are at a source-to-detector distance of 12.5\,cm. When installed with a dedicated Compton and background suppression shield, the source-to-detector distance of the LaBr$_3$(Ce) is 13.5\,cm. The coordinate system is defined in Section \ref{sec:coordinate}.}
\label{tab:LaBr_coordinates}
\small
\begin{tabular}{ccc}
\hline
Array   &	$\theta$  & $\phi$  \\
Position  &	(deg)   & (deg)  \\
\hline
1& 54.7 & 22.5 \\
2& 54.7 & 112.5 \\
3& 54.7 & 202.5 \\
4& 54.7 & 292.5 \\
\\
5& 125.3 & 22.5 \\
6& 125.3 & 112.5 \\
7& 125.3 & 202.5 \\
8& 125.3 & 292.5 \\
\hline
\end{tabular}
\end{table}

\subsection{Compton and Background Suppression Shields}
The overall sensitivity of the spectrometer can be greatly improved by surrounding the HPGe detectors with suppression shields fabricated from a high-efficiency scintillator to detect $\gamma$ rays either escaping from, or entering through, the sides and backs of the HPGe crystals. 
Signals from these suppression shields can then be used to veto the unwanted signals that Compton-scattered events and environmental backgrounds produce in the HPGe detectors. GRIFFIN was thus designed from the outset to include a set of Compton and background suppression shields that are currently being incorporated into the spectrometer.

The suppression shields for the GRIFFIN HPGe $\gamma$-ray detectors are formed from three components; a side shield, a back catcher, and a front shield, each fabricated from the high-density scintillator bismuth germanate (BGO). The geometry follows the design of the TIGRESS suppression shields \cite{Schumaker2007}.
The side shields continue the 22.5-degree taper of the front of the GRIFFIN HPGe clover detector until a BGO thickness of 20\,mm is reached. Each of the four sides of the shield contain two optically-isolated BGO crystals to allow crystal-specific suppression of the HPGe detector. Each of these BGO crystals is read out by two 19\,mm PMTs for a total of 16 PMTs per side shield. The back catcher consists of four crystals of 37\,mm thick BGO (one behind each HPGe crystal), each of which is read out by two low-profile 25\,mm PMTs. The front shields are constructed of four separate tapered trapezoidal BGO plates of 10\,mm thickness, each consisting of two optically isolated crystals read out by two 13\,mm PMTs. The four front shield plates will be separately mounted to the GRIFFIN mechanical support structure on rods that will enable their insertion and withdrawal at 22.5\,$^{\circ}$ relative to the axis of the corresponding HPGe detector. 

This design was chosen in order to operate the GRIFFIN spectrometer in both a ``maximum efficiency'' and an ``optimal peak-to-total'' configuration for different experiments, and the desire to be able to switch between these configurations in a matter of hours as dictated by the particular experimental conditions. In the maximum efficiency configuration, the BGO front shields are withdrawn and the 16 HPGe clover detectors are close-packed around the target location with their front faces 110\,mm from the centre of the array (upper panel of Figure \ref{fig:GRIFFIN_overview}). In the optimal peak-to-total configuration, the HPGe detectors are withdrawn to 145\,mm and the BGO front shield plates are inserted to 110\,mm to form a complete suppression shield around each HPGe detector (lower panel of Figure \ref{fig:GRIFFIN_overview}).

Suppression shields can be installed around each of the LaBr$_3$(Ce) detectors mounted in the ancillary detector positions of the array. These shields have a cylindrical geometry with a 20\,mm thick, 110\,mm long annulus of BGO formed by 3 optically isolated crystals surrounding a central cylindrical bore of 65\,mm diameter to accommodate the GRIFFIN LaBr$_3$(Ce) detector. Each crystal is read out by two 19\,mm PMTs for a total of six per ancillary detector suppression shield unit. The front of the ancillary detector suppression shield is tapered at 12.8\,$^{\circ}$ in order to close-pack with the HPGe suppression shields within the triangular faces of the rhombicuboctahedral geometry. 

A high-density tungsten alloy collimator of 10\,mm thickness is mounted to the front face of both the LaBr$_3$(Ce) and HPGe suppression shields to reduce the incident flux of photons originating directly from the beam-implantation location which would otherwise cause false-coincidence events in the BGO crystals.

The performance of the GRIFFIN Compton and background suppression shields will be described in a subsequent publication.

\subsection{DESCANT}
\label{sec:DESCANT}
\begin{figure}
\centering
\includegraphics[width=1.0\linewidth]{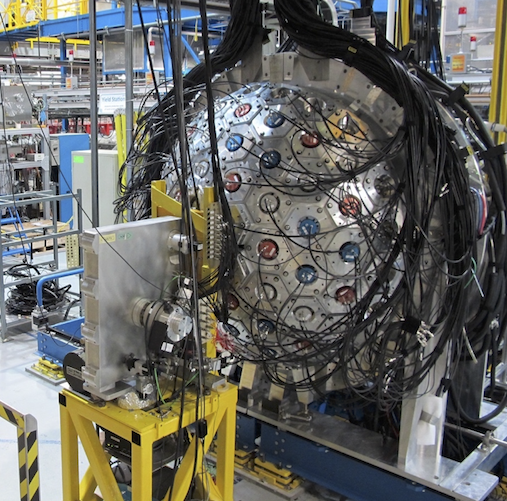}
\caption{A photograph of the DESCANT array of neutron detectors coupled to the GRIFFIN spectrometer. The tape box can be seen in the foreground.}
\label{fig:DESCANT}
\end{figure}

The deuterated scintillator array for neutron tagging (DESCANT) \cite{Garrett2014,Bildstein2015} is designed to be coupled with both the TIGRESS \cite{Hackman2014}, and GRIFFIN spectrometers. 
With GRIFFIN, it will be used for neutron tagging in studies of nuclei that undergo $\beta$-delayed neutron emission.
The 70 detector units each contain $\sim$2\,litres of deuterated benzene liquid scintillator, BC537, and were fabricated by the Bicron division of St. Gobain. The deuterated scintillator offers the opportunity to use pulse-shape discrimination (PSD) to distinguish between neutron and $\gamma$-ray interactions, as well as provides a relationship between incident neutron energy and signal pulse-height \cite{Bildstein2013}.
As can be seen in Figure \ref{fig:DESCANT}, the DESCANT array replaces the four downstream HPGe clover detectors, and covers a solid angle of 1.08\,$\pi$sr with a maximum angle of 65.5\,$^{\circ}$ with respect to the beam axis. The source-to-detector distance is 50\,cm and the individual detector cans are 15\,cm thick. The output signal from each PMT is processed directly using a fast-sampling digitizer integrated with the GRIFFIN DAQ system.
A detailed description of DESCANT can be found in References \cite{Garrett2014,Bildstein2015,Bildstein2013}.

\section{Data Analysis Techniques and Array Performance}
\label{sec:analysis-performance}
\subsection{HPGe Clovers}
When $\gamma$ rays are emitted from excited nuclear states, they possess several observable properties which can be used to reveal details of the nucleus. These observables are the energy, intensity, coincidence relations, spatial direction and the degree of polarization of the radiation. The 64 crystals of the 16 HPGe clover detectors of the GRIFFIN spectrometer arranged in a rhombicuboctahedral geometry around the source form a powerful arrangement which is ideally suited to measure all of these observable properties and elucidate features of the underlying nuclear structure.

\subsubsection{Cross-talk Correction}
\begin{figure}[!ht]
\centering
    \subfigure{
        {\includegraphics[width=\linewidth]{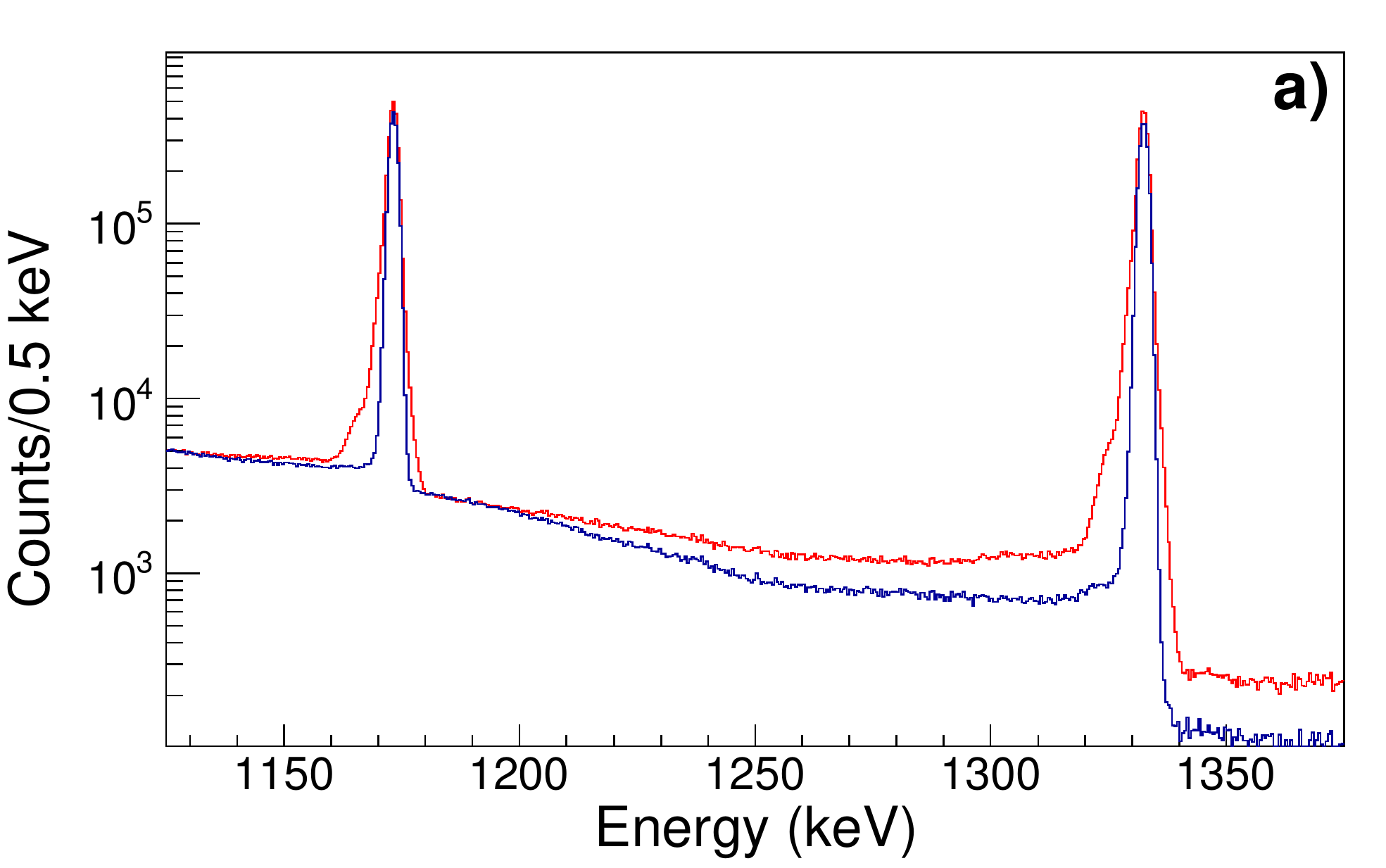}
        \label{fig:preCTsum}}
    }\\
    \subfigure{
        {\includegraphics[width=\linewidth]{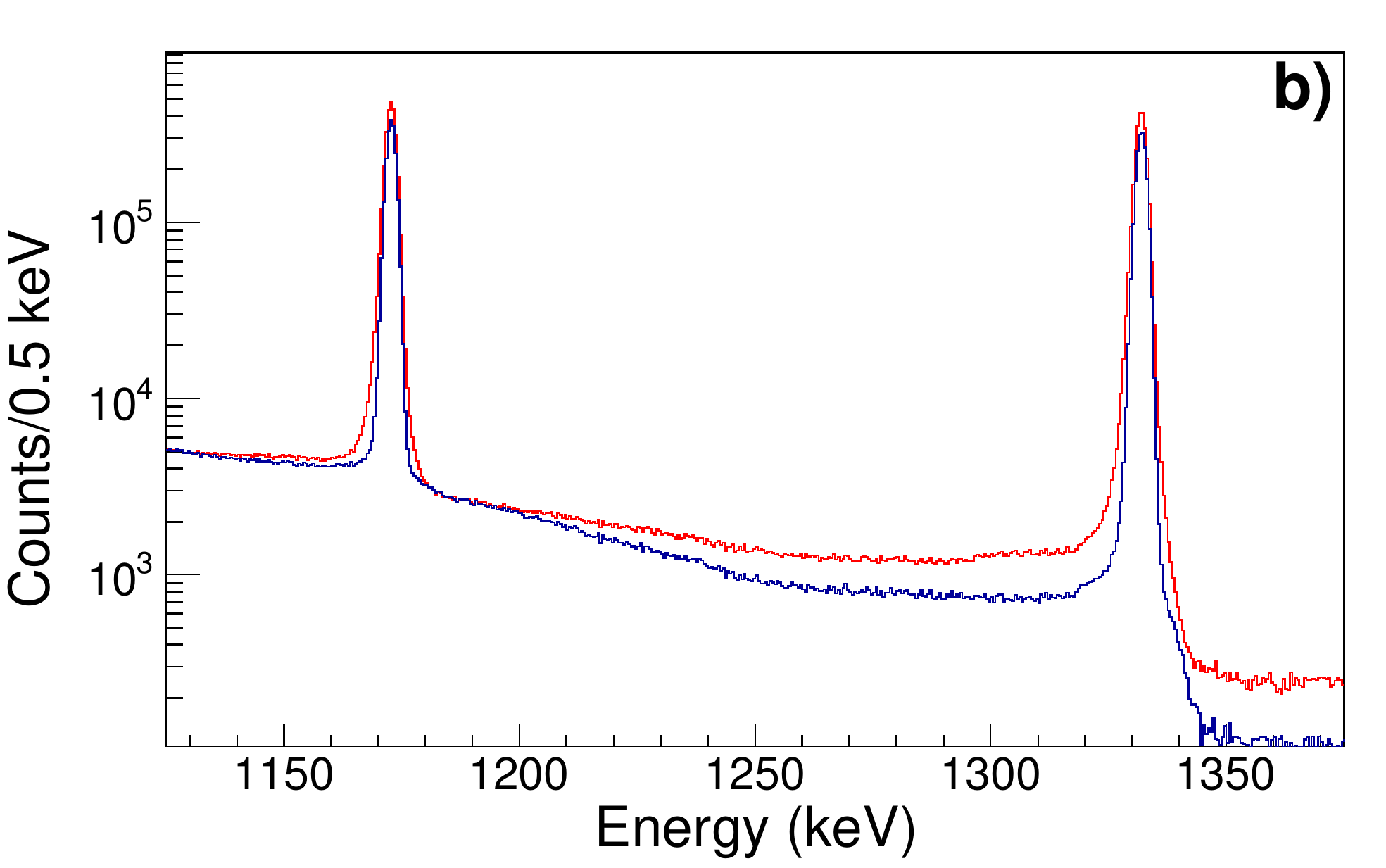}
        \label{fig:postCTsum}}
    }\\
\caption{(Blue) Singles and (Red) Addback $\gamma$-ray energy spectra of a $^{60}$Co source collected with the GRIFFIN HPGe clover detectors. The energy resolution in addback is reduced due to cross-talk in the upper panel, (a). In the lower panel (b) the cross-talk correction is applied to recover the energy resolution. See text for details.}
\label{fig:CTsum}
\end{figure}

When a $\gamma$ ray deposits energy in one crystal of a GRIFFIN clover, there is an observable modification of the apparent energy collected when a coincident, or scattered, $\gamma$ ray interacts in the other crystals of that clover. This so-called cross-talk is dependent on the energy deposited in the neighboring crystals of a clover. It is clear from Figure \ref{fig:preCTsum} that one must correct for the effects of cross-talk for an accurate determination of a peak area and centroid. This statement is especially true when using clover addback algorithms (Section \ref{sec:addback}) for the recovery of Compton-scattered $\gamma$ rays because these events are guaranteed to concurrently deposit energy into two crystals and be affected by cross-talk.

In order to account for cross-talk in the GRIFFIN setup, the following assumptions are made:
\begin{enumerate}
	\item The magnitude of the cross-talk effect is linearly proportional to the energy deposited in a neighboring crystal.
    \item The magnitude of the cross-talk effect is independent of the energy deposited in the crystal itself.
    \item The magnitude of the cross-talk effect in events involving three crystals is the direct sum of the cross-talk effect from each of the two crystals on the third.
\end{enumerate}

Based on these assumptions, for energy depositions in two crystals of a clover, one expects the following:
\begin{eqnarray}
E_1^\prime & = & E_1 + \alpha_{12}E_2 \nonumber \\
E_2^\prime & = & E_2 + \alpha_{21}E_1,
\end{eqnarray}
where $E_i$ represents the true energy deposited in the $i$th crystal, $E_i^\prime$ is the resulting cross-talk affected energy, and $\alpha_{ij}$ is the coefficient that determines how much the energy deposited in crystal $j$ changes the observed energy in crystal $i$. Note that $\alpha$ is assumed to be constant across the energy range.

\begin{figure}[!ht]
	\centering
	\includegraphics[width=\linewidth]{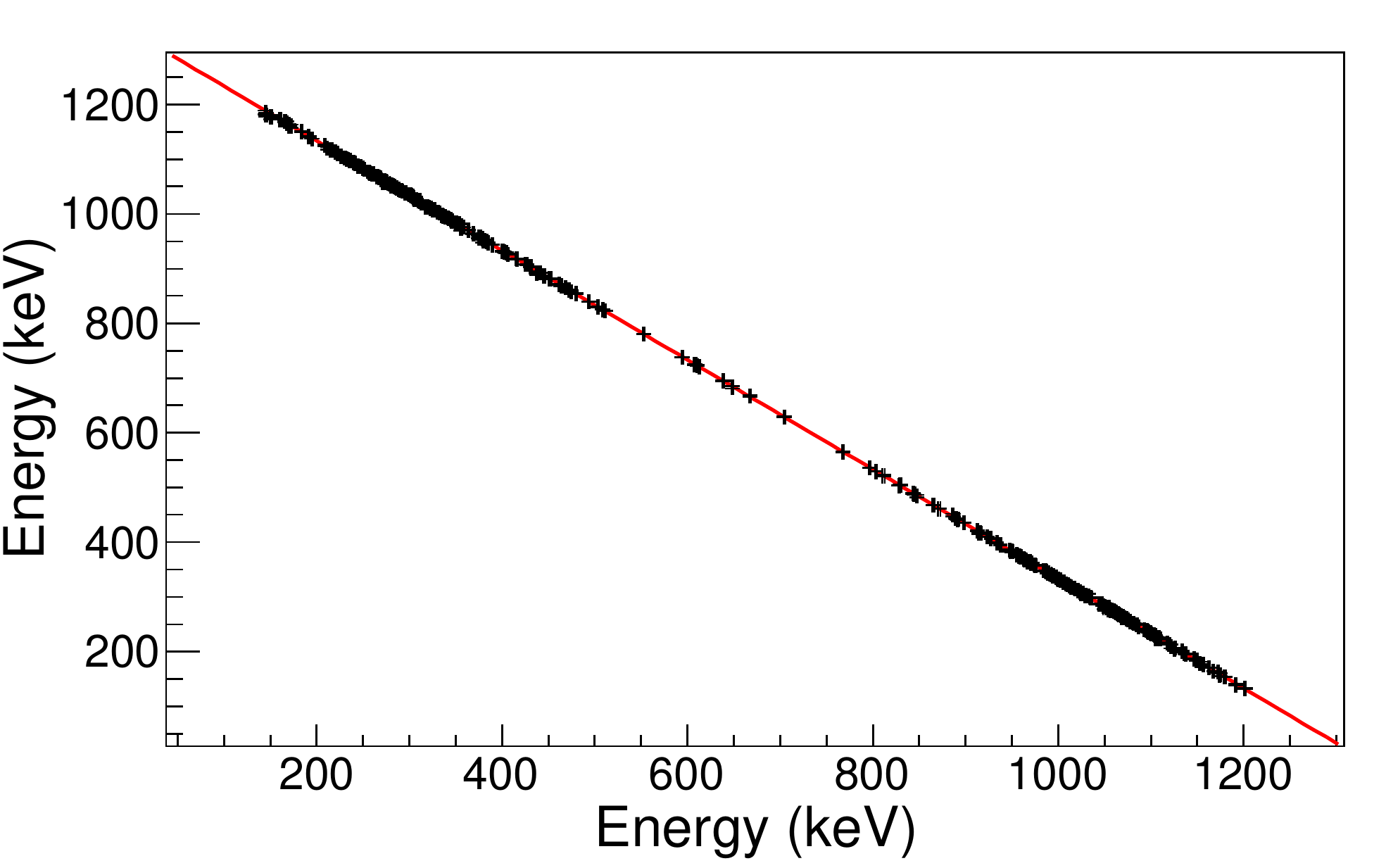}
    \caption{The fit to the Compton scattering of a 1332\,keV $\gamma$ ray used to determine the cross-talk coefficients for each crystal combination.} 
    \label{fig:fitCT}
\end{figure}

The $\alpha$ parameters in the above equations have been determined by taking advantage of Compton-scattered $\gamma$ rays that are completely detected within one clover. Such events create the characteristic ``scatter'' peaks in a $\gamma$-$\gamma$ energy coincidence matrix where the two energies, $E_1$ and $E_2$, sum to the full photo-peak energy, $E=E_1+E_2$.
This relationship can be used to define the following fit function:
\begin{equation}
E_2^\prime = E\frac{1-\alpha_{12}\alpha_{21}}{1-\alpha_{12}} -E_1^\prime\frac{1-\alpha_{21}}{1-\alpha_{12}} 
\end{equation}
The $\alpha$ coefficients for each crystal pairing can be determined by fitting the line generated by Compton scattering. As an example, the fit for Compton-scattering of the 1332~keV $\gamma$ ray from a $^{60}$Co source for the 1 and 2 crystals in a particular HPGe Clover detector is shown in Figure \ref{fig:fitCT}. Here the coefficients are found to be $\alpha_{12} = 0.0024$ and $\alpha_{21} = -0.0092$. This spectrum includes all pairs of coincident $\gamma$ rays that sum to $1332\pm15$~keV. The width of this coincident-energy-sum window has little effect on the values of the extracted coefficients. It is important to note that the cross-talk between different pairs of crystals can not only differ in magnitude, but also in sign. This observation emphasizes the need to determine the coefficients between every crystal pair combination within a clover.

Applying this cross-talk correction to each clover restores the excellent energy resolution of the HPGe detectors as shown in Figure \ref{fig:postCTsum}. Note that even after the cross-talk correction, the energy resolution for add-back events will always remain slightly worse than for single-crystal photo-peak events because the addback events include contributions from more than one crystal.

\subsubsection{Addback Methods}
\label{sec:addback}
\begin{figure}[!ht]
\centering
\includegraphics[width=1\linewidth]{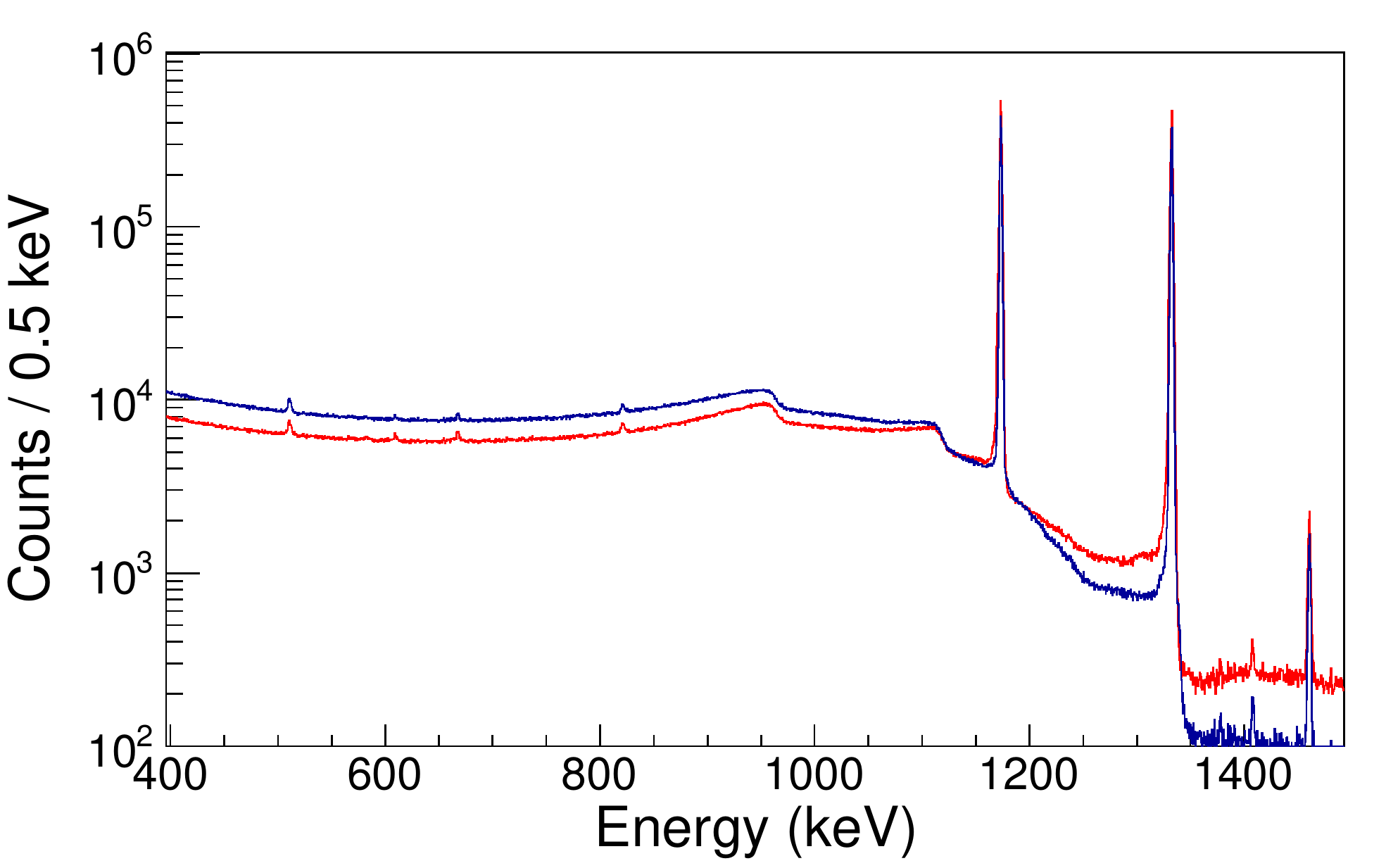}
\caption{\label{fig:peak-to-total} (Blue) Single crystal and, (Red) clover addback $\gamma$-ray energy spectra of a $^{60}$Co source collected with the GRIFFIN spectrometer with the unsuppressed HPGe detectors at a source-to-detector distance of 11\,cm. The peak-to-total ratio of the spectrum is improved due to the recovery of crystal-to-crystal Compton scattering into full energy photo-peaks. }
\end{figure}

The clover configuration of the GRIFFIN detector crystals allows for the application of various addback algorithms where the energies detected in multiple crystals are summed together. This process typically increases the photo-peak efficiency of the GRIFFIN spectrometer, while reducing the Compton scattered $\gamma$-ray background (Figure \ref{fig:peak-to-total}).

\begin{figure}[!ht]
\centering
\includegraphics[width=1\linewidth]{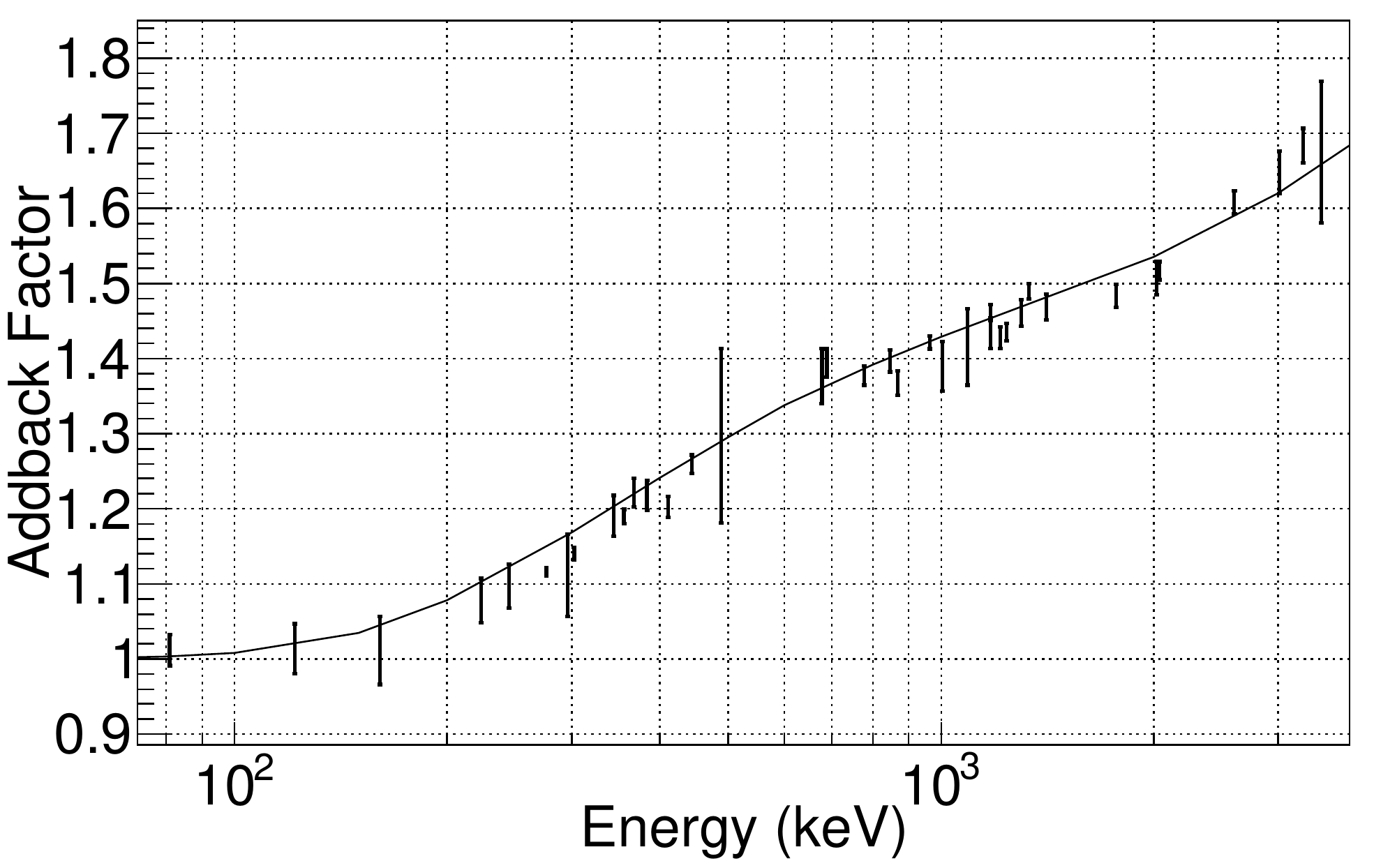}
\caption{\label{fig:ABFactor}The addback factor for 16 GRIFFIN clover detectors at a source-to-detector distance of 11\,cm. The data presented were determined using $^{133}$Ba, $^{56}$Co, $^{60}$Co, and $^{152}$Eu calibration sources. The line shows the simulated addback factor from GEANT4.
}
\end{figure}

The addback method presented here is the procedure by which the energy detected within a temporal coincidence window in each of the four crystals of a clover are added together. The relative amount that the efficiency is increased at a specific energy is defined as the ``addback factor''. The addback factor as a function of energy for 16 clovers at a source-to-detector distance of 11\,cm is shown for clover addback in Figure \ref{fig:ABFactor} along with the expectation from a GEANT4 simulation.

\subsubsection{Summing Corrections}
\label{sec:summing}
\begin{figure}[!ht]
\centering
\includegraphics[width=1\linewidth]{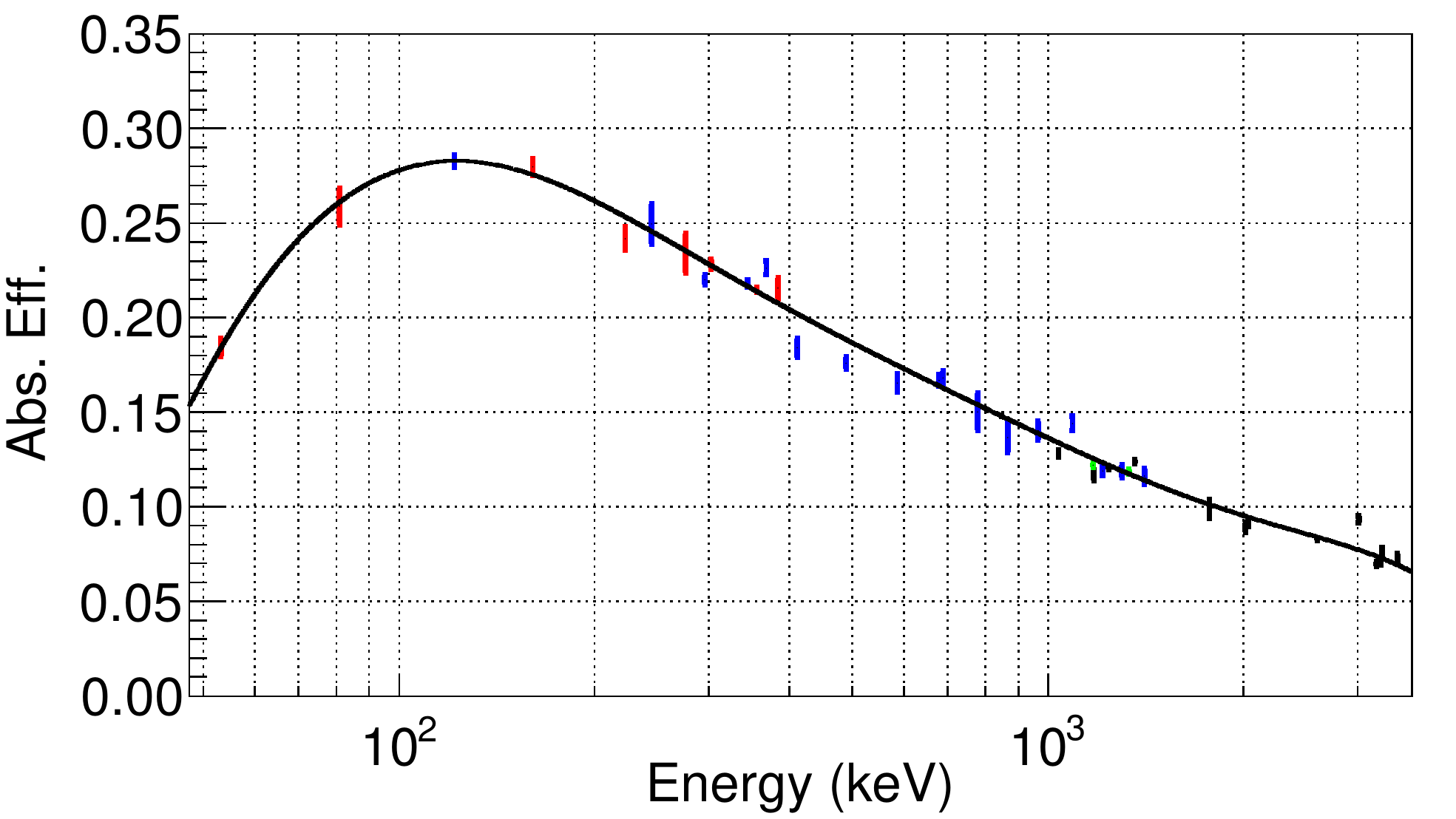}
\caption{\label{fig:ABEffNoSumming}The addback photo-peak efficiency for the full 16 clovers in the GRIFFIN array {\bf without summing corrections} applied. As can be seen, there is a very large amount of scatter in efficiencies without summing corrections. There is also an overall loss of efficiency due to the large probability of summing-out of each photopeak. {\it Color online:} (Red) $^{133}$Ba, (Blue)$^{152}$Eu, (Green) $^{60}$Co, (Black)$^{56}$Co.}
\end{figure}

The impressive $\gamma$-ray photo-peak efficiency of the GRIFFIN spectrometer is in part due to the large volume of the HPGe crystals, which also have large $\gamma$-ray interaction cross-sections. When multiple $\gamma$ rays are emitted in the same decay event, there is a non-zero probability that more than one $\gamma$ ray will interact with the same crystal. This is the process of coincidence summing and must be taken into account for an accurate determination of $\gamma$-ray intensities. 
When the full energy of one $\gamma$ ray is deposited in a HPGe crystal, and a second $\gamma$ ray interacts in the same crystal, there are two possible results:
\begin{enumerate}
	\item The full energy of the second $\gamma$ ray is not completely deposited in the crystal due to an interaction by Compton scattering or pair production. The full energy of the first $\gamma$ ray and the partial energy of the second $\gamma$ ray are summed together. This scenario lowers the number of counts observed for the first $\gamma$ ray that would have otherwise been detected at the full photo-peak energy.
    \item Both $\gamma$ rays are completely absorbed in the crystal. This process creates a sum peak at an energy equal to the sum of the two individual $\gamma$ rays and lowers the number of counts observed for the first and second $\gamma$ rays that would have otherwise been detected as their full photo-peak energies.
    In addition, this adds counts at the summed energy which could be the full photo-peak energy of an alternative decay branch from the upper energy level.
\end{enumerate}
Coincidence summing will not only affect the intensities measured in experiments but must also be taken into account when determining the absolute and relative efficiency of the detectors from the intensities of full photo-peaks observed from calibration sources. For example, Figure \ref{fig:ABEffNoSumming} shows the clover addback absolute efficiency data when summing effects have {\it not} been taken into account. When summing is not accounted for, there can be a large effect on the efficiency at each measured energy (from 1-10\% depending on the $\gamma$-ray multiplicity). The data points also display a large scatter around the fitted curve. As the probability of a $\gamma$-ray summing out of a photo-peak tends to be larger than the probability of summing in, there is effectively an overall drop in the absolute efficiency determined when summing effects are neglected.

We have developed an empirical method to correct for summing effects by taking advantage of the symmetry in the angular correlation about 90\,$^{\circ}$ of $\gamma$ rays emitted in the de-excitation of a nucleus. This symmetry means that the number of individual coincident gamma rays observed in a 180\,$^{\circ}$ crystal pair is equal to the number which is summed in a single crystal (i.e. an angular difference of 0\,$^{\circ}$ between the two emitted gamma rays).
The correction method involves constructing coincidence spectra between crystal pairs separated by 180\,$^{\circ}$. The summing out from the photo-peak of the $\gamma$ ray of interest should be equal, within counting statistics, to the total number of events in the 180$^{\circ}$ coincidence spectrum when gated on the $\gamma$ ray of interest. As all the HPGe crystals in GRIFFIN have a similar response, this coincidence spectrum will also accurately account for all events in which the second $\gamma$ ray does not deposit its full energy.

The level of summing into a photo-peak can be determined by examining the number of counts in each of the photo-peaks for the two individual $\gamma$ rays involved in the sum in the 180$^{\circ}$ coincidence spectra constructed by gating on each of the individual $\gamma$ rays.
The area measured in this case would be statistically equal to the number of events in which the two $\gamma$ rays deposited their full energy in the same detector. 
This method has the advantage that it can also be used to determine the summing contributions from the characteristic X rays that are emitted following electron capture and $\beta$ decay. This process is especially advantageous as the summing of X rays is in general difficult to correctly reproduce in simulation. Examples of the summing corrections are shown in Tables \ref{tab:sumout} and \ref{tab:sumin}.

\begin{table}[!t]
\begin{center}
    \caption{Examples of the contribution of summing out of a photo-peak for $\gamma$ rays from a $^{152}$Eu source with the HPGe detectors at a source-to-detector distance of 11\,cm using the SCEPTAR chamber. Clover addback is applied to the data. 
    } 
    \begin{tabular}{cccc}
    \hline
    Energy		& Area & Total $180^{\circ}$  & Sum out \\ 
    (keV) & $\times10^6$ & counts $\times10^4$ & \% \\
    \hline
    121.8 	& 5.35(12) & 31(3) & 5.48(12) \\ 
    344.3 	& 4.03(6) & 12.9(11) & 3.124(5) \\
    1408.0	& 1.844(10) & 6.55(19)  & 3.438(19) \\
    \hline
    \end{tabular}
    \label{tab:sumout}
\end{center}
\end{table}

In general, when using clover addback, there is a roughly 4 times higher probability to sum out of the photo-peak compared to the single-crystal photo-peaks. This observation can be explained by the roughly 4 times larger interaction cross-section when using all 4 crystals in a clover compared to a single crystal.

\begin{table}[!t]
\begin{center}
    \caption{Example of the contribution of summing into a photo-peak for the 366.5\,keV $\gamma$ ray in $^{152}$Eu with the HPGe at a source-to-detector distance of 11\,cm surrounding the SCEPTAR chambers. Clover addback is applied to the data.} 
    \begin{tabular}{ll}
    \hline
    Area of 244.7\,keV	& 1.3032(54)$\times10^6$ \\ \hline
    Efficiency at 121.8\,keV & 0.270(3) \\\hline
    Conversion coefficient $\alpha_{121}$ & 1.155 \\\hline
    Expected summing & 10201(121) \\\hline
    Using 180$^\circ$ coincidence & 9904(164) \\\hline
    Rel. summing into 366.5\,keV & 7.6(7)\%\\\hline
    \end{tabular}
    \label{tab:sumin}
\end{center}
\end{table}

As shown in Table~\ref{tab:sumin}, this empirical method performs well in determining the level of summing from experimental data. The 366.5\,keV photo-peak can be created with the sum of the 121.8\,keV and 244.7\,keV $\gamma$ rays. The expected summing in Table~\ref{tab:sumin} was calculated with the formula $A_{245}\frac{\epsilon_{122}}{16}\frac{1}{1+\alpha_{122}}$ where $\alpha_{122}$ is the total internal conversion coefficient of the 121.8\,keV transition.
It is important to note that this empirical method does not require extensive knowledge of the level scheme used to estimate the summing corrections. Using the efficiency method based on the known level scheme, if there were no knowledge of the conversion coefficient of the 121.8\,keV transition, the amount of summing would have been calculated to be 17\% of the total 366.5\,keV photo-peak intensity. This value is a very large overestimation and emphasizes the advantage of the empirical 180$^{\circ}$ coincidence-based method for summing corrections.

\subsubsection{Absolute Efficiency}
\label{sec:HPGeAbsEff}
The efficiencies of the GRIFFIN HPGe crystals are measured using standard radioactive sources placed in a source holder at the implantation position located in the center of the array. Typically, a relative efficiency curve is constructed using $^{133}$Ba, $^{152}$Eu and $^{56}$Co sources, which cover the $\gamma$-ray energy range from 0.05 to 3.5\,MeV. Following summing corrections described in Section \ref{sec:summing}, the efficiency data are then fit with either a piece-wise quadratic function in ln($\epsilon$) vs ln($E_\gamma$) \cite{Rizwan2015} or, for experiments also interested in high-energy ( $> \sim$3\,MeV) $\gamma$ rays, the 8th order polynomial function \cite{Kis1998}:

\begin{equation}
\text{ln}(\epsilon) = \sum^{8}_{i=0}a_i[\text{ln}(E_\gamma)]^i,\label{eq:efficiency_log}
\end{equation}

\noindent where $a_i$ are the fitted parameters and $E_\gamma$ is the $\gamma$-ray energy in MeV.

\begin{figure*}
\centering
    \subfigure{
        {\includegraphics[width=1.0\linewidth]{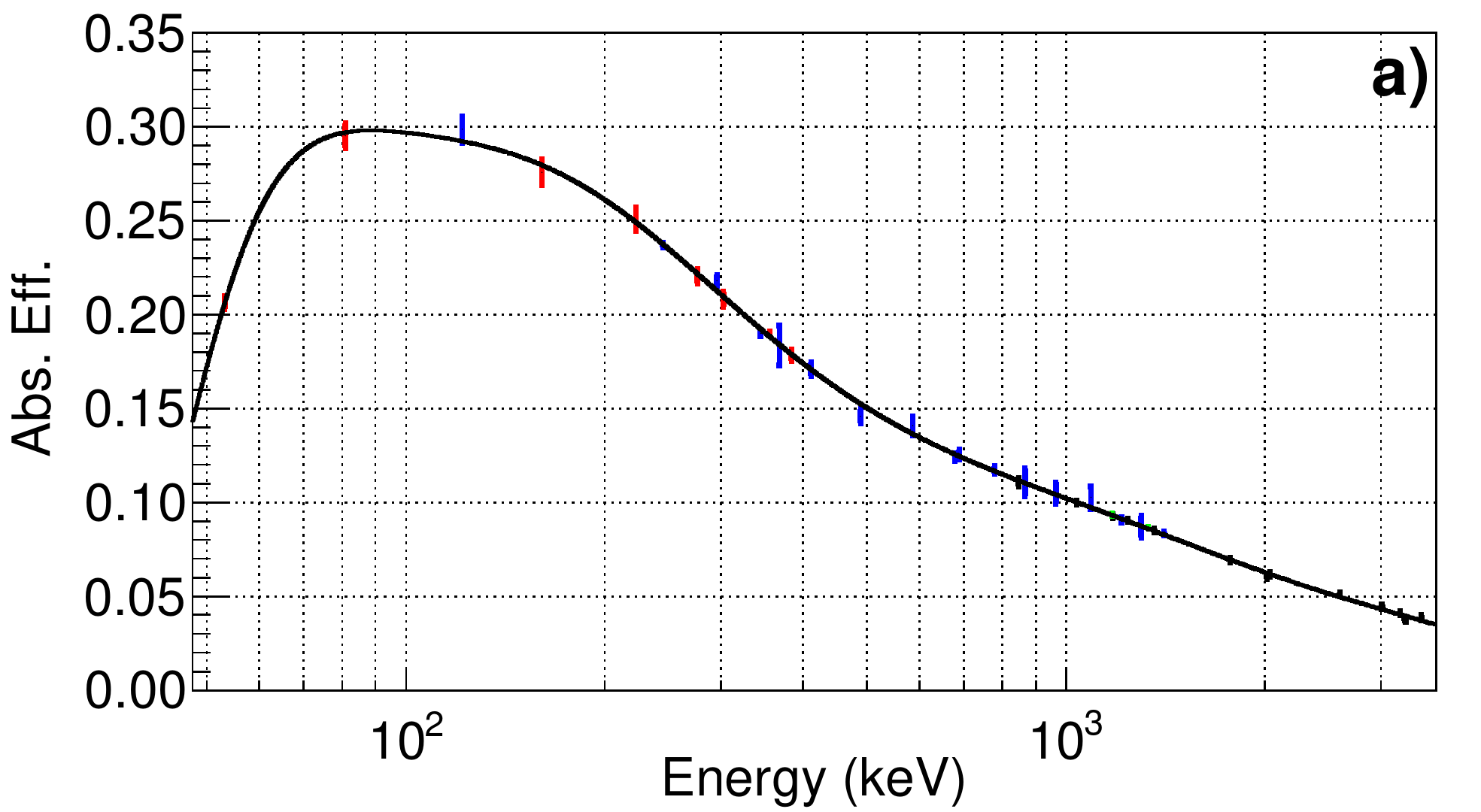}
        \label{fig:GeEff}}
    }\\
    \subfigure{
        {\includegraphics[width=1.0\linewidth]{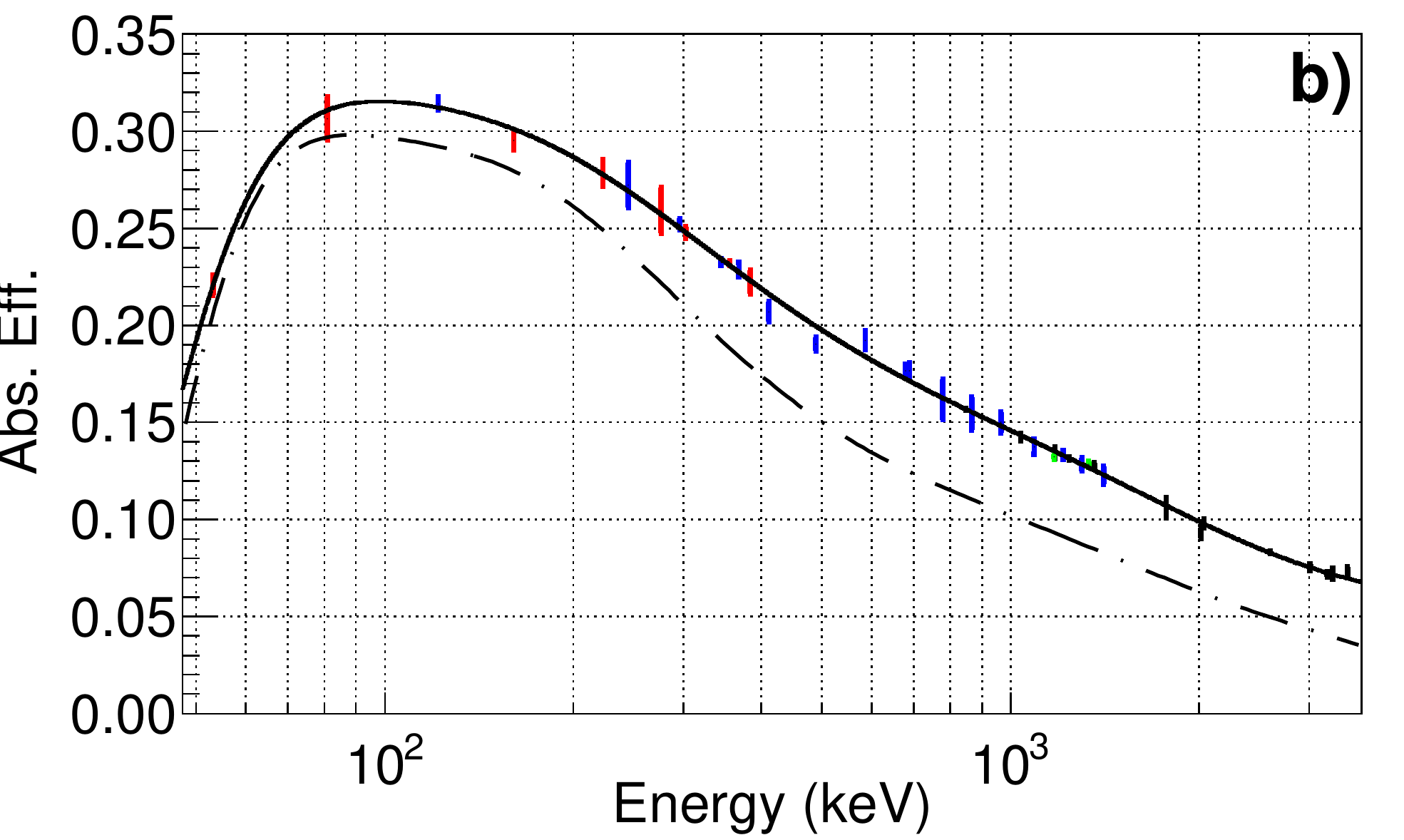}
        \label{fig:ABEff}}
    }\\
\caption{The absolute photo-peak efficiency {\bf a):} from 60 individual crystals, and {\bf b):} Clover-addback mode for 15 clovers in the GRIFFIN spectrometer. The HPGe detectors were at a source-to-detector distance of 11\,cm with no Delrin absorber in combination with the PACES and fast $\beta$ scintillator ancillary detectors. The data have been corrected for summing effects. The dashed line represents the fit to the individual crystal efficiencies for comparison purposes. {\it Color online:} (Red) $^{133}$Ba, (Blue)$^{152}$Eu, (Green) $^{60}$Co, (Black)$^{56}$Co.}
\label{fig:GeEfficiency}
\end{figure*}

The absolute photo-peak efficiency is measured with a $^{60}$Co source with an activity certified to $\pm 1\%$. Summing corrections are also applied to the $^{60}$Co data. The relative efficiency curve is scaled to the absolute efficiency measurements using the $^{60}$Co source. Figure \ref{fig:GeEff} shows the GRIFFIN absolute photo-peak efficiency curve for HPGe detectors at 11\,cm around the PACES and fast $\beta$ scintillator chambers for the $\gamma$-ray energy range from 0.05 to 3.5\,MeV. The efficiency at 1\,MeV in this configuration is measured to be 10.06(11)\%.

The same process is followed for determining the photo-peak efficiency with any of the various addback algorithms. The results for clover-addback with the same experimental setup are shown in Figure \ref{fig:ABEff} where the efficiency at 1\,MeV is measured to be 14.20(16)\%.

\begin{figure}
\centering
    \subfigure{
        {\includegraphics[width=1.0\linewidth]{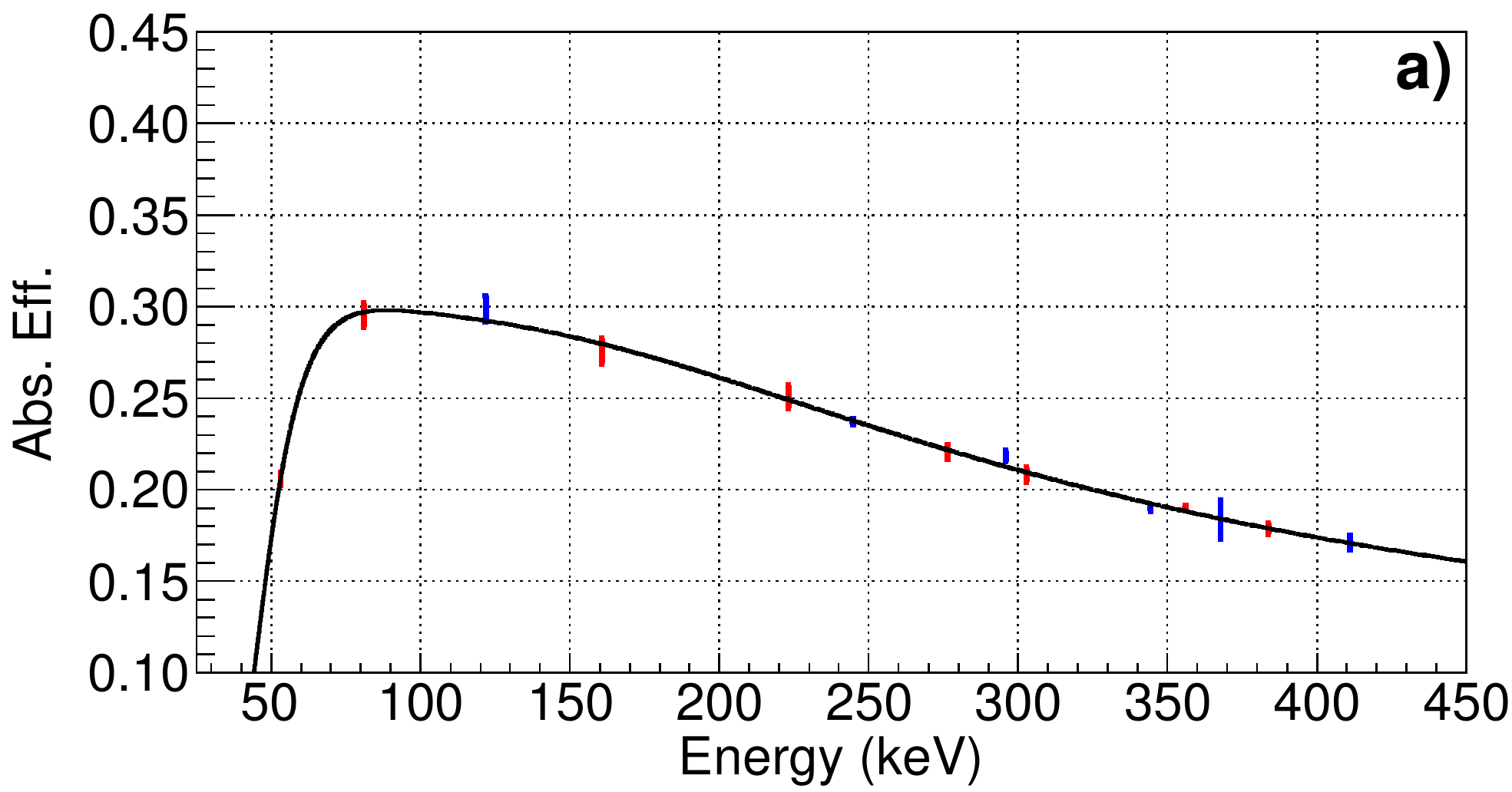}
        \label{fig:delrin_eff_20mm}}
    }\\
    \subfigure{
        {\includegraphics[width=1.0\linewidth]{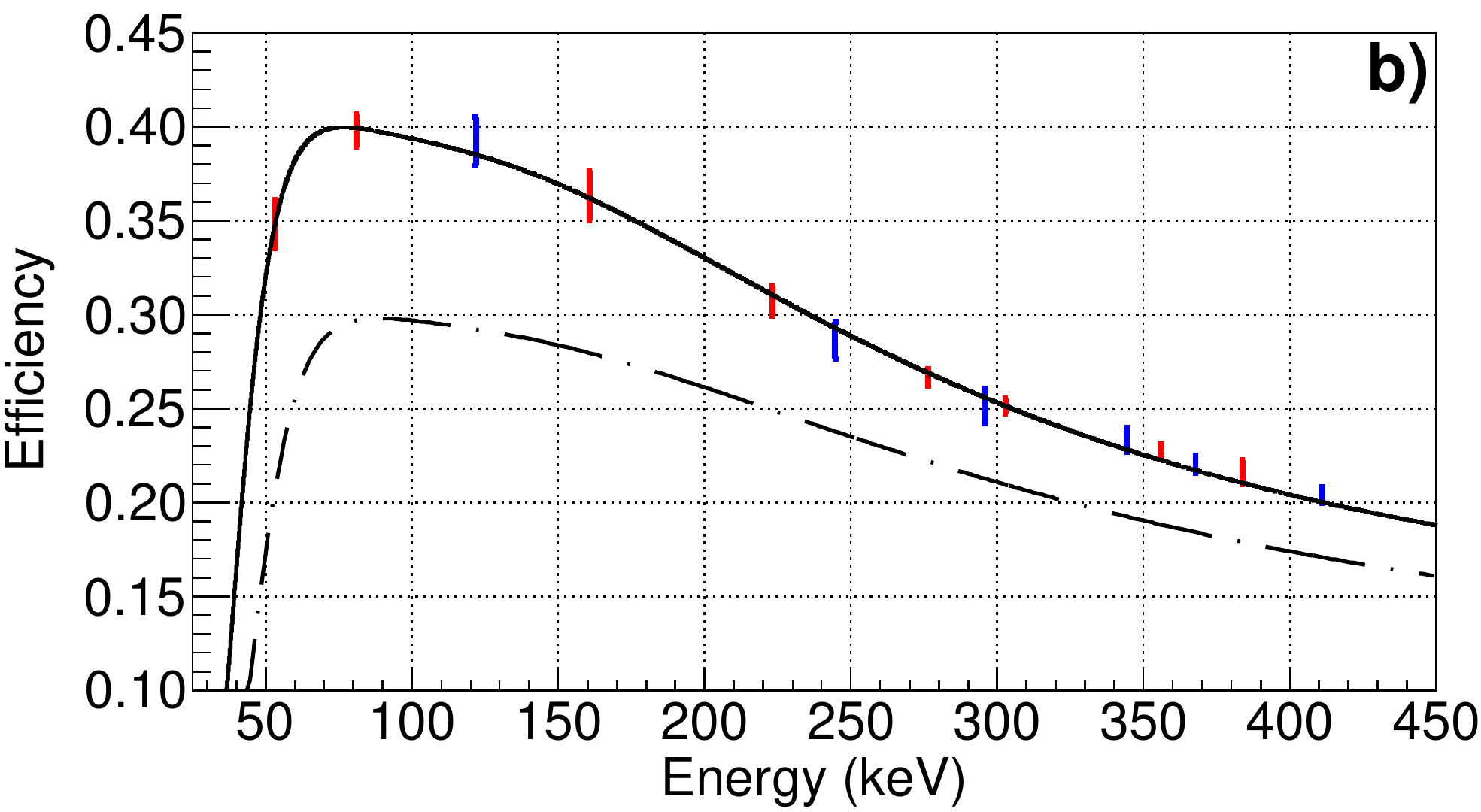}
        \label{fig:delrin_eff_0mm}}
    }\\
\caption{The low-energy singles photo-peak efficiency for the GRIFFIN array including the SCEPTAR vacuum chamber. {\bf a):} 16 clovers with 20 mm Delrin absorber, compared to {\bf b):} 16 clovers with 0 mm of Delrin around the SCEPTAR vacuum chamber. The data have been corrected for summing effects. The 20~mm Delrin efficiency is shown as a dashed line for comparison. {\it Color online:} (Red) $^{133}$Ba, (Blue)$^{152}$Eu, (Green) $^{60}$Co, (Black)$^{56}$Co.}
\label{fig:lowEEffDelrin}
\end{figure}

The addition of the Delrin absorbers around the vacuum chamber for stopping high-energy $\beta$ particles from reaching the HPGe crystals has the effect of reducing the $\gamma$-ray efficiency, especially at low energies. Figure \ref{fig:lowEEffDelrin} shows the difference in GRIFFIN photo-peak efficiency when the 20\,mm Delrin absorbers are added around the vacuum chamber.

\subsubsection{$\gamma-\gamma$ Angular Correlations}
\label{sec:gamma-gamma-correlations}

The assignment of spins and parities to excited nuclear states is important in studies of nuclear structure. In a $\gamma-\gamma$ cascade emitted from a series of excited nuclear states, an anisotropy is generally observed in the spatial distribution of the second $\gamma$ ray with respect to the first $\gamma$ ray. The anisotropy depends on the sequence of spin values for the nuclear states involved, the multipolarities, and the mixing ratios of the emitted $\gamma$ rays. These $\gamma-\gamma$ angular correlations can be used for the assignment of spins and parities to the nuclear states, and thus provide a powerful means to explore the structure of nuclei. In general, $\gamma-\gamma$ angular correlations take the form:

\begin{equation}
W(\theta) = \sum_{k=0,\textup{even}}^{k_{max}}{a_{k} P_{k}(\cos \Theta)}
\end{equation}

\noindent where $a_{k}$ are coefficients dependent on the nuclear spins, multipolarities and the mixing ratios involved in the cascade \cite{Rose1967}, $P_{k}(\cos \Theta)$ are the Legendre polynomials with $\Theta$ being the angle between the emitted $\gamma$ rays, and $k_{max}$ determined from the minimum value of either the spin of the intermediate state or the $\gamma$-ray multipolarities of each transition involved in the cascade.

In the GRIFFIN geometry, there are 64 crystals making 4096 crystal pairs, including summing crystals with themselves at 0$^\circ$, i.e., sum peaks. These pairs form 52 unique angular differences each with a corresponding weight or number of occurrences, as shown in Table \ref{tab:openingAngles}. The angular differences range from zero degrees (two $\gamma$-ray interactions in the same crystal) to 180$^\circ$. The large number and wide range of angular differences provides an excellent sensitivity to $\gamma-\gamma$ angular correlations with the GRIFFIN spectrometer.

\begin{table}[ht]
\caption{Angular differences between HPGe crystal pairs in the GRIFFIN geometry with the HPGe detectors at a source-to-detector distance of 11\,cm. Two independent sets of crystal pairs at 86.2\,$^{\circ}$ and 93.8\,$^{\circ}$ have different geometries, but angular differences that are the same to four decimal places.} 
\centering 
\begin{tabular}{cc|cc} 
\hline 
 & Num. of &  & Num. of \\
Angle $^{\circ}$ & Pairs & Angle $^{\circ}$ & Pairs\\ [0.5ex] 
\hline 
0.0 & 64 & 91.5 & 128 \\ 
18.8 & 128 & 93.8 & 48 \\
25.6 & 64 & 93.8 & 64 \\
26.7 & 64 & 97.0 & 64 \\
31.9 & 64 & 101.3 & 64 \\
33.7 & 48 & 103.6 & 96 \\ 
44.4 & 128 & 106.9 & 64 \\
46.8 & 96 & 109.1 & 96 \\
48.6 & 128 & 110.1 & 64 \\
49.8 & 96 & 112.5 & 64 \\ 
53.8 & 48 & 113.4 & 64 \\
60.2 & 96 & 115.0 & 96 \\
62.7 & 48 & 116.9 & 64 \\
63.1 & 64 & 117.3 & 48 \\ 
65.0 & 96 & 119.8 & 96 \\
66.5 & 64 & 126.2 & 48 \\
67.5 & 64 & 130.2 & 96 \\
69.9 & 64 & 131.4 & 128 \\ 
70.9 & 96 & 133.2 & 96 \\
73.1 & 64 & 135.6 & 128 \\
76.4 & 96 & 146.3 & 48 \\
78.7 & 64 & 148.1 & 64 \\ 
83.0 & 64 & 152.3 & 64 \\
86.2 & 64 & 154.4 & 64 \\
86.2 & 48 & 160.2 & 128 \\
88.5 & 128 & 180.0 & 64 \\ [1ex] 
\hline 
\end{tabular}
\label{tab:openingAngles} 
\end{table}

Detailed $\gamma-\gamma$ angular correlation analysis methods have been established for the GRIFFIN spectrometer and are described in Reference \cite{Smith2018}. This technique includes a treatment of the data using an event-mixing approach to account for variations in individual crystal efficiencies. An example of one method used for assigning nuclear spins involves making a direct comparison between the experimental data and a series of GEANT4 simulations. By using the exact GRIFFIN geometry in the simulation, the effects of finite solid angle due to the large size of the crystals are properly taken into account. This approach allows for a direct comparison between the simulated and experimental data sets.

\begin{figure}[ht]
\centering
\includegraphics[width=\columnwidth]{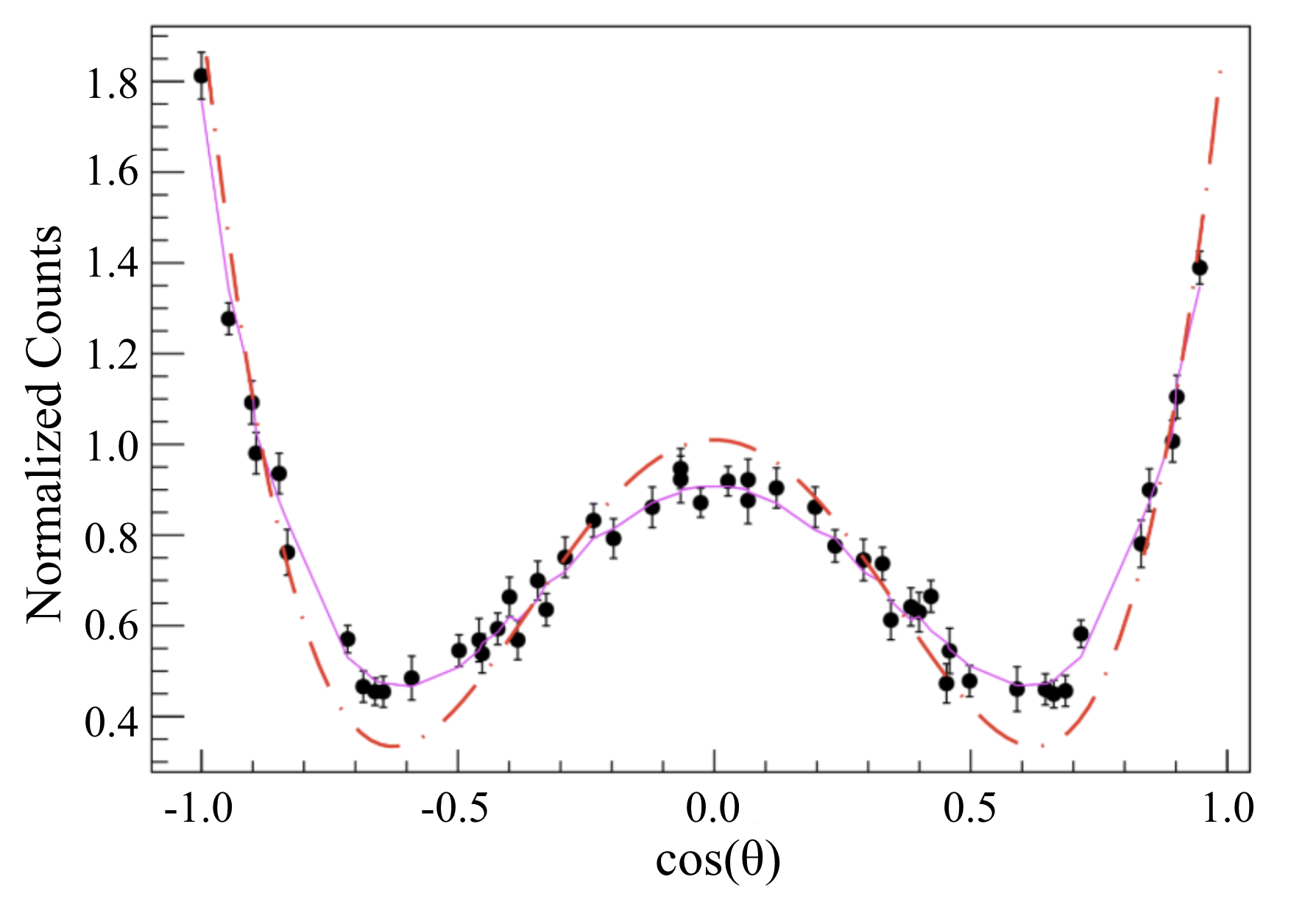}
\caption{\label{fig:Ga020_templatefit}The $\gamma - \gamma$ angular correlation of the 1333.1-1039.2\,keV, $J^\pi_i\rightarrow J^\pi_x\rightarrow J^\pi_f=0^+\rightarrow 2^+\rightarrow 0^+$, pure $E2$ cascade in $^{66}$Zn. The dashed red line shows the theoretical angular correlation for this cascade without solid angle corrections. The reduced $\chi^2$ between a full GEANT4 simulation (magenta solid line) and the experimental data (black points) is 1.01 indicating that the simulation correctly accounts for the attenuation effects from the finite size of the detectors.}
\end{figure}

As an example of a $\gamma - \gamma$ angular correlation measurement with GRIFFIN, Figure \ref{fig:Ga020_templatefit} presents data from the decay of a $^{66}$Ga radioactive ion beam implanted into the tape at the centre of the GRIFFIN spectrometer. 
Shown in Figure \ref{fig:Ga020_templatefit} is the $\gamma - \gamma$ angular correlation of the 1333.1-1039.2\,keV, $J^\pi_i\rightarrow J^\pi_x\rightarrow J^\pi_f=0^+\rightarrow 2^+\rightarrow 0^+$ cascade. As these are pure $E2$ transitions, there is no uncertainty introduced from potential multipole admixtures. The experimental data points are compared to the dashed red line representing the theoretical angular correlation, and the result of the GEANT4 simulation shown as the solid magenta line. A clear attenuation of the angular correlation is observed in comparison with the theoretical values, but is correctly accounted for in the simulation. The reduced $\chi^2$ for the comparison of the simulation and experimental data is 1.01.
To assign spins and determine mixing ratios, a $\chi^2$ analysis is performed for various combinations of nuclear spins in the cascade, and for different values of multipole mixing ratios. 

As mentioned, Reference \cite{Smith2018} provides a full description of the analysis methods established for detailed $\gamma-\gamma$ angular correlation measurements with the GRIFFIN spectrometer. The analogous $\gamma$-e$^{-}$ angular correlations involving transitions which decay by internal conversion electron emission are discussed in Section \ref{sec:g-e-ang-corr}. With the high efficiency of the array, it is possible to obtain $\gamma-\gamma$ angular correlation measurements for nuclei that have low production yields, allowing for detailed information to be obtained for nuclei across the nuclear chart and far from the line of $\beta$ stability.

\subsubsection{Compton Polarimetry}
$\gamma-\gamma$ angular correlations can be used to identify the multipolarity of transitions and the spins of states involved in cascades. In some cases, this can leave ambiguity in the assignment of parity to the states involved, unless other arguments can be invoked. The linear polarization of the $\gamma$ rays in the cascade is an observable that can identify the electric or magnetic nature of the radiation and therefore be used to unambiguously assign the parities of the states \cite{Fagg1959}. The high coincidence efficiency and high granularity of the GRIFFIN spectrometer makes Compton polarimetry a useful tool in the detailed study of atomic nuclei. Here we present a demonstration of this technique in GRIFFIN by adapting the recently developed quasi-continuous method of Compton polarimetry \cite{Alikhani2012} to an array of clover detectors. This approach increases the sensitivity of the spectrometer as a Compton polarimeter by a factor of two over classical analysis methods employed with arrays of HPGe clover detectors \cite{Schlitt1994,Hutter2002}.

The quasi-continuous method of Compton polarimetry was recently developed by Alikhani {\it et al.} \cite{Alikhani2012} for use with highly segmented HPGe detectors and $\gamma$-ray tracking arrays such as AGATA \cite{AGATA} and GRETINA \cite{GRETINA}. The polarization is determined from a measurement of the angle-dependent difference in the scattering of polarized, $W^{CE}_{pol}(\Theta,\vartheta,\xi)$, and unpolarized, $W^{CE}_{unpol}(\Theta,\vartheta)$, $\gamma$ rays, where the superscript $CE$ means ``Compton Effect''. The asymmetry is expressed as,

\begin{equation}
A^{CE}_{pol}(\Theta,\vartheta,\xi)=1-\frac{W^{CE}_{pol}(\Theta,\vartheta,\xi)}{W^{CE}_{unpol}(\Theta,\vartheta)}
\end{equation}

\noindent where the angles in the GRIFFIN system are defined in the following ways;\\
$\Theta$ is the angle between the two coincident $\gamma$ rays in the cascade (the vectors of these $\gamma$ rays also define the {\it coincidence plane}),\\
$\vartheta$ is the angle by which the $\gamma$ ray scatters with respect to its initial direction (this defines the {\it scattering plane}), and\\
$\xi$ is the {\it azimuthal scattering angle} between the scattering and coincidence planes.\\

In highly segmented HPGe and $\gamma$-ray tracking detectors, granularity in the azimuthal scattering angle is achieved through accurate knowledge of the interaction positions within the detectors by pulse-shape analysis. In arrays of composite detectors, such as GRIFFIN, one can take advantage of the large number of pairs of crystals at different angles in the array. Compton scattered events are considered only between two crystals within the same clover, but the azimuthal scattering angle of these events is determined by the orientation of those two crystals with respect to the coincident $\gamma$ ray detected elsewhere in the array. This application provides 49 unique azimuthal scattering angles, analogous to the number of unique angles available for $\gamma-\gamma$ angular correlations discussed in Section \ref{sec:gamma-gamma-correlations} but excluding coincidences involving crystals in the same clover. The polarized $\gamma$ rays are obtained from the normal coincident events involving 3 crystals (with 2 crystals in the same clover and the coincident $\gamma$ ray in a different clover). The unpolarized $\gamma$ rays come from a similar event-mixing approach used in the $\gamma$-$\gamma$ angular correlation analysis \cite{Smith2018}.

The method is practically applied in data analysis by redefining the asymmetry as,

\begin{equation}
A(\xi)=\frac{1}{2}Q(E_\gamma)P_\gamma(\Theta)\cos(2\xi)
\end{equation}

\noindent where the Quality factor $Q(E_\gamma)$ is the polarization sensitivity of the detector setup under the present experimental conditions for the $\gamma$-ray energy under investigation, and $P_\gamma$ is the degree of linear polarization of the $\gamma$ radiation incident on the detector.

In order to optimize the sensitivity to the Compton effect, the further restriction of considering only coincidence events between clovers separated by 90\,$^\circ$ is imposed. This geometry corresponds to crystal coincidence angles between 68 and 113\,$^\circ$ in order to select only those events where the angle between the coincidence plane and the electric field plane lies close to 90\,$^{\circ}$, where the analyzing power is maximal.

\begin{figure}
\centering
\includegraphics[width=1.0\linewidth]{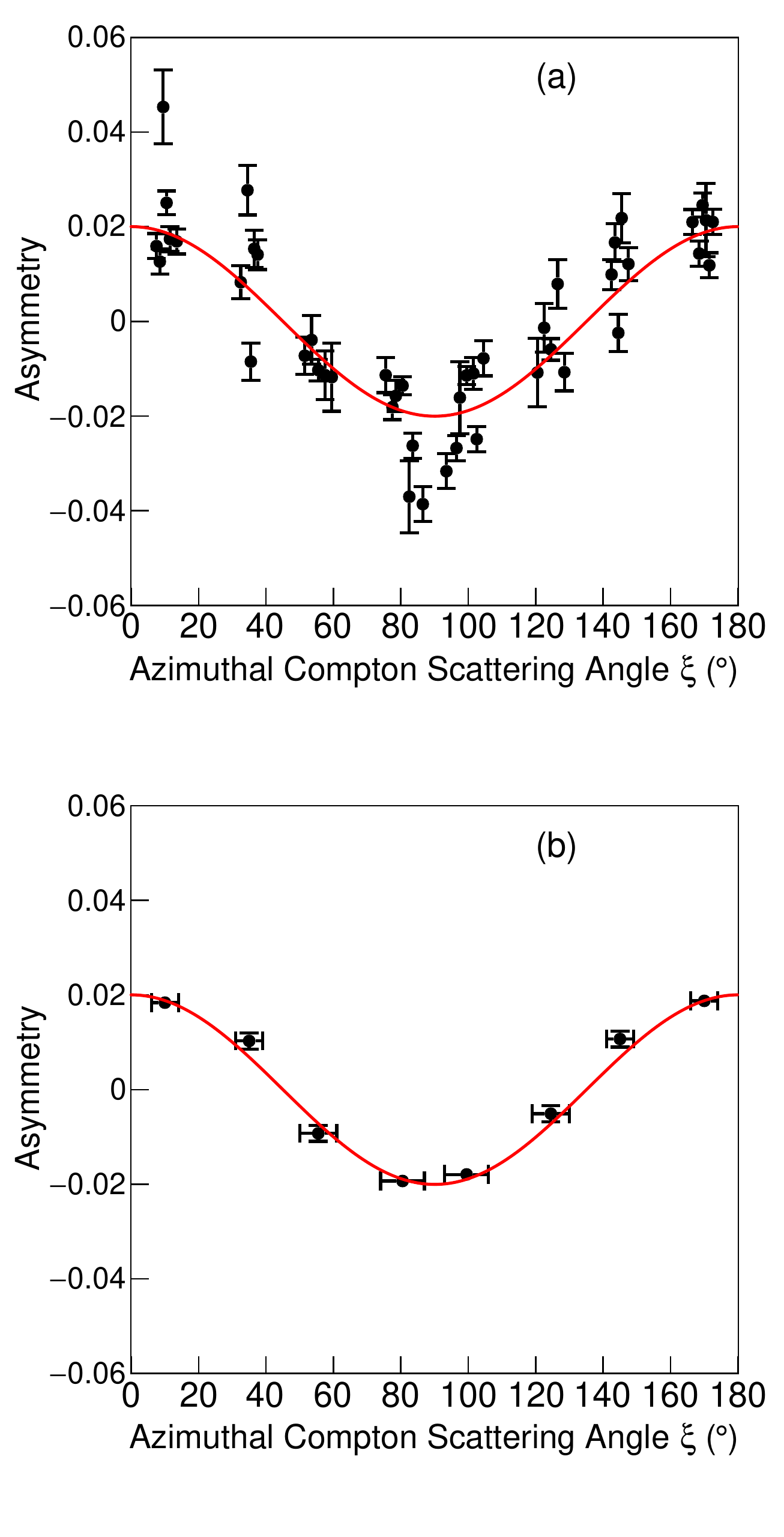}
\caption{Asymmetry of the 1332\,keV $\gamma$ ray from $^{60}$Co as a function of azimuthal scattering angle used to determine the quality factor $(Q(E_\gamma))$ for Compton polarimetry. (a) shows the data without grouping whereas (b) shows the selective grouping described in the text. The red line is the theoretical asymmetry scaled to fit the data.}
\label{fig:CompPol_60Co}
\end{figure}

The technique is calibrated with a measurement using a $^{60}$Co source for which the polarization value is well known ($P(90^\circ)=\frac{1}{6}=0.167$ for the 1332\,keV $E2$ $\gamma$ ray). The data are shown in Figure \ref{fig:CompPol_60Co} with (b) and without (a) grouping. The red line is the theoretical asymmetry scaled to fit the data. The difference in the observed and expected asymmetry for this well known case determines the value of the quality factor, $(Q(E_\gamma))$.

\begin{figure}
\centering
\includegraphics[width=1.0\linewidth]{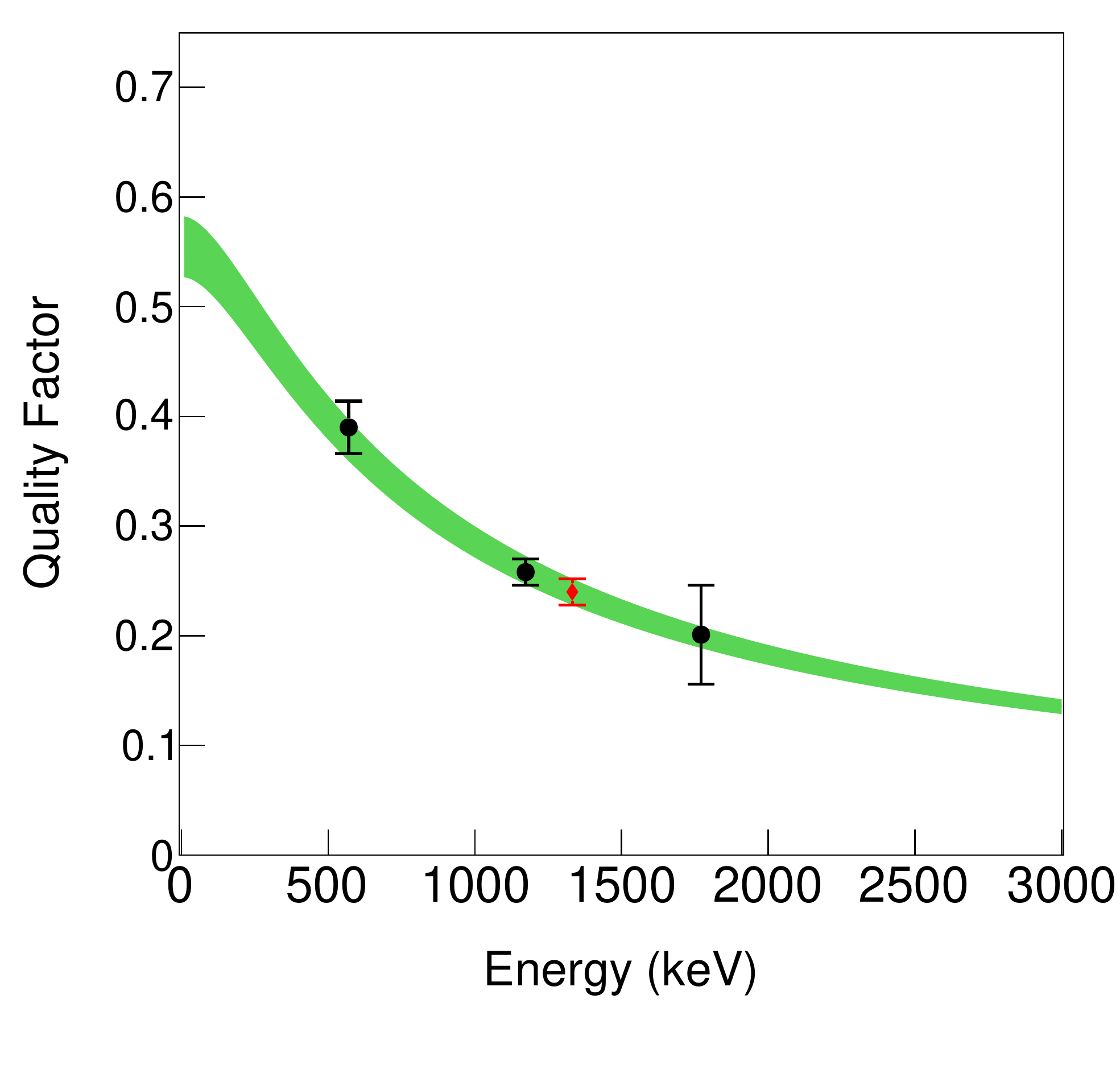}
\caption{The $\gamma$-ray energy dependence of the quality factor ($Q(E_\gamma)$) of the GRIFFIN spectrometer for Compton polarimetry, scaled from the measurement of 0.240(12) at 1332\,keV (red diamond marker). The width of the green band represents the uncertainty in the scaled value. The other measurements (black square markers) agree well with the scaled value. See text for details.}
\label{fig:ComPol_ScaledQ}
\end{figure}

A quality factor of 0.240(12) is measured at 1332\,keV for 16 clover detectors in the GRIFFIN spectrometer. This factor does not change if the number of clovers used for the measurement is reduced. However, reducing the number of clovers will reduce the level of statistics and the ultimate precision of measurements.

\begin{figure}
\centering
\includegraphics[width=1.0\linewidth]{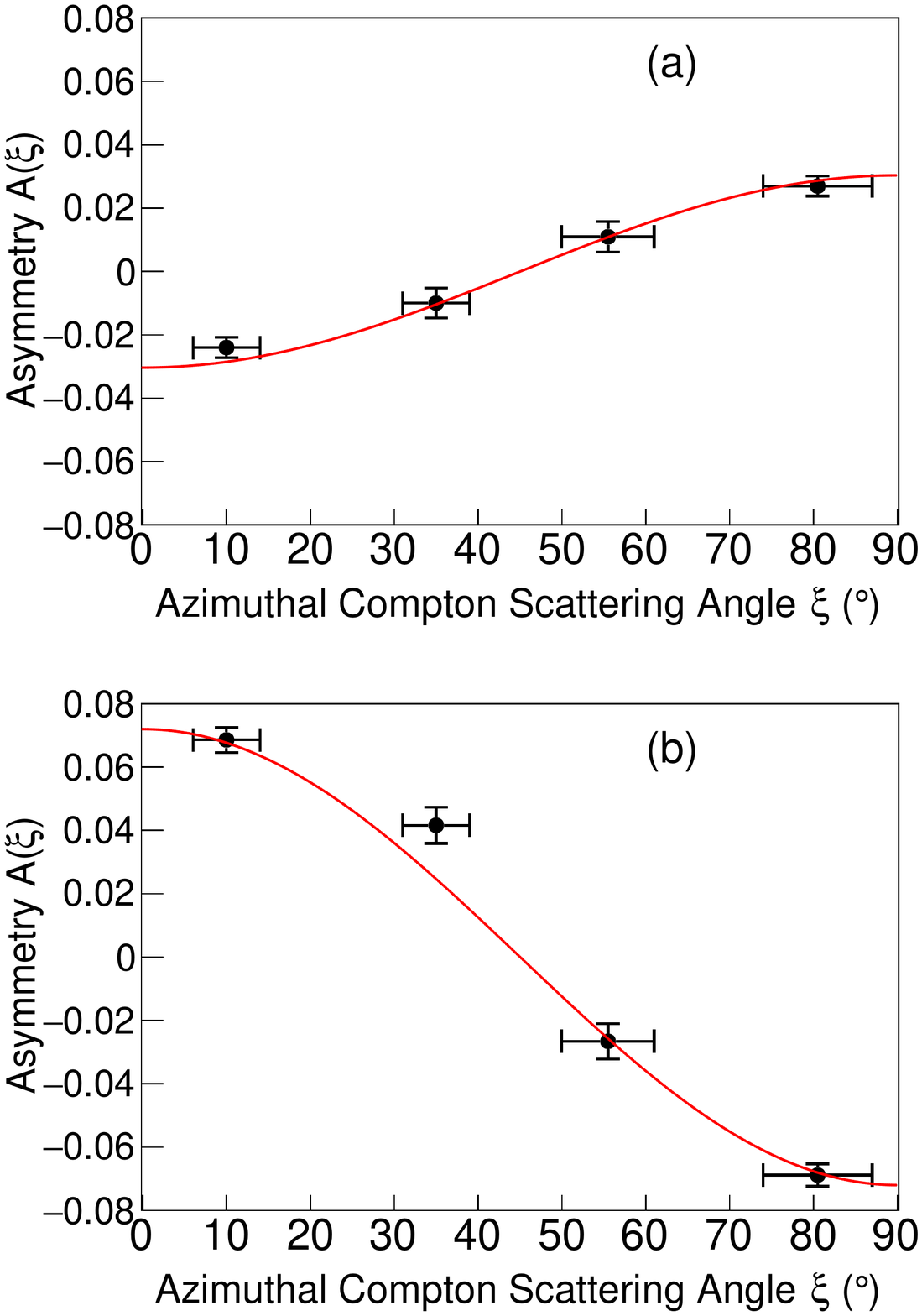}
\caption{The Compton polarimetry asymmetry of the $\gamma$ rays emitted from $^{207}$Pb as a function of azimuthal scattering angle. The red lines represent the theoretical distributions scaled by $Q(E_\gamma)$. (a): The 1064\,keV $M4(+E5)$ transition. (b): The 570\,keV $E2$ transition.}
\label{fig:ComPol_207Bi}
\end{figure}

Figure \ref{fig:CompPol_60Co}b) shows the same data as in (a) but the data have been grouped into eight angular bins. The x error bar represents the range of angles included in the bin. In this binning procedure, several angular combinations have been excluded because they represent a low number or individual crystal pairs lying at angles that are well separated from any others. The statistical uncertainty of these points is significantly larger than in the bins constructed of many angular pairs.

The $Q(E_\gamma)$ factor at the energy under investigation is scaled from 1332\,keV following the procedure described in Ref. \cite{Alikhani2012}. The behavior of the energy-dependent scaling can be seen in Figure \ref{fig:ComPol_ScaledQ}. Measurements of the quality factor at other energies using transitions with a well known polarization value ($E_\gamma$, $Q(E_\gamma)$, $Nucleus$; 570\,keV, 0.390(24), $^{207}$Pb; 1173\,keV, 0.258(12), $^{60}$Ni; 1771\,keV, 0.201(45), $^{56}$Fe) match well with the line scaled from 1332\,keV.

In Figure \ref{fig:ComPol_207Bi}, this Compton polarimetry method has been applied to $\gamma$ rays observed from a $^{207}$Bi source. The (a) panel shows the Compton scatter of the 1064\,keV $M4$(+$E5$) $\gamma$ ray in coincidence with the 570\,keV $\gamma$ ray. The (b) panel shows the Compton scatter of the 570\,keV $E2$ $\gamma$ ray in coincidence with the 1064\,keV $\gamma$ ray.
Here the data have been folded about 90$^\circ$ in comparison to Figure \ref{fig:CompPol_60Co}. The red lines represent the theoretical distributions calculated from the spins, parities and multipolaries involved in the cascade and the $Q(E_\gamma)$ factor scaled from 1332\,keV.  The agreement between the experimental data from the $^{207}$Bi source and the expected distributions is excellent. 

In the case of a transition with an admixture of multipolarities, a goodness-of-fit analysis can be performed on the experimental data in order to make a measurement of the mixing ratio. This technique is complementary to the use of $\gamma-\gamma$ angular correlations described in Section \ref{sec:gamma-gamma-correlations}. The combination of these two methods can offer an additional experimental handle on the assignment of spins, parities and multipole mixing ratios. The high efficiency, large number of crystal pairs and angles of the GRIFFIN array enable this powerful approach for detailed spectroscopic studies of nuclei away from the line of $\beta$ stability.

\subsection{SCEPTAR}
\begin{figure*}
\centering
\includegraphics[width=1.0\linewidth]{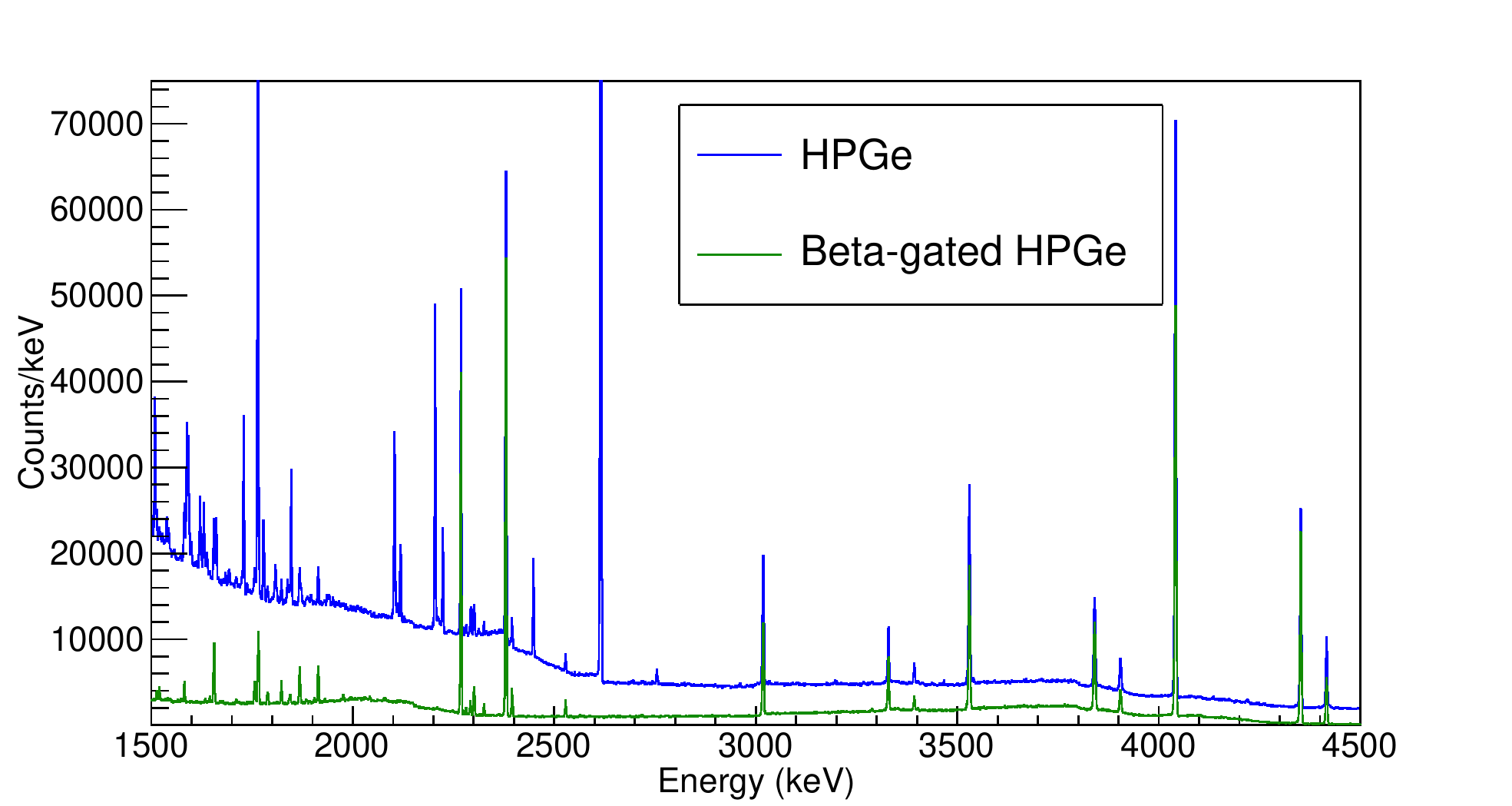}
\caption{$\gamma$-ray energy spectrum of $^{132}$Sn following the $\beta$ decay of $^{132}$In with and without the requirement of a coincident detection of a $\beta$ particle in SCEPTAR.}
\label{fig:SCEPTAR_example_132Sn}
\end{figure*}

The plastic scintillators of SCEPTAR subtend roughly 80\% of $4\pi$sr and are used to tag $\beta$ particles emitted in the decay of a parent nucleus. The absolute efficiency is 80\% when determined from the ratio of counts in a $\gamma$-ray singles peak and the counts in a $\gamma$-ray spectrum generated in coincidence with an event in SCEPTAR.
This temporal coincidence requirement provides a reduction of the $\gamma$-ray random room background observed in the spectra by many orders of magnitude as can be seen in the example shown in Figure \ref{fig:SCEPTAR_example_132Sn} for the $\beta$ decay of $^{132}$In into $^{132}$Sn. This strong suppression of room background is particularly important in studies of exotic nuclei very far from the line of $\beta$ stability, which are generally delivered with very low beam intensities ($\leq$ 1\,ion/s).

\subsection{PACES}

High-resolution electron spectroscopy is a powerful technique for the detailed study of atomic nuclei. Further sensitivity can be achieved by the combination of $\gamma$-ray and electron detectors for coincidence spectroscopy. The process of internal conversion is an alternative mode to $\gamma$-ray emission in the decay of excited states and can provide key information not always available from $\gamma$-ray observation alone. The internal conversion coefficient (ICC), equal to the intensity of internal conversion electron emission divided by the intensity of $\gamma$-ray emission, is sensitive to the multipolarity of the transition, or admixture of multipolarities.
In the case of electric monopole ($E0$) transitions between $0^+$ states, single $\gamma$-ray emission is forbidden meaning that electron (internal conversion or internal pair formation) spectroscopy must be utilized in the study of $E0$ transition strengths \cite{Wood1999}.

\begin{figure*}
\centering
\includegraphics[width=1.0\linewidth]{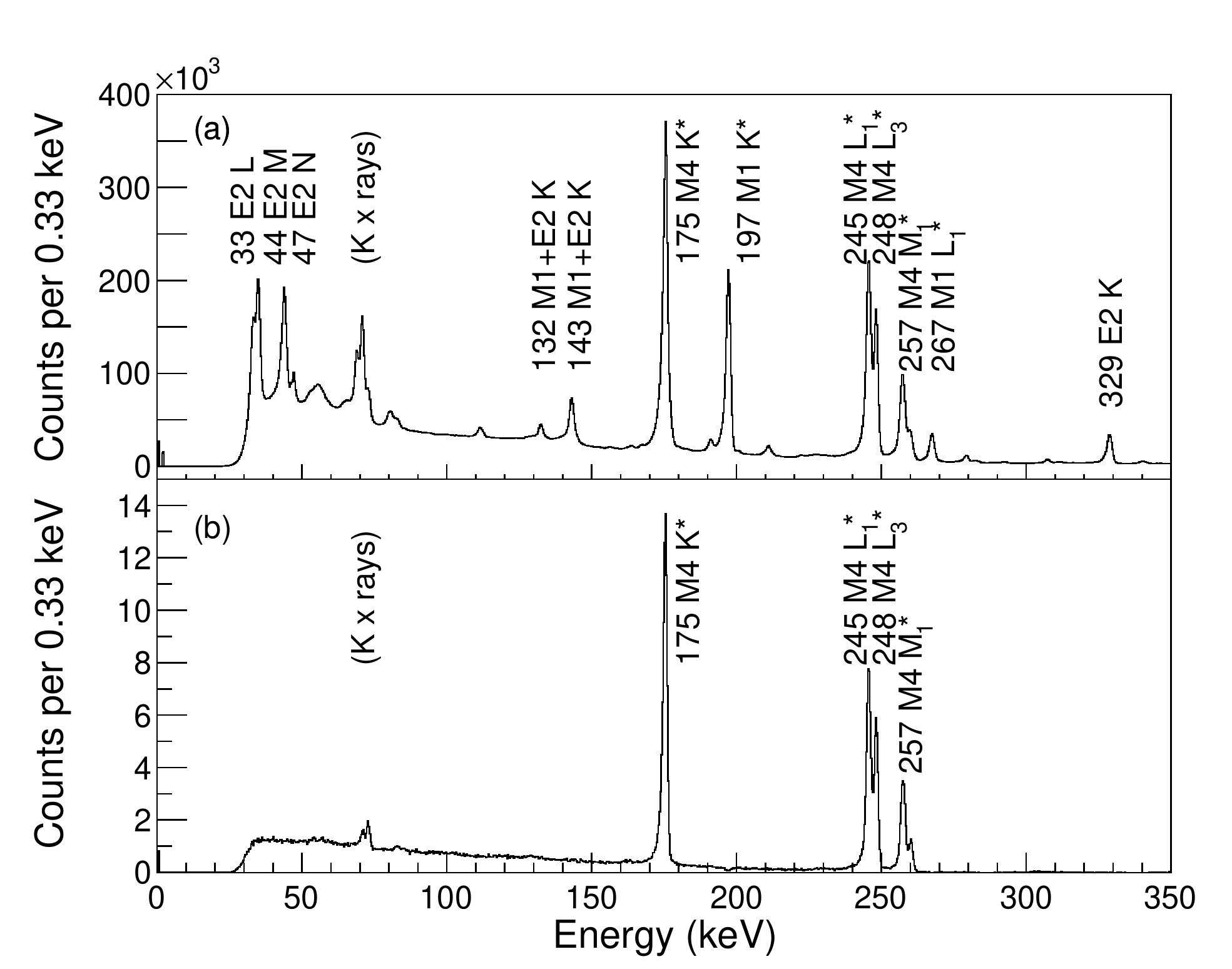}
\caption{PACES electron energy spectrum obtained during delivery of a $^{198}$Tl and $^{198m}$Tl beam to GRIFFIN observed in (a) singles and (b) coincidence with a 282\,keV (3$^-\rightarrow 7^+$ transition in $^{198}$Tl) $\gamma$ ray detected in the HPGe. Conversion electrons emitted following internal transitions in $^{198}$Tl (marked with an asterisk) and $^{198}$Hg are labeled showing the electron energy in keV, multipolarity and atomic electron sub-shell.}
\label{fig:PACES_energy}
\end{figure*}

Figure \ref{fig:PACES_energy} (a) shows an energy spectrum from singles events observed in PACES during delivery of a beam containing $^{198}$Tl and $^{198m}$Tl to GRIFFIN. The spectrum shown in Figure \ref{fig:PACES_energy} (b) is in coincidence with a 282\,keV (3$^-\rightarrow 7^+$ transition in $^{198}$Tl) $\gamma$ ray detected in the HPGe which highlights the decay of the $E_{x}$=543\,keV, 7$^+$ isomeric state. Peaks associated with conversion in each of the atomic subshells up to M can be resolved, offering the opportunity to examine sub-shell ratios for the determination of transition multipolarity.

As the PACES detectors are located inside the vacuum chamber, the peak-to-total ratio of the spectrum can sometimes suffer because the detectors are sensitive to not just internal conversion electrons (ICE), but also $\gamma$ rays, X rays and $\beta$ electrons. The severity of the impact on the spectral quality depends strongly on the nature of the radioactivity present in the chamber.

\subsubsection{Absolute Efficiency}

\begin{figure}
\centering
\includegraphics[width=1.0\linewidth]{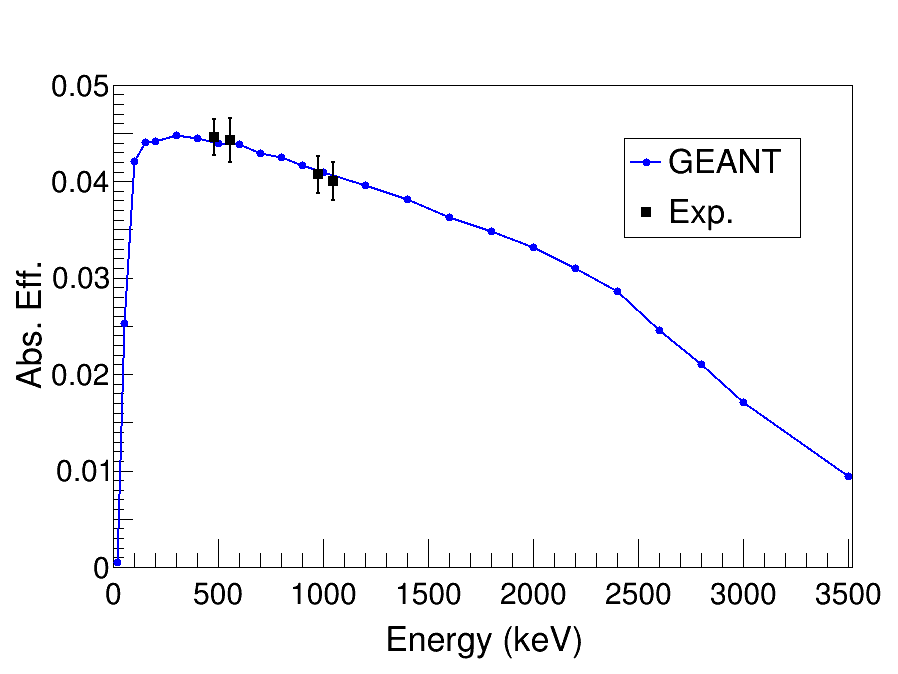}
\caption{Absolute full-energy electron detection efficiency of PACES in the region from 0.02 to 3.5\,MeV. The efficiency was simulated in GEANT4 (blue line and points). Simulated data are scaled to the experimentally measured efficiency (black points) obtained with a $^{207}$Bi source. See text for details.}
\label{fig:PACES_efficiency}
\end{figure}

The absolute efficiency of PACES has been determined using a 25\,kBq $^{207}$Bi source placed at the implantation location of the beam under high vacuum conditions. 
The source activity is electro-plated in a $\sim$5\,mm diameter spot at the center of a 25\,mm metallic disc without any protective layer in order to avoid energy straggling of the emitted electrons. 
K and L shell ICEs with energies 481.7 and 553.8\,keV (5/2$^{-}$ $\rightarrow$ 1/2$^{-}$ transition in $^{207}$Pb) and 975.6 and 1047.8\,keV (13/2$^{+}$ $\rightarrow$ 5/2$^{-}$), respectively, were used during the calibration, shown in Figure \ref{fig:PACES_efficiency} (black points). Results are compared to a GEANT4 simulation (blue) which has been normalized to the experimental data points. The simulation assumes a 0.04\,mm dead layer at the front surface of each Si(Li) detector. 

The GEANT4 simulation was useful to understand the features of the PACES efficiency curve.
The dead-layer thickness affects the detection efficiency for electrons up to a few hundred keV and the location of the sharp decline in efficiency at low energies is strongly dependent on this property of the detectors.
As the electron energy increases above 300\,keV, there is an approximately linear decrease in efficiency that is associated with an increase in probability for scattering out of the active volume at the sides of the detector, and the production of bremsstrahlung photons which also escape the crystal. 
Beyond 2\,MeV, electrons entering the detector have sufficient energy to punch-through the 5\,mm thickness at the rear of the detector. This punch-through effect, combined with an ever increasing probability of electrons scattering out of the sides and bremsstrahlung photons escaping the active volume, leads to an even steeper decline in efficiency at higher electron energies.

The full-energy peak areas measured in PACES also require coincidence summing corrections. Unfortunately, the empirical method used for the HPGe detectors, and described in Section \ref{sec:summing}, cannot be utilized because the PACES detector pairs only involve relative angular differences up to a maximum of $\approx$112\,$^{\circ}$. In general, these summing corrections are not straightforward for PACES, where several types of particles (photons, ICEs, and $\beta^{\pm}$ particles) can potentially lead to an apparent loss (or gain) in full-energy peak efficiency. Summing corrections can still be performed analytically by combining spectroscopic information (e.g., the decay scheme and branching ratios) with knowledge of the full-energy peak and \textit{total} detection efficiencies from simulation, the latter accounting for summing with Compton-scattered photons. The complexity of summing corrections scales with the complexity of the decay scheme being studied. In practice, it is usual in electron spectroscopy to develop a self-consistent efficiency calibration using transitions with well known internal conversion coefficients (such as pure $E2$ transitions) observed in the same dataset. In this way, an average value of the summing corrections is captured within the efficiency calibration.

In the case presented in Figure \ref{fig:PACES_efficiency} for the EC decay of $^{207}$Bi, the summing corrections are minimal. This decay produces relatively few discrete radiation emissions and no $\beta$ particles, being dominated by the 13/2$^{+}$ $\rightarrow$ 5/2$^{-}$ $\rightarrow$ 1/2$^{-}$ cascade in $^{207}$Pb. Summing-out corrections applied to the corresponding 1063.7 and 569.7\,keV $\gamma$ rays observed in the HPGe detectors account for $<$1$\%$ of the full-energy peak areas.

The self-consistent efficiency approach from in-beam data is always followed in the analysis of PACES data in GRIFFIN experiments and the GEANT4 simulation or source calibrations are used only as an approximate guide. This usually allows for the efficiency at an energy of interest to be interpolated rather than extrapolated. In this way the uncertainty in the efficiency calibration depends on the statistics obtained in the transitions used for the calibration, but are typically better than 5\,\% of the efficiency value.

\subsubsection{Gamma-Electron Angular Correlations}
\label{sec:g-e-ang-corr}

In the case of two transitions emitted sequentially between initial, intermediate and final nuclear states (\textit{J$_{i}$},\textit{J$_{x}$},\textit{J$_{f}$}) where one or both of the transitions is emitted as an internal conversion electron, the ($\gamma$-e$^{-}$) angular correlation has the form: 

\begin{equation} 
W_{\Theta} = \sum_{k=even}^{2L} b_{k}A_{k}P_{k}\cos{\Theta},	
\label{eq:egamma_corr}  
\end{equation}
\noindent
where $A_k$ the correlation coefficient, \textit{L} is the transition angular momentum, and the Legendre polynomial, $P_k\cos{\Theta}$, are modified by the particle parameter, \textit{b$_{k}$}. The expanded form of \textit{b$_{k}$A$_{k}$} corresponding to each transition is provided by Becker and Steffen for both $\gamma$ rays and ICEs \cite{Becker1969}.
In the case of transitions with mixed multipolarity, the ICE mixing ratio, \textit{$\delta_{ICE}$}, must also be taken into account and is usually calculated \cite{Kibedi2008}. Tabulations and procedures for determining the particle parameter \textit{b$_{k}$} are available in Ref. \cite{Hager1968}. 

Gamma-electron ($\gamma$-e$^{-}$) coincidence matrices form the basis for the angular correlation measurements. Individual matrices can be produced for each unique combination of detectors, as demonstrated for $\gamma$-$\gamma$ angular correlations. In the fully populated GRIFFIN spectrometer composed of 5 PACES detectors and 15 HPGe clovers (one HPGe clover must be removed to accommodate the PACES dewar), 300 individual detector combinations are available. This setup provides coincidence angles between 4.7 and 175.3\,$^{\circ}$ for a source-to-HPGe distance of 110\,mm.
Due to the relative placement of the HPGe and PACES detectors, the number of detector combinations peaks at angular differences of around 90\,$^{\circ}$, with fewer detector combinations available at the extreme angles. In the analysis, similar detection angles are combined or grouped.

The experimental $\gamma$-e$^{-}$ angular correlation is expressed as

\begin{equation} %
W_{\Theta}^{exp} = \frac{N_{\Theta}^{C}}{N_{\Theta}^{EM}}\times\frac{\sum_{i}^{n} N_{\Theta i}^{EM}}{\sum_{i}^{n} N_{\Theta i}^{C}},	
\label{eq:egamma_corr_exp}  
\end{equation}
\noindent
where $N_{\Theta}^{C}$ is the number of $\gamma-e^{-}$ coincidences observed in true (time-correlated) coincidence for HPGe-PACES detectors at an angular difference of $\Theta$ (for \textit{n} different angular bins) and $N_{\Theta}^{EM}$ is the number of \textit{event-mixed} (uncorrelated) coincidences. The latter is crucial to the normalization of the angular correlation and accounts for the variation in relative efficiency between different HPGe-PACES detector combinations. This accounting includes different detection efficiencies as well as differences in the attenuation of $\gamma$ rays by PACES itself (especially important for those HPGe crystals which are shadowed by the PACES crystals). 
The event-mixed coincidences, $N_{\Theta}^{EM}$, are derived from coincident data constructed from uncorrelated decay events with timestamp differences between 2\,$\mu$s and 200\,$\mu$s. Both $N_{\Theta}^{C}$ and $N_{\Theta}^{EM}$ are found from the number of counts in the $\gamma$-ray or e$^{-}$ full-energy peak with appropriate background subtraction.

\begin{figure}
\centering
\includegraphics[width=1.0\linewidth]{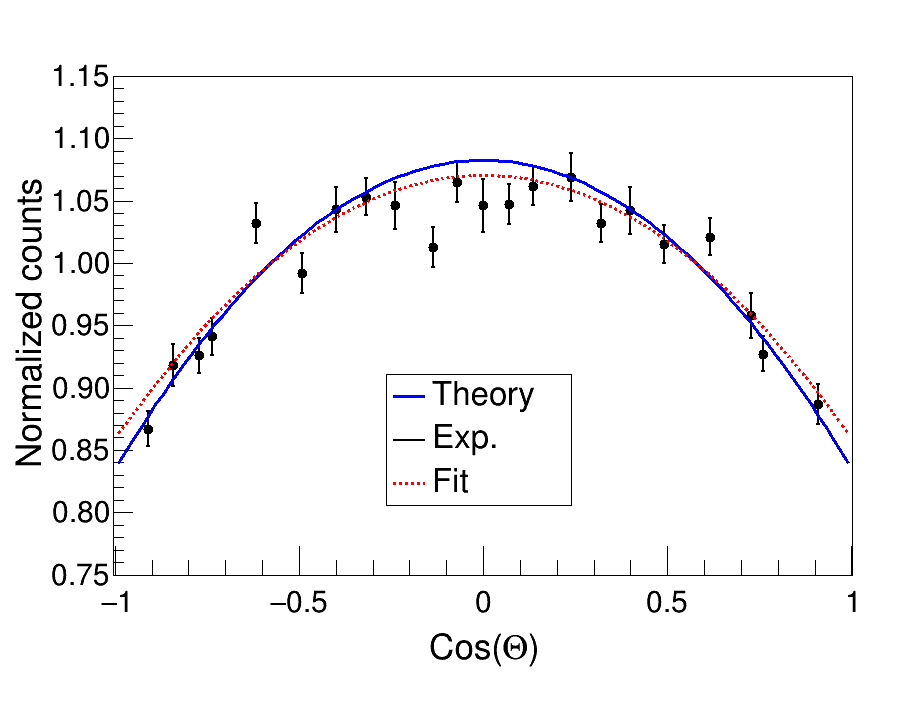}
\caption{Gamma-electron angular correlation observed using GRIFFIN and PACES between the 261\,keV $7^+\rightarrow3^-$ ($M4$ K e$^{-}$) and 283\,keV $3^-\rightarrow2^-$ ($M1$ $\gamma$ ray) transitions in $^{198}$Tl. Experimental data (black points) are compared to the expected theoretical angular correlation (blue line) which does not include any attenuation effects due to the finite size and geometry of the detectors. The red dashed line is the best fit to the data. See text for details.}
\label{fig:PACES_angular_correlation}
\end{figure}

The measured $\gamma$-e$^{-}$ angular correlation of the 283\,keV $M1$ $\gamma$ ray and 175\,keV $M4$ K electron of the cascade depopulating the $T_{1/2}$=1.9\,h, $J^\pi=7^+$ isomer in $^{198}$Tl \cite{NDS198} is shown in Figure \ref{fig:PACES_angular_correlation} with the associated energy spectra shown in Figure \ref{fig:PACES_energy}. 
The total number of coincidences observed in the full-energy peak for this example was 2.1$\times$10$^{5}$. 
Similar angular difference combinations within $\pm$5\,$^{\circ}$ have been grouped together so that each bin has a similar number of detector pairs. 
The measured asymmetry is in good qualitative agreement with that expected for the $M4-M1$ cascade calculated using Equation \ref{eq:egamma_corr} with the mixing ratios for both transitions set to zero. This scenario corresponds to $b_{2}A_{2}$ = -0.165 ($b_{4}A_{4}$ = 0). The best fit to the data is obtained with a reduced $\chi^{2}$ value of 1.94 for $b_{2}A_{2}$ = -0.141(8) and $b_{4}A_{4}$ fixed at zero. Thus an attenuation of the expected asymmetry of $\approx$85$\%$ is observed. A GEANT4 simulation of this 15 unsuppressed HPGe clovers and 5 PACES detector arrangement is under development in order to fully characterize the attenuation effects due to finite detector size for $\gamma$-e$^{-}$ angular correlations as has been done for $\gamma$-$\gamma$ angular correlations \cite{Smith2018}.

\subsection{LaBr$_3$(Ce) and Fast $\beta$ Scintillator}

Transition probabilities between nuclear states are powerful probes for studying the structure of atomic nuclei. With the 8 LaBr$_3$(Ce) ancillary detectors and the fast plastic scintillator located at zero degrees for $\beta$-particle tagging, GRIFFIN is able to simultaneously apply both the general advanced time delayed $\beta \gamma \gamma$(t) (ATD) method \cite{Mach1989,Moszynski1989,Mach1991} and the more general Generalized Centroid Difference Method (GCDM) \cite{Regis2013} to $\beta$-decay studies with radioactive-ion beams. The specific applications of these methods and performance in the GRIFFIN system are discussed in the remainder of this Section.

\subsubsection{Fast-Timing Using the Convolution Fit Method}

When the lifetime to be measured is long compared to the timing resolution of the system, extracting the half-life can be done with the convolution method. In this case, the time distribution between two events (either $\beta\gamma$ or $\gamma\gamma$) is no longer a symmetric Gaussian, but the decaying side of the distribution will present an exponential decay with a slope equal to the decay constant, $\lambda$, of the level. This distribution can be fitted with the convolution of a Gaussian and the exponential decay component:

\begin{equation}
  F(t_j) =\gamma \int_A^{+\infty} e^{- \delta (t_j - t)^2} e^{-\lambda t}dt \label{eq:convolution_method}
\end{equation}

\noindent where $\gamma$ is the normalization factor, $\delta$ is a parameter related to the width of the Gaussian prompt distribution and $A$ is the centroid of the Gaussian which is related to the position of a prompt transition of the same energy.

\begin{figure}
\centering
\includegraphics[width=1.0\linewidth]{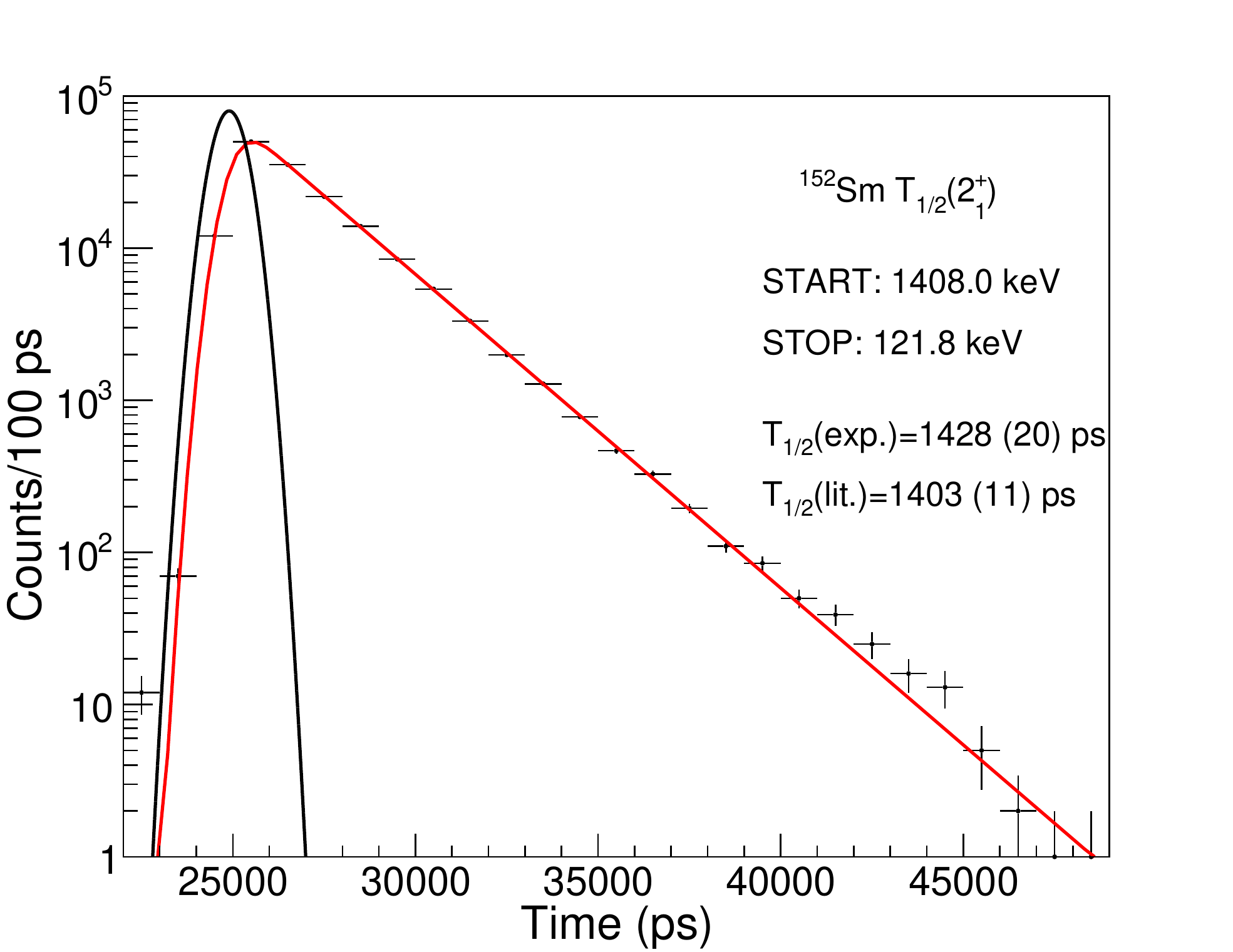}
\caption{Example of the convolution fit method for $^{152}$Eu source data collected with the LaBr$_3$(Ce) in the GRIFFIN array to measure the lifetime of the $2^+_1$ state in $^{152}$Sm. See text for details.}
\label{fig:lifetime_convolution_method}
\end{figure}

Figure \ref{fig:lifetime_convolution_method} illustrates the convolution fit method for the $2^+_1$ state in $^{152}$Sm populated in the electron-capture decay of a $^{152}$Eu calibration source. It shows the time-to-amplitude Converter (TAC) spectrum (time difference) of events detected in the LaBr$_3$(Ce) array with gates on the 1408.0\,keV (START) and 121.8\,keV (STOP) energy peaks. The entire time range has been fitted with Equation~\ref{eq:convolution_method}. The result quoted here of T$_{1/2}$=1428(20)\,ps compares well to the literature value of 1403(11)\,ps~\cite{NDS152}.

\subsubsection{Fast-Timing Using the Centroid Difference Method} \label{sec:centroid_shift}

When the lifetime to be measured is short (below the timing resolution of the system), no slope will be visible in the TAC spectra. However, the centroid of the resulting Gaussian distribution will be shifted from the distribution created by a prompt transition (a transition decaying from a level with $\tau < 1$\,ps) by its mean lifetime, $\tau$. Since a reference prompt transition is not always available, the lifetime information can be extracted using the mirror centroid difference method \cite{Regis2010}. As the STOP signal in the TACs is delayed by $\sim 25$\,ns (in practice this means that the TAC range will be from -25\,ns to +25\,ns, instead of the usual 0-50\,ns), by reversing the gates to the START (feeding transition) - STOP (decaying transition) to START (decaying transition) - STOP (feeding transition), a ``negative" lifetime can be measured. In this way, a measurement of the centroid shift between the decaying-feeding and feeding-decaying will result in:

\begin{equation}
  \text{C}_\text{feed} - \text{C}_\text{decay} = \Delta\text{C} = \tau - (-\tau) = 2\tau\label{eq:centroid_shift1}
\end{equation}

\noindent where $\text{C}_\text{feed}$ is the centroid of the timing distribution in which START is the decaying transition and STOP the feeding transition, and $\text{C}_\text{decay}$ is the reverse. The centroid difference method measures twice the mean lifetime of the level and suppresses systematic errors.

A common issue encountered when using the centroid shift method is that the feeding and decaying transitions do not have the same energy, and, despite the use of a CFD module in the signal processing, there is a significant time-walk (a dependence of the timing response with the photon energy). The time-walk can be carefully calibrated for each detector by using transitions with precisely known lifetimes. When adding together all the time-walks from the different crystals of an array a ``mean prompt response difference'' (PRD) can be constructed. An additional term must be introduced to Equation~\ref{eq:centroid_shift1} to account for this effect, of the form PRD$(\Delta \text{E}_\gamma ) = \text{PRD}(\text{E}_\text{feed}) - \text{PRD}(\text{E}_\text{decay})$, where PRD is the centroid of the time response of a prompt transition at a given energy:

\begin{equation}
	\Delta\text{C} = \text{PRD}(\Delta \text{E}_\gamma ) + 2\tau . 
    \label{eq:centroid_shift2}
\end{equation}

Thus the ability of the method to measure lifetimes is primarily limited by the ability to establish the centroid of a timing distribution and the time-walk of the experimental setup. Making use of standard calibration sources, the time-walk of a LaBr$_3$(Ce) array can be determined down to 2-5\,ps precision in the 0.1-1.4\,MeV range (see Reference~\cite{Regis2016} and Section~\ref{sec:timing_performance}). The statistical error when determining the centroid of a Gaussian distribution is given by:

\begin{equation}
	\delta\text{C} = \frac{\sigma}{\sqrt{n}} \simeq \frac{\text{FWHM}}{2.355\cdot\sqrt{n}}\label{eq:centroid_error}
\end{equation}

\noindent where FWHM is the full-width at half maximum timing resolution of the system and $n$ is the number of detected coincidences. With the typical timing resolution of the GRIFFIN LaBr$_3$(Ce)-LaBr$_3$(Ce) crystals coincidence of $\sim 300$\,ps, the collection of 10,000 coincidence counts can reduce the uncertainty in determining the centroid to about $\sim 1$\,ps and the final uncertainty of the measured lifetime will be dominated by the precision of the time-walk correction.

\begin{figure}
\centering
\includegraphics[width=1.0\linewidth]{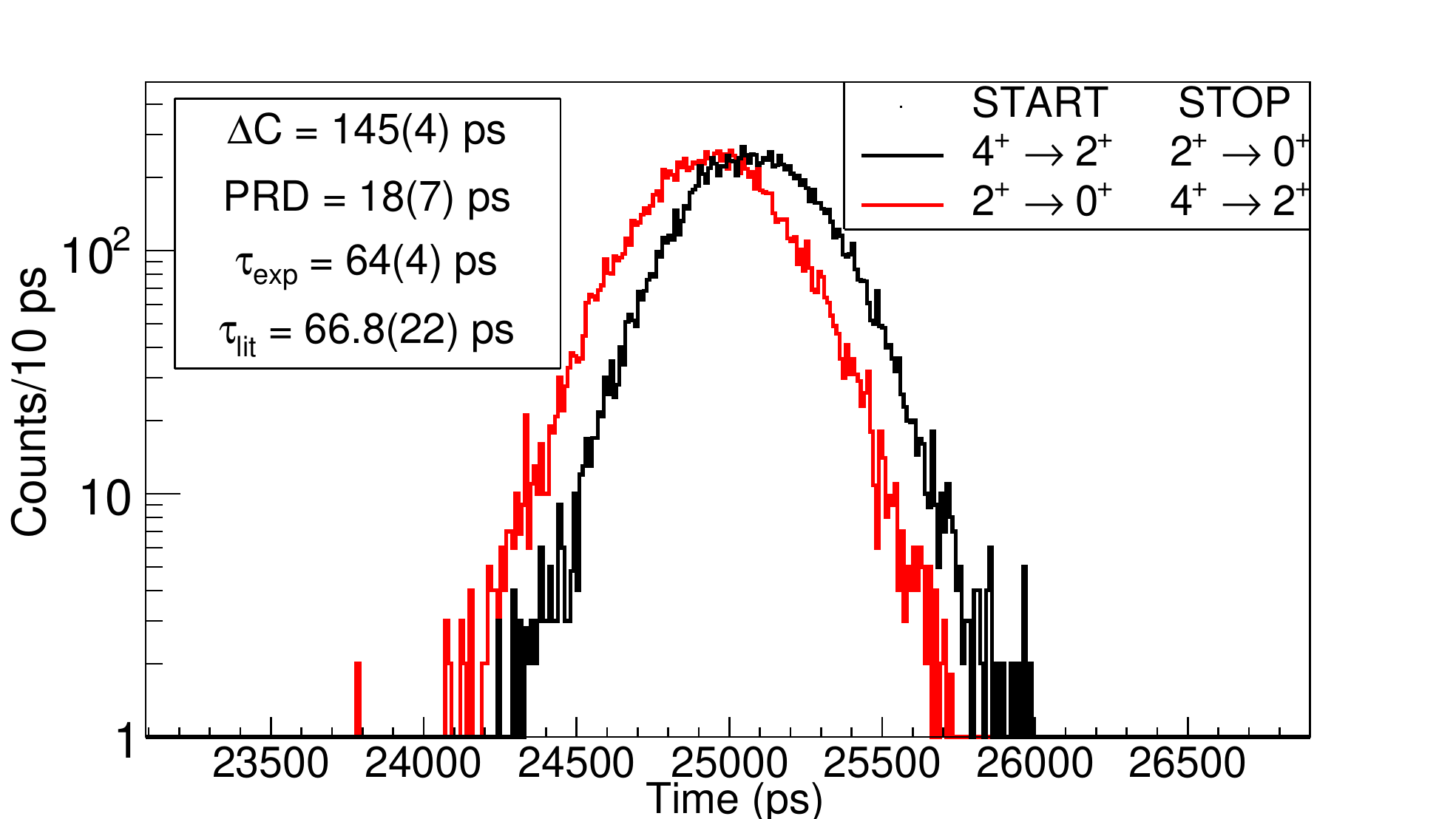}
\caption{Mean lifetime of the $2^+_1$ state in $^{200}$Hg measured using the centroid difference method. See text for details.}
\label{fig:lifetime_centroid_shift_method}
\end{figure}

Figure~\ref{fig:lifetime_centroid_shift_method} illustrates the centroid difference method for the $2^+_1$ state in $^{200}$Hg ($\tau_\text{lit}=66.8(22)$\,ps~\cite{Pritychenko2016}). The delayed spectrum (black) uses the $4^+_1 \rightarrow 2^+_1$ transition as START and the $2^+_1 \rightarrow 0^+_1$ transition as STOP. The anti-delayed spectrum (red) uses the $2^+_1 \rightarrow 0^+_1$ transition as START and the $4^+_1 \rightarrow 2^+_1$ transition as STOP. In both cases, the $3^+_3 \rightarrow 4^+_1$ transition was selected in the HPGe detectors to reduce background contributions to the spectrum. The centroid shift between the delayed and anti-delayed distributions was corrected by the PRD. Due to the very low level of Compton background, no further corrections were needed. The result $\tau_\text{exp}=64(4)$\,ps is in excellent agreement with the literature value \cite{Pritychenko2016}, demonstrating the ability of the setup to measure lifetimes down to $\sim 10$\,ps given sufficient statistics.

\subsubsection{Timing Performance}\label{sec:timing_performance}

As detailed in Section~\ref{sec:centroid_shift}, the ability to measure lifetimes with a LaBr$_3$(Ce) array is given by the timing resolution of the crystals and the time-walk of the system. While it is paramount to optimize these two features, there is a strong co-dependence between them and as such only one is optimized at the cost of the other one. When using small setups with only 1 or 2 inorganic scintillators and a fast plastic scintillator, for example, there is a small number of combinations and each one can be analyzed independently. In such a case, the timing resolution can be prioritized at the cost of having very different time-walk in each detector combination. In larger arrays, such as GRIFFIN, the number of combinations grows geometrically with the number of crystals, thus making it inefficient to analyze every combination independently. Instead, the time-walk optimization is prioritized, making it as flat as possible and similar for every crystal. When this is achieved, all the detector combinations behave in a similar manner and can be added together. While the final result has slightly worse resolution than the individual ones, the increase in statistics more than compensates for the loss.

The main parameters varied when optimizing the time-walk were the high voltage (HV) bias of the PMT and the pole-zero correction (Z) and external delay of the CFD. The HV bias of each detector was adjusted so that the signal amplitude for the 662\,keV photopeak of a $^{137}$Cs source matched that of a reference detector operated at -1400\,V. The resulting HV for the other detectors ranged from -1350 to -1500\,V. The pole-zero corrections were optimized individually for each crystal using an oscilloscope and found to be in the -0.3 to -0.4\,mV range. An external delay of 20\,ns was chosen for use with all CFDs.

\begin{figure}
\centering
\includegraphics[width=1.0\linewidth]{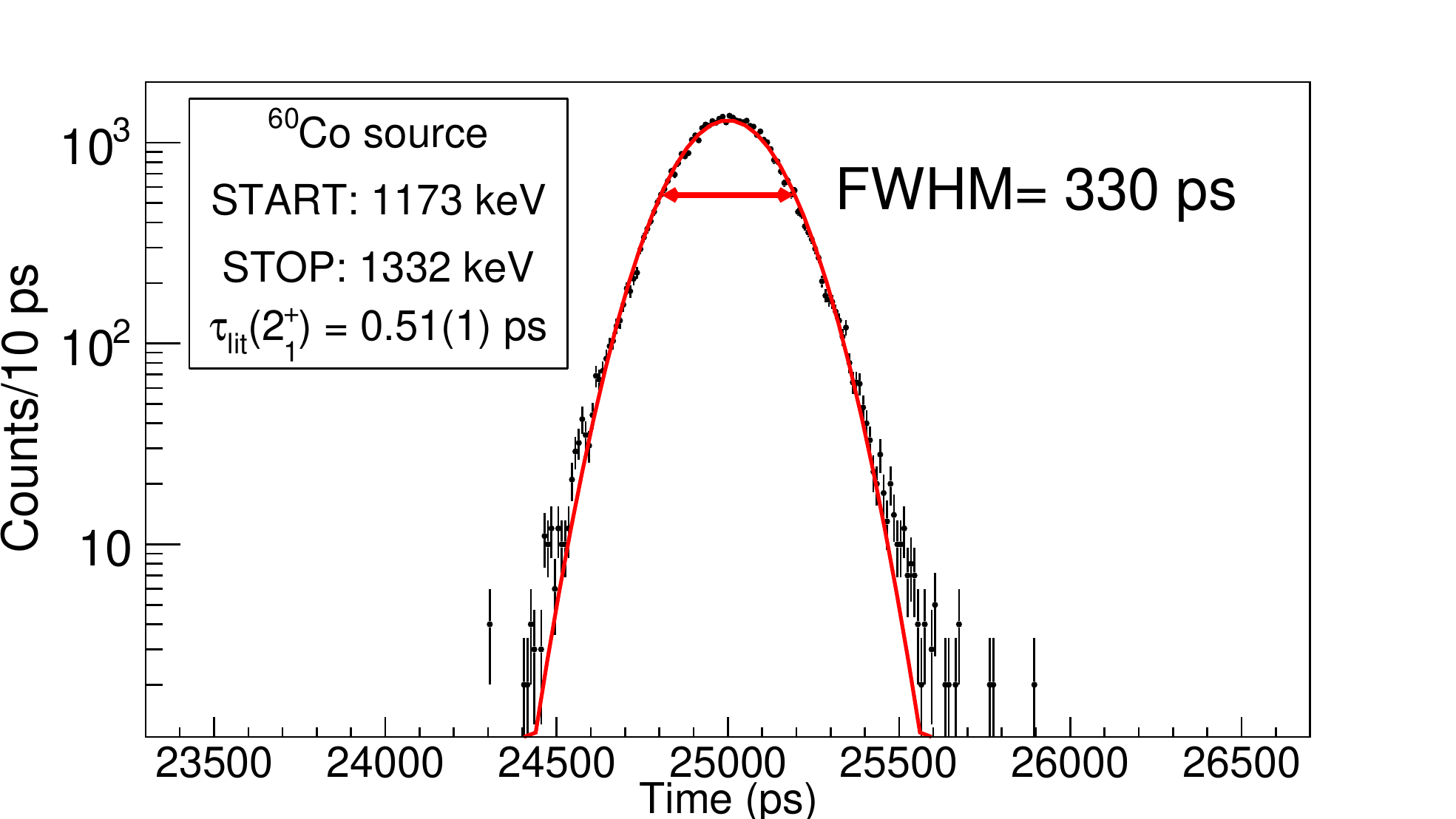}
\caption{Timing resolution of the full 8-detector LaBr$_3$(Ce) array in GRIFFIN measured with a $^{60}$Co source.}
\label{fig:timing-resolution}
\end{figure}

With this set of optimized settings, the whole system (every LaBr$_3$(Ce)-LaBr$_3$(Ce) combination added together) had a timing resolution of $\approx 330$\,ps as shown in Figure~\ref{fig:timing-resolution}. It should be noted that the timing resolution of individual combinations ranged from 280 to 400\,ps. It is speculated that the variation from crystal to crystal could arise from each one originating from a different manufacturing batch and each one being of a different age since manufacture.

The full-energy-peak time-walk of the LaBr$_3$(Ce) array was studied using a standard $^{152}$Eu calibration source. Due to the precisely known lifetimes, the time walk was measured to be $\pm 60$\,ps for photons with energy in the 244 to 1299\,keV range. This value is similar to those reported by other hybrid HPGe-LaBr$_3$(Ce) arrays~\cite{Regis2012,Regis2016,Bucurescu2016,Rudigier2017}. 

\subsubsection{Absolute Efficiency}

The absolute photo-peak efficiency of the 8-detector LaBr$_3$(Ce) array in combination with the PACES and ZDS detectors in the chamber (the usual configuration used in fast-timing experiments), was measured using standard $^{60}$Co, $^{133}$Ba and $^{152}$Eu calibration sources. The 15 HPGe detectors were at the standard 11~cm and the LaBr$_3$(Ce) at 12.5~cm distance from the source. The sources were placed in the usual source holder, positioning the activity at the beam-implantation point at the central focus of the detector arrays. No Delrin absorbers were present for these measurements.

\begin{figure}
\centering
\includegraphics[width=1.0\linewidth]{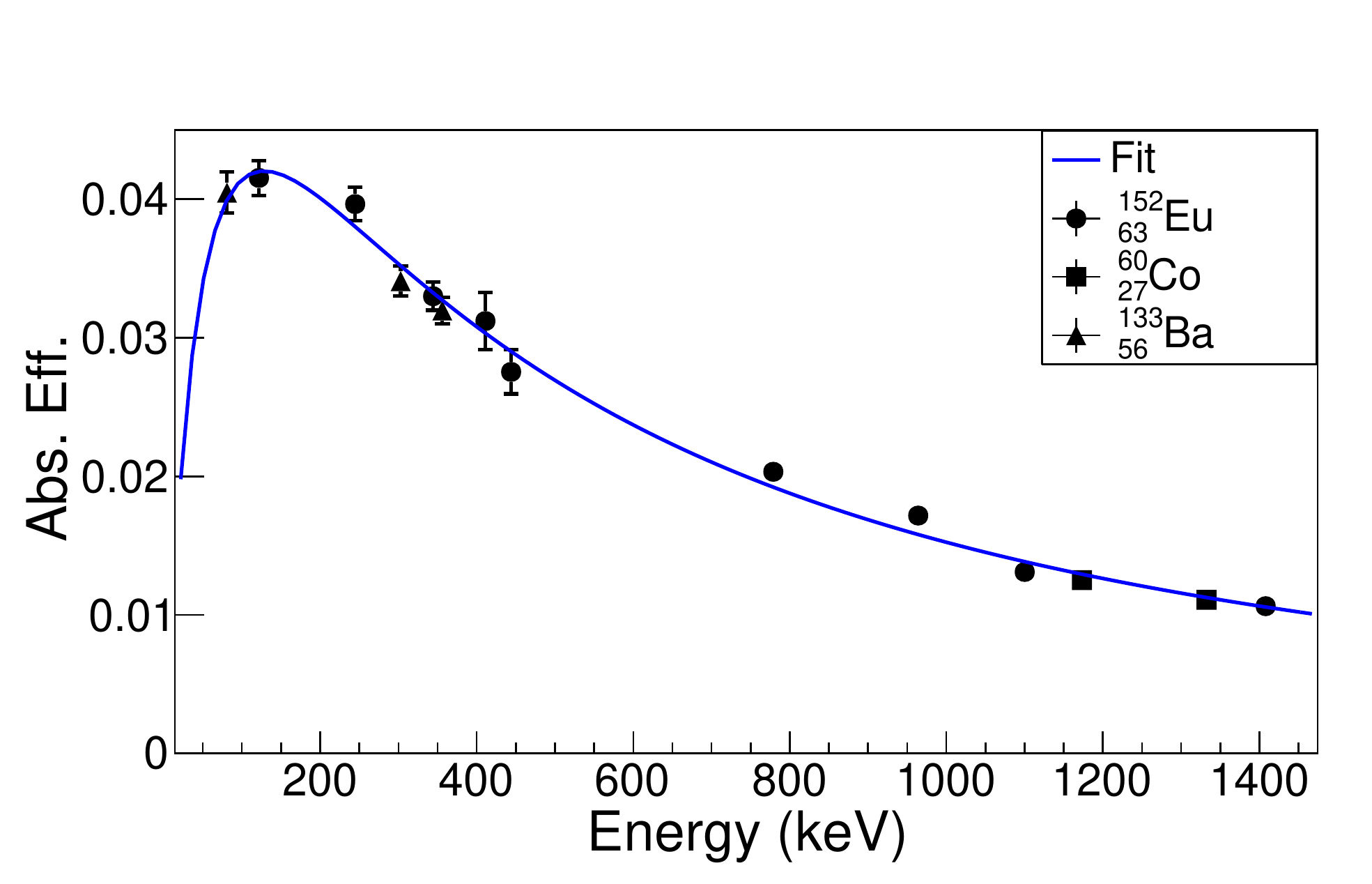}
\caption{Absolute photo-peak efficiency of the 8-detector LaBr$_3$(Ce) array in GRIFFIN, with PACES and the ZDS around the implantation point and no Delrin absorbers. The data were taken with $^{60}$Co (\textit{squares}), $^{133}$Ba (\textit{triangles)} and $^{152}$Eu (\textit{circles}) calibration sources.}
\label{fig:LaBr-efficiency}
\end{figure}

The activity of each source was known with good precision ($\sim 1 \%$) and the efficiency using each source was measured independently (no renormalization was applied). The combined data of all three sources is shown in Figure~\ref{fig:LaBr-efficiency}. The values have been corrected for the summing effects described in Section~\ref{sec:summing}. The data were fitted using Equation~\ref{eq:efficiency_log}, but only up to the quadratic term. The eight LaBr$_3$(Ce) crystals have an absolute photo-peak efficiency of $1.8(1) \%$ at 1\,MeV and the array has a maximum efficiency of $4.3(1) \%$ at $\sim 130$\,keV.

\section{Potential Future Developments}
\label{sec:Future}
The current location of GRIFFIN at ISAC-I allows for some flexibility in the LEBT beamlines~\cite{Baartman2014} that can be used to deliver radioactive beams to the spectrometer. In its current configuration, the beam to GRIFFIN is delivered directly from the ISOL target with no intervening experimental setup to modify the beam besides electrostatic ion optical components. With the addition of a bender and a short section of high-vacuum beamline, the ISAC polarizer beamline \cite{Levy2013}, which is a LEBT beamline parallel to the current RIB delivery to GRIFFIN, avails the possibility of transporting nuclear-spin polarized beam or beam that is isotope and/or isomer purified using resonant ionization. This section briefly outlines the developments in beam delivery and the associated experiments that will be made possible. It also describes the future upgrade of the SCEPTAR detector array that would be needed for the polarized beam experiments.

\subsection{Polarized Beam}
The polarizer beamline \cite{Levy2013} has the capability of optically pumping the ground state of an ionic or atomic species to preferentially populate magnetic substates in the electronic and nuclear levels coupled through the hyperfine interaction. This process leads to a nuclear-spin polarized beam which then needs to be transported to the point of implantation (for example the GRIFFIN tape system) while maintaining the polarization. To achieve this, the beamline transporting the polarized ions will be surrounded by magnetic coils that will supply a weak guide field (1\,mT).

Decay spectroscopy using nuclear-spin polarized beams would enable the assignment of spins and parities of daughter levels by utilizing the asymmetric $\beta$ decay of the polarized parent nucleus. This method is complimentary to assignments using $\gamma-\gamma$ angular correlations (outlined in Section \ref{sec:gamma-gamma-correlations} and Reference \cite{Smith2018}) and will usually provide access to additional states. This capability can be advantageous in certain situations where the high statistics may not be available, or the level is not involved in a suitable $\gamma-\gamma$ cascade. The method is outlined in Reference \cite{Shimoda2013} in the study of $^{31}$Na$\rightarrow ^{31}$Mg decay using a different experimental setup at TRIUMF. Performing such experiments with the GRIFFIN spectrometer will provide a more effective and dedicated decay-spectroscopy setup with significantly enhanced $\gamma$-detection efficiency. Spin and parity assignments to excited states in nuclei are an essential piece of experimental information for detailed tests of theoretical models.

One modification that will be needed for such experiments is an upgrade of the SCEPTAR detector for $\beta$ tagging due to the need for determining the $\beta$ asymmetry. The upgrade of SCEPTAR is described in Section \ref{sec:beta-upgrade}.

\subsection{Resonant Ionization}
The ability to transport radioactive beam from the polarizer beamline to the GRIFFIN setup would also open up the possibility of conducting highly selective decay-spectroscopy experiments using laser-assisted decay spectroscopy \cite{Lynch2015,Lynch2012}. This powerful technique combined with the detection sensitivity provided by the GRIFFIN spectrometer and its ancillary detector suites would enable experiments where clean decay pathways and branching ratios would be determined from specific ground or isomeric states in the parent nucleus. The polarizer beamline is currently used for collinear laser spectroscopy experiments with optical detection to study nuclear moments and nuclear charge radii from the collected hyperfine structure spectra. The collinear setup would be augmented by a second high-power laser, which will be used to ionize the resonantly excited atoms. To efficiently achieve resonant ionization, localised to the region where the resonantly excited atoms will interact with the high-power ionizing laser, the vacuum in the polarizer beamline will be enhanced to reduce the background dominating collisional ionisation with the residual gas in the beamline. The beamline extending from the polarizer beamline to the GRIFFIN setup would not require any vacuum enhancement from additional pumping since it will already be transporting ions. 

\subsection{$\beta$ detectors}
\label{sec:beta-upgrade}
An upgrade to the SCEPTAR detector is foreseen in order to optimize the geometry to match that of the GRIFFIN spectrometer.
A new design will follow the pattern laid out by the faces of the individual GRIFFIN HPGe crystals resulting in the same rhombicuboctahedral geometry as the superstructure that holds the GRIFFIN detectors. Each crystal face will have a scintillator in front of it, following the design of the original SCEPTAR detector that mapped the faces of the 8$\pi$ detectors, to provide a veto for the bremsstrahlung photons that would otherwise contribute to the $\gamma$-ray background. In addition, eight triangular scintillators would occupy the space in front of each ancillary detector position. The technology to be employed for the new design will likely take advantage of compact silicon-photomultiplier (SiPM) arrays attached directly to thin plastic scintillators in place of the current photomultiplier tubes connected through complex light guides.

\section{Summary}
\label{sec:summary}
Gamma-Ray Infrastructure For Fundamental Investigations of Nuclei, GRIFFIN, is a new high-efficiency $\gamma$-ray spectrometer designed for use in decay spectroscopy experiments with stopped low-energy radioactive ion beams
provided by TRIUMF's Isotope Separator and Accelerator (ISAC-I) radioactive
ion beam facility. 
GRIFFIN is composed of sixteen Compton-suppressed large-volume 
HPGe clover detectors \cite{Rizwan2016} combined with a suite of ancillary detection systems coupled to a custom digital data acquisition system \cite{Garnsworthy2017}. 
The infrastructure and detectors of the spectrometer have been described in detail. The performance characteristics and the general analysis techniques applied to the experimental data are discussed with examples.
The performance of the Compton and background suppression shields will be described in a subsequent publication. 

The system is now fully commissioned and in operation at TRIUMF-ISAC-I. Examples of early measurements performed with the device are half-life measurements of astrophysical interest in the neutron-rich Cd isotopes \cite{Dunlop2016}, spectroscopy of $^{50}$Sc related to the rare magnetic octupole transition \cite{Garnsworthy2017-50Sc}, and in support of a precision half-life measurement of the $^{22}$Mg Fermi superallowed $\beta$ emitter \cite{Dunlop2017}.
The GRIFFIN facility will support a broad spectrum of research in the areas of nuclear structure, nuclear astrophysics and fundamental symmetries.

\section{Acknowledgements}
The first phase of the GRIFFIN infrastructure has been funded jointly by the Canada Foundation for Innovation, TRIUMF and the University of Guelph. The second phase of the GRIFFIN Compton and background suppression shields infrastructure has been funded jointly by the Canada Foundation for Innovation, the British Columbia Knowledge Development Fund (BCKDF) and the Ontario Ministry of Research and Innovation (ON-MRI).
TRIUMF receives funding through a contribution agreement through the National Research Council Canada. C.E.S. acknowledges support from the Canada Research Chairs program. This work was supported by the Natural Sciences and Engineering Research Council of Canada.

\bibliographystyle{apsrev}
\bibliography{GRIFFIN_Array_NIM_Paper}

\begin{thebibliography}{67}
\expandafter\ifx\csname natexlab\endcsname\relax\def\natexlab#1{#1}\fi
\expandafter\ifx\csname bibnamefont\endcsname\relax
  \def\bibnamefont#1{#1}\fi
\expandafter\ifx\csname bibfnamefont\endcsname\relax
  \def\bibfnamefont#1{#1}\fi
\expandafter\ifx\csname citenamefont\endcsname\relax
  \def\citenamefont#1{#1}\fi
\expandafter\ifx\csname url\endcsname\relax
  \def\url#1{\texttt{#1}}\fi
\expandafter\ifx\csname urlprefix\endcsname\relax\def\urlprefix{URL }\fi
\providecommand{\bibinfo}[2]{#2}
\providecommand{\eprint}[2][]{\url{#2}}

\bibitem[{\citenamefont{Nishimura}(2012)}]{EURIKA}
\bibinfo{author}{\bibfnamefont{S.}~\bibnamefont{Nishimura}},
  \bibinfo{journal}{Progress of Theoretical and Experimental Physics}
  \textbf{\bibinfo{volume}{2012}}, \bibinfo{pages}{03C006}
  (\bibinfo{year}{2012}),
  \eprint{/oup/backfile/content_public/journal/ptep/2012/1/10.1093_ptep_pts078/1/pts078.pdf},
  \urlprefix\url{http://dx.doi.org/10.1093/ptep/pts078}.

\bibitem[{\citenamefont{Prisciandaro et~al.}(2003)\citenamefont{Prisciandaro,
  Morton, and Mantica}}]{NSCL-BCS}
\bibinfo{author}{\bibfnamefont{J.~I.} \bibnamefont{Prisciandaro}},
  \bibinfo{author}{\bibfnamefont{A.~C.} \bibnamefont{Morton}},
  \bibnamefont{and} \bibinfo{author}{\bibfnamefont{P.~F.}
  \bibnamefont{Mantica}}, \bibinfo{journal}{Nuclear Instruments and Methods in
  Physics Research Section A: Accelerators, Spectrometers, Detectors and
  Associated Equipment} \textbf{\bibinfo{volume}{505}}, \bibinfo{pages}{140 }
  (\bibinfo{year}{2003}), ISSN \bibinfo{issn}{0168-9002},
  \bibinfo{note}{proceedings of the tenth Symposium on Radiation Measurements
  and Applications},
  \urlprefix\url{http://www.sciencedirect.com/science/article/pii/S0168900203010374}.

\bibitem[{\citenamefont{Mitchell et~al.}(2014)\citenamefont{Mitchell, Bertone,
  DiGiovine, Lister, Carpenter, Chowdhury, Clark, D'Olympia, Deo, Kondev
  et~al.}}]{XARRAY}
\bibinfo{author}{\bibfnamefont{A.~J.} \bibnamefont{Mitchell}},
  \bibinfo{author}{\bibfnamefont{P.~F.} \bibnamefont{Bertone}},
  \bibinfo{author}{\bibfnamefont{B.}~\bibnamefont{DiGiovine}},
  \bibinfo{author}{\bibfnamefont{C.~J.} \bibnamefont{Lister}},
  \bibinfo{author}{\bibfnamefont{M.~P.} \bibnamefont{Carpenter}},
  \bibinfo{author}{\bibfnamefont{P.}~\bibnamefont{Chowdhury}},
  \bibinfo{author}{\bibfnamefont{J.~A.} \bibnamefont{Clark}},
  \bibinfo{author}{\bibfnamefont{N.}~\bibnamefont{D'Olympia}},
  \bibinfo{author}{\bibfnamefont{A.~Y.} \bibnamefont{Deo}},
  \bibinfo{author}{\bibfnamefont{F.~G.} \bibnamefont{Kondev}},
  \bibnamefont{et~al.}, \bibinfo{journal}{Nuclear Instruments and Methods in
  Physics Research Section A: Accelerators, Spectrometers, Detectors and
  Associated Equipment} \textbf{\bibinfo{volume}{763}}, \bibinfo{pages}{232 }
  (\bibinfo{year}{2014}), ISSN \bibinfo{issn}{0168-9002},
  \urlprefix\url{http://www.sciencedirect.com/science/article/pii/S0168900214007311}.

\bibitem[{\citenamefont{Fynbo et~al.}(2017)\citenamefont{Fynbo, Kirseboom, and
  Tengblad}}]{ISOLDE-IDS}
\bibinfo{author}{\bibfnamefont{H.}~\bibnamefont{Fynbo}},
  \bibinfo{author}{\bibfnamefont{O.~S.} \bibnamefont{Kirseboom}},
  \bibnamefont{and} \bibinfo{author}{\bibfnamefont{O.}~\bibnamefont{Tengblad}},
  \bibinfo{journal}{Journal of Physics G: Nuclear and Particle Physics}
  \textbf{\bibinfo{volume}{44}}, \bibinfo{pages}{044005}
  (\bibinfo{year}{2017}),
  \urlprefix\url{http://stacks.iop.org/0954-3899/44/i=4/a=044005}.

\bibitem[{\citenamefont{He et~al.}(2014)\citenamefont{He, Li, Hua, Li, Jiang,
  Ye, Zhang, Han, Xu, Wang et~al.}}]{Lanzhou-DecayStation}
\bibinfo{author}{\bibfnamefont{C.}~\bibnamefont{He}},
  \bibinfo{author}{\bibfnamefont{X.}~\bibnamefont{Li}},
  \bibinfo{author}{\bibfnamefont{H.}~\bibnamefont{Hua}},
  \bibinfo{author}{\bibfnamefont{Z.}~\bibnamefont{Li}},
  \bibinfo{author}{\bibfnamefont{D.}~\bibnamefont{Jiang}},
  \bibinfo{author}{\bibfnamefont{Y.}~\bibnamefont{Ye}},
  \bibinfo{author}{\bibfnamefont{S.}~\bibnamefont{Zhang}},
  \bibinfo{author}{\bibfnamefont{R.}~\bibnamefont{Han}},
  \bibinfo{author}{\bibfnamefont{C.}~\bibnamefont{Xu}},
  \bibinfo{author}{\bibfnamefont{E.}~\bibnamefont{Wang}}, \bibnamefont{et~al.},
  \bibinfo{journal}{Nuclear Instruments and Methods in Physics Research Section
  A: Accelerators, Spectrometers, Detectors and Associated Equipment}
  \textbf{\bibinfo{volume}{747}}, \bibinfo{pages}{52 } (\bibinfo{year}{2014}),
  ISSN \bibinfo{issn}{0168-9002},
  \urlprefix\url{http://www.sciencedirect.com/science/article/pii/S016890021400179X}.

\bibitem[{\citenamefont{Grinyer et~al.}(2014)\citenamefont{Grinyer, Thomas,
  Blank, Bouzomita, Austin, Ball, Bucaille, Delahaye, Finlay, Fr\'emont
  et~al.}}]{GANIL-ID-Station}
\bibinfo{author}{\bibfnamefont{G.~F.} \bibnamefont{Grinyer}},
  \bibinfo{author}{\bibfnamefont{J.~C.} \bibnamefont{Thomas}},
  \bibinfo{author}{\bibfnamefont{B.}~\bibnamefont{Blank}},
  \bibinfo{author}{\bibfnamefont{H.}~\bibnamefont{Bouzomita}},
  \bibinfo{author}{\bibfnamefont{R.~A.~E.} \bibnamefont{Austin}},
  \bibinfo{author}{\bibfnamefont{G.~C.} \bibnamefont{Ball}},
  \bibinfo{author}{\bibfnamefont{F.}~\bibnamefont{Bucaille}},
  \bibinfo{author}{\bibfnamefont{P.}~\bibnamefont{Delahaye}},
  \bibinfo{author}{\bibfnamefont{P.}~\bibnamefont{Finlay}},
  \bibinfo{author}{\bibfnamefont{G.}~\bibnamefont{Fr\'emont}},
  \bibnamefont{et~al.}, \bibinfo{journal}{Nuclear Instruments and Methods in
  Physics Research Section A: Accelerators, Spectrometers, Detectors and
  Associated Equipment} \textbf{\bibinfo{volume}{741}}, \bibinfo{pages}{18 }
  (\bibinfo{year}{2014}), ISSN \bibinfo{issn}{0168-9002},
  \urlprefix\url{http://www.sciencedirect.com/science/article/pii/S0168900213016719}.

\bibitem[{\citenamefont{Pietri et~al.}(2007)\citenamefont{Pietri, Regan,
  Podoly\'{a}k, Rudolph, Steer, Garnsworthy, Werner-Malento, Hoischen,
  G\'{o}rska, Gerl et~al.}}]{RISING-Isomer}
\bibinfo{author}{\bibfnamefont{S.}~\bibnamefont{Pietri}},
  \bibinfo{author}{\bibfnamefont{P.~H.} \bibnamefont{Regan}},
  \bibinfo{author}{\bibfnamefont{Z.}~\bibnamefont{Podoly\'{a}k}},
  \bibinfo{author}{\bibfnamefont{D.}~\bibnamefont{Rudolph}},
  \bibinfo{author}{\bibfnamefont{S.}~\bibnamefont{Steer}},
  \bibinfo{author}{\bibfnamefont{A.~B.} \bibnamefont{Garnsworthy}},
  \bibinfo{author}{\bibfnamefont{E.}~\bibnamefont{Werner-Malento}},
  \bibinfo{author}{\bibfnamefont{R.}~\bibnamefont{Hoischen}},
  \bibinfo{author}{\bibfnamefont{M.}~\bibnamefont{G\'{o}rska}},
  \bibinfo{author}{\bibfnamefont{J.}~\bibnamefont{Gerl}}, \bibnamefont{et~al.},
  \bibinfo{journal}{Nuclear Instruments and Methods in Physics Research Section
  B: Beam Interactions with Materials and Atoms}
  \textbf{\bibinfo{volume}{261}}, \bibinfo{pages}{1079 }
  (\bibinfo{year}{2007}), ISSN \bibinfo{issn}{0168-583X}, \bibinfo{note}{the
  Application of Accelerators in Research and Industry},
  \urlprefix\url{http://www.sciencedirect.com/science/article/pii/S0168583X07010294}.

\bibitem[{\citenamefont{Kumar et~al.}(2009)\citenamefont{Kumar, Molina, Pietri,
  Casarejos, Algora, Benlliure, Doornenbal, Gerl, G\'orska, Kojouharov
  et~al.}}]{RISING-ActiveStopper}
\bibinfo{author}{\bibfnamefont{R.}~\bibnamefont{Kumar}},
  \bibinfo{author}{\bibfnamefont{F.~G.} \bibnamefont{Molina}},
  \bibinfo{author}{\bibfnamefont{S.}~\bibnamefont{Pietri}},
  \bibinfo{author}{\bibfnamefont{E.}~\bibnamefont{Casarejos}},
  \bibinfo{author}{\bibfnamefont{A.}~\bibnamefont{Algora}},
  \bibinfo{author}{\bibfnamefont{J.}~\bibnamefont{Benlliure}},
  \bibinfo{author}{\bibfnamefont{P.}~\bibnamefont{Doornenbal}},
  \bibinfo{author}{\bibfnamefont{J.}~\bibnamefont{Gerl}},
  \bibinfo{author}{\bibfnamefont{M.}~\bibnamefont{G\'orska}},
  \bibinfo{author}{\bibfnamefont{I.}~\bibnamefont{Kojouharov}},
  \bibnamefont{et~al.}, \bibinfo{journal}{Nuclear Instruments and Methods in
  Physics Research Section A: Accelerators, Spectrometers, Detectors and
  Associated Equipment} \textbf{\bibinfo{volume}{598}}, \bibinfo{pages}{754 }
  (\bibinfo{year}{2009}), ISSN \bibinfo{issn}{0168-9002},
  \urlprefix\url{http://www.sciencedirect.com/science/article/pii/S0168900208013259}.

\bibitem[{\citenamefont{Podoly\'ak}(2008)}]{DESPEC}
\bibinfo{author}{\bibfnamefont{Z.}~\bibnamefont{Podoly\'ak}},
  \bibinfo{journal}{Nuclear Instruments and Methods in Physics Research Section
  B: Beam Interactions with Materials and Atoms}
  \textbf{\bibinfo{volume}{266}}, \bibinfo{pages}{4589 }
  (\bibinfo{year}{2008}), ISSN \bibinfo{issn}{0168-583X},
  \bibinfo{note}{proceedings of the XVth International Conference on
  Electromagnetic Isotope Separators and Techniques Related to their
  Applications},
  \urlprefix\url{http://www.sciencedirect.com/science/article/pii/S0168583X08007829}.

\bibitem[{\citenamefont{Svensson and Garnsworthy}(2014)}]{Svensson14}
\bibinfo{author}{\bibfnamefont{C.~E.} \bibnamefont{Svensson}} \bibnamefont{and}
  \bibinfo{author}{\bibfnamefont{A.~B.} \bibnamefont{Garnsworthy}},
  \bibinfo{journal}{Hyperfine Interact.} \textbf{\bibinfo{volume}{225}},
  \bibinfo{pages}{127} (\bibinfo{year}{2014}).

\bibitem[{\citenamefont{Dilling and Kr\"{u}cken}(2014)}]{Dilling14}
\bibinfo{author}{\bibfnamefont{J.}~\bibnamefont{Dilling}} \bibnamefont{and}
  \bibinfo{author}{\bibfnamefont{R.}~\bibnamefont{Kr\"{u}cken}},
  \bibinfo{journal}{Hyperfine Interact.} \textbf{\bibinfo{volume}{225}},
  \bibinfo{pages}{111} (\bibinfo{year}{2014}).

\bibitem[{\citenamefont{Rizwan et~al.}(2016)\citenamefont{Rizwan, Garnsworthy,
  Andreoiu, Ball, Chester, Domingo, Dunlop, Hackman, Rand, Smith
  et~al.}}]{Rizwan2016}
\bibinfo{author}{\bibfnamefont{U.}~\bibnamefont{Rizwan}},
  \bibinfo{author}{\bibfnamefont{A.~B.} \bibnamefont{Garnsworthy}},
  \bibinfo{author}{\bibfnamefont{C.}~\bibnamefont{Andreoiu}},
  \bibinfo{author}{\bibfnamefont{G.~C.} \bibnamefont{Ball}},
  \bibinfo{author}{\bibfnamefont{A.}~\bibnamefont{Chester}},
  \bibinfo{author}{\bibfnamefont{T.}~\bibnamefont{Domingo}},
  \bibinfo{author}{\bibfnamefont{R.}~\bibnamefont{Dunlop}},
  \bibinfo{author}{\bibfnamefont{G.}~\bibnamefont{Hackman}},
  \bibinfo{author}{\bibfnamefont{E.~T.} \bibnamefont{Rand}},
  \bibinfo{author}{\bibfnamefont{J.~K.} \bibnamefont{Smith}},
  \bibnamefont{et~al.}, \bibinfo{journal}{Nucl. Instruments Methods Phys. Res.
  Sect. A Accel. Spectrometers, Detect. Assoc. Equip.}
  \textbf{\bibinfo{volume}{820}}, \bibinfo{pages}{126} (\bibinfo{year}{2016}),
  ISSN \bibinfo{issn}{01689002},
  \urlprefix\url{http://www.sciencedirect.com/science/article/pii/S0168900216300341}.

\bibitem[{\citenamefont{Garnsworthy et~al.}(2015)}]{Garnsworthy2015}
\bibinfo{author}{\bibfnamefont{A.~B.} \bibnamefont{Garnsworthy}}
  \bibnamefont{et~al.}, \bibinfo{journal}{EPJ Web of Conferences}
  \textbf{\bibinfo{volume}{93}}, \bibinfo{pages}{01032} (\bibinfo{year}{2015}).

\bibitem[{\citenamefont{Garrett et~al.}(2015)}]{Garrett2015}
\bibinfo{author}{\bibfnamefont{P.~E.} \bibnamefont{Garrett}}
  \bibnamefont{et~al.}, \bibinfo{journal}{Journal of Physics Conference Series}
  \textbf{\bibinfo{volume}{639}}, \bibinfo{pages}{012006}
  (\bibinfo{year}{2015}).

\bibitem[{\citenamefont{Garnsworthy and Garrett}(2014)}]{Garnsworthy2014}
\bibinfo{author}{\bibfnamefont{A.~B.} \bibnamefont{Garnsworthy}}
  \bibnamefont{and} \bibinfo{author}{\bibfnamefont{P.~E.}
  \bibnamefont{Garrett}}, \bibinfo{journal}{Hyperfine Interact.}
  \textbf{\bibinfo{volume}{225}}, \bibinfo{pages}{121} (\bibinfo{year}{2014}).

\bibitem[{\citenamefont{Ball et~al.}(2012)}]{Ball2012}
\bibinfo{author}{\bibfnamefont{G.~C.} \bibnamefont{Ball}} \bibnamefont{et~al.},
  \bibinfo{journal}{J. Phys. Conf. Ser.} \textbf{\bibinfo{volume}{387}},
  \bibinfo{pages}{012014} (\bibinfo{year}{2012}).

\bibitem[{\citenamefont{Garrett et~al.}(2007)}]{Garrett2007}
\bibinfo{author}{\bibfnamefont{P.~E.} \bibnamefont{Garrett}}
  \bibnamefont{et~al.}, \bibinfo{journal}{Nucl. Instruments Methods Phys. Res.
  Sect. B} \textbf{\bibinfo{volume}{261}}, \bibinfo{pages}{1084}
  (\bibinfo{year}{2007}).

\bibitem[{\citenamefont{Zganjar et~al.}(2007)}]{Zganyar2007}
\bibinfo{author}{\bibfnamefont{E.~F.} \bibnamefont{Zganjar}}
  \bibnamefont{et~al.}, \bibinfo{journal}{Acta. Phys. Pol. B}
  \textbf{\bibinfo{volume}{38}}, \bibinfo{pages}{1179} (\bibinfo{year}{2007}).

\bibitem[{\citenamefont{Ball et~al.}(2005)}]{Ball2005}
\bibinfo{author}{\bibfnamefont{G.~C.} \bibnamefont{Ball}} \bibnamefont{et~al.},
  \bibinfo{journal}{J. Phys. G (London)} \textbf{\bibinfo{volume}{31}},
  \bibinfo{pages}{S1491} (\bibinfo{year}{2005}).

\bibitem[{\citenamefont{Svensson et~al.}(2003)\citenamefont{Svensson, Austin,
  Ball, Finlay, Garrett, Grinyer, Hackman, Osborne, Sarazin, Scraggs
  et~al.}}]{Svensson2003}
\bibinfo{author}{\bibfnamefont{C.~E.} \bibnamefont{Svensson}},
  \bibinfo{author}{\bibfnamefont{R.~A.~E.} \bibnamefont{Austin}},
  \bibinfo{author}{\bibfnamefont{G.~C.} \bibnamefont{Ball}},
  \bibinfo{author}{\bibfnamefont{P.}~\bibnamefont{Finlay}},
  \bibinfo{author}{\bibfnamefont{P.~E.} \bibnamefont{Garrett}},
  \bibinfo{author}{\bibfnamefont{G.~F.} \bibnamefont{Grinyer}},
  \bibinfo{author}{\bibfnamefont{G.~S.} \bibnamefont{Hackman}},
  \bibinfo{author}{\bibfnamefont{C.~J.} \bibnamefont{Osborne}},
  \bibinfo{author}{\bibfnamefont{F.}~\bibnamefont{Sarazin}},
  \bibinfo{author}{\bibfnamefont{H.~C.} \bibnamefont{Scraggs}},
  \bibnamefont{et~al.}, \bibinfo{journal}{Nuclear Instruments and Methods in
  Physics Research Section B: Beam Interactions with Materials and Atoms}
  \textbf{\bibinfo{volume}{204}}, \bibinfo{pages}{660 } (\bibinfo{year}{2003}),
  ISSN \bibinfo{issn}{0168-583X}, \bibinfo{note}{14th International Conference
  on Electromagnetic Isotope Separators and Techniques Related to their
  Applications},
  \urlprefix\url{http://www.sciencedirect.com/science/article/pii/S0168583X0202147X}.

\bibitem[{\citenamefont{Garrett}(2014)}]{Garrett2014}
\bibinfo{author}{\bibfnamefont{P.~E.} \bibnamefont{Garrett}},
  \bibinfo{journal}{Hyperfine Interact.} \textbf{\bibinfo{volume}{225}},
  \bibinfo{pages}{137} (\bibinfo{year}{2014}).

\bibitem[{\citenamefont{Bildstein et~al.}(2015)\citenamefont{Bildstein,
  Garrett, Ashley, Ball, Bianco, Bandyopadhyay, Bangay, Crider, Demand, Deng
  et~al.}}]{Bildstein2015}
\bibinfo{author}{\bibfnamefont{V.}~\bibnamefont{Bildstein}},
  \bibinfo{author}{\bibfnamefont{P.}~\bibnamefont{Garrett}},
  \bibinfo{author}{\bibfnamefont{S.~F.} \bibnamefont{Ashley}},
  \bibinfo{author}{\bibfnamefont{G.~C.} \bibnamefont{Ball}},
  \bibinfo{author}{\bibfnamefont{L.}~\bibnamefont{Bianco}},
  \bibinfo{author}{\bibfnamefont{D.}~\bibnamefont{Bandyopadhyay}},
  \bibinfo{author}{\bibfnamefont{J.}~\bibnamefont{Bangay}},
  \bibinfo{author}{\bibfnamefont{B.~P.} \bibnamefont{Crider}},
  \bibinfo{author}{\bibfnamefont{G.}~\bibnamefont{Demand}},
  \bibinfo{author}{\bibfnamefont{G.}~\bibnamefont{Deng}}, \bibnamefont{et~al.},
  \bibinfo{journal}{EPJ Web of Conferences} \textbf{\bibinfo{volume}{93}},
  \bibinfo{pages}{07005} (\bibinfo{year}{2015}).

\bibitem[{\citenamefont{Garnsworthy
  et~al.}(2017{\natexlab{a}})\citenamefont{Garnsworthy, Pearson, Bishop, Shaw,
  Smith, Bowry, Bildstein, Hackman, Garrett, Linn et~al.}}]{Garnsworthy2017}
\bibinfo{author}{\bibfnamefont{A.~B.} \bibnamefont{Garnsworthy}},
  \bibinfo{author}{\bibfnamefont{C.~J.} \bibnamefont{Pearson}},
  \bibinfo{author}{\bibfnamefont{D.}~\bibnamefont{Bishop}},
  \bibinfo{author}{\bibfnamefont{B.}~\bibnamefont{Shaw}},
  \bibinfo{author}{\bibfnamefont{J.~K.} \bibnamefont{Smith}},
  \bibinfo{author}{\bibfnamefont{M.}~\bibnamefont{Bowry}},
  \bibinfo{author}{\bibfnamefont{V.}~\bibnamefont{Bildstein}},
  \bibinfo{author}{\bibfnamefont{G.}~\bibnamefont{Hackman}},
  \bibinfo{author}{\bibfnamefont{P.~E.} \bibnamefont{Garrett}},
  \bibinfo{author}{\bibfnamefont{Y.}~\bibnamefont{Linn}}, \bibnamefont{et~al.},
  \bibinfo{journal}{Nucl. Instruments Methods Phys. Res. Sect. A Accel.
  Spectrometers, Detect. Assoc. Equip.} \textbf{\bibinfo{volume}{853}},
  \bibinfo{pages}{85} (\bibinfo{year}{2017}{\natexlab{a}}).

\bibitem[{\citenamefont{Baartman}(2014)}]{Baartman2014}
\bibinfo{author}{\bibfnamefont{R.}~\bibnamefont{Baartman}},
  \bibinfo{journal}{Hyperfine Interact.} \textbf{\bibinfo{volume}{225}},
  \bibinfo{pages}{69} (\bibinfo{year}{2014}).

\bibitem[{\citenamefont{Biersack and Haggmark}(1980)}]{SRIM-2013}
\bibinfo{author}{\bibfnamefont{J.~P.} \bibnamefont{Biersack}} \bibnamefont{and}
  \bibinfo{author}{\bibfnamefont{L.}~\bibnamefont{Haggmark}},
  \bibinfo{journal}{Nucl. Instruments and Methods}
  \textbf{\bibinfo{volume}{174}}, \bibinfo{pages}{257} (\bibinfo{year}{1980}).

\bibitem[{BOS()}]{BOSCH}
\emph{\bibinfo{title}{Bosch-rexroth aluminum framing resource center}},
  \urlprefix\url{http://www13.boschrexroth-us.com/framing_shop/Default.aspx}.

\bibitem[{NIM()}]{NIM}
\emph{\bibinfo{title}{Standard nim instrumentation system (doe/er-0457t)}},
  \urlprefix\url{http://www.osti.gov/energycitations/servlets/purl/7120327-MV8wop/7120327.PDF}.

\bibitem[{CAE()}]{CAEN}
\emph{\bibinfo{title}{Caen - costruzioni apparecchiature elettroniche nucleari
  s.p.a.}}, \urlprefix\url{http://www.caen.it}.

\bibitem[{ise()}]{iseg}
\emph{\bibinfo{title}{iseg spezialelektronik gmbh}},
  \urlprefix\url{http://iseg-hv.com/}.

\bibitem[{\citenamefont{Ritt and Amaudruz}(1997)}]{MIDAS}
\bibinfo{author}{\bibfnamefont{S.}~\bibnamefont{Ritt}} \bibnamefont{and}
  \bibinfo{author}{\bibfnamefont{P.-A.} \bibnamefont{Amaudruz}},
  \bibinfo{journal}{Proc. 10th IEEE Real Time Conf. (Beaune)} p.
  \bibinfo{pages}{309} (\bibinfo{year}{1997}).

\bibitem[{\citenamefont{Svensson et~al.}(2005)\citenamefont{Svensson, Amaudruz,
  Andreoiu, Andreyev, Austin, Ball, Bandyopadhyay, Boston, Chakrawarthy, Chen
  et~al.}}]{Svensson2005}
\bibinfo{author}{\bibfnamefont{C.~E.} \bibnamefont{Svensson}},
  \bibinfo{author}{\bibfnamefont{P.}~\bibnamefont{Amaudruz}},
  \bibinfo{author}{\bibfnamefont{C.}~\bibnamefont{Andreoiu}},
  \bibinfo{author}{\bibfnamefont{A.}~\bibnamefont{Andreyev}},
  \bibinfo{author}{\bibfnamefont{R.~A.~E.} \bibnamefont{Austin}},
  \bibinfo{author}{\bibfnamefont{G.~C.} \bibnamefont{Ball}},
  \bibinfo{author}{\bibfnamefont{D.}~\bibnamefont{Bandyopadhyay}},
  \bibinfo{author}{\bibfnamefont{A.~J.} \bibnamefont{Boston}},
  \bibinfo{author}{\bibfnamefont{R.~S.} \bibnamefont{Chakrawarthy}},
  \bibinfo{author}{\bibfnamefont{A.~A.} \bibnamefont{Chen}},
  \bibnamefont{et~al.}, \bibinfo{journal}{J. Phys. G}
  \textbf{\bibinfo{volume}{31}}, \bibinfo{pages}{S1663} (\bibinfo{year}{2005}),
  \urlprefix\url{http://stacks.iop.org/0954-3899/31/i=10/a=050}.

\bibitem[{\citenamefont{Hackman and Svensson}(2014)}]{Hackman2014}
\bibinfo{author}{\bibfnamefont{G.}~\bibnamefont{Hackman}} \bibnamefont{and}
  \bibinfo{author}{\bibfnamefont{C.~E.} \bibnamefont{Svensson}},
  \bibinfo{journal}{Hyperfine Interact.} \textbf{\bibinfo{volume}{225}},
  \bibinfo{pages}{241} (\bibinfo{year}{2014}).

\bibitem[{\citenamefont{Schumaker et~al.}(2007)\citenamefont{Schumaker,
  Hackman, Pearson, Svensson, Andreoiu, Andreyev, Austin, Ball, Bandyopadhyay,
  Boston et~al.}}]{Schumaker2007}
\bibinfo{author}{\bibfnamefont{M.~A.} \bibnamefont{Schumaker}},
  \bibinfo{author}{\bibfnamefont{G.}~\bibnamefont{Hackman}},
  \bibinfo{author}{\bibfnamefont{C.~J.} \bibnamefont{Pearson}},
  \bibinfo{author}{\bibfnamefont{C.~E.} \bibnamefont{Svensson}},
  \bibinfo{author}{\bibfnamefont{C.}~\bibnamefont{Andreoiu}},
  \bibinfo{author}{\bibfnamefont{A.}~\bibnamefont{Andreyev}},
  \bibinfo{author}{\bibfnamefont{R.~A.~E.} \bibnamefont{Austin}},
  \bibinfo{author}{\bibfnamefont{G.~C.} \bibnamefont{Ball}},
  \bibinfo{author}{\bibfnamefont{D.}~\bibnamefont{Bandyopadhyay}},
  \bibinfo{author}{\bibfnamefont{A.~J.} \bibnamefont{Boston}},
  \bibnamefont{et~al.}, \bibinfo{journal}{Nuclear Instruments and Methods in
  Physics Research Section A: Accelerators, Spectrometers, Detectors and
  Associated Equipment} \textbf{\bibinfo{volume}{570}}, \bibinfo{pages}{437 }
  (\bibinfo{year}{2007}), ISSN \bibinfo{issn}{0168-9002},
  \urlprefix\url{http://www.sciencedirect.com/science/article/pii/S0168900206019024}.

\bibitem[{\citenamefont{Bildstein et~al.}(2013)\citenamefont{Bildstein,
  Garrett, Wong, Bandyopadhyay, Bangay, Bianco, Hadinia, Leach,
  Sumithrarachchi, Ashley et~al.}}]{Bildstein2013}
\bibinfo{author}{\bibfnamefont{V.}~\bibnamefont{Bildstein}},
  \bibinfo{author}{\bibfnamefont{P.}~\bibnamefont{Garrett}},
  \bibinfo{author}{\bibfnamefont{J.}~\bibnamefont{Wong}},
  \bibinfo{author}{\bibfnamefont{D.}~\bibnamefont{Bandyopadhyay}},
  \bibinfo{author}{\bibfnamefont{J.}~\bibnamefont{Bangay}},
  \bibinfo{author}{\bibfnamefont{L.}~\bibnamefont{Bianco}},
  \bibinfo{author}{\bibfnamefont{B.}~\bibnamefont{Hadinia}},
  \bibinfo{author}{\bibfnamefont{K.}~\bibnamefont{Leach}},
  \bibinfo{author}{\bibfnamefont{C.}~\bibnamefont{Sumithrarachchi}},
  \bibinfo{author}{\bibfnamefont{S.}~\bibnamefont{Ashley}},
  \bibnamefont{et~al.}, \bibinfo{journal}{Nuclear Instruments and Methods in
  Physics Research Section A: Accelerators, Spectrometers, Detectors and
  Associated Equipment} \textbf{\bibinfo{volume}{729}}, \bibinfo{pages}{188 }
  (\bibinfo{year}{2013}),
  \urlprefix\url{http://www.sciencedirect.com/science/article/pii/S0168900213009285}.

\bibitem[{\citenamefont{Rizwan et~al.}(2015)\citenamefont{Rizwan, Chester,
  Domingo, Starosta, Williams, and Voss}}]{Rizwan2015}
\bibinfo{author}{\bibfnamefont{U.}~\bibnamefont{Rizwan}},
  \bibinfo{author}{\bibfnamefont{A.}~\bibnamefont{Chester}},
  \bibinfo{author}{\bibfnamefont{T.}~\bibnamefont{Domingo}},
  \bibinfo{author}{\bibfnamefont{K.}~\bibnamefont{Starosta}},
  \bibinfo{author}{\bibfnamefont{J.}~\bibnamefont{Williams}}, \bibnamefont{and}
  \bibinfo{author}{\bibfnamefont{P.}~\bibnamefont{Voss}},
  \bibinfo{journal}{Nuclear Instruments and Methods in Physics Research Section
  A: Accelerators, Spectrometers, Detectors and Associated Equipment}
  \textbf{\bibinfo{volume}{802}}, \bibinfo{pages}{102 } (\bibinfo{year}{2015}),
  ISSN \bibinfo{issn}{0168-9002},
  \urlprefix\url{http://www.sciencedirect.com/science/article/pii/S0168900215010359}.

\bibitem[{\citenamefont{Kis et~al.}(1998)\citenamefont{Kis, Fazekas, {\"
  O}st{\" o}r, R{\' e}vayb, Belgya, Moln{\' a}r, and Koltay}}]{Kis1998}
\bibinfo{author}{\bibfnamefont{Z.}~\bibnamefont{Kis}},
  \bibinfo{author}{\bibfnamefont{B.}~\bibnamefont{Fazekas}},
  \bibinfo{author}{\bibfnamefont{J.}~\bibnamefont{{\" O}st{\" o}r}},
  \bibinfo{author}{\bibfnamefont{Z.}~\bibnamefont{R{\' e}vayb}},
  \bibinfo{author}{\bibfnamefont{T.}~\bibnamefont{Belgya}},
  \bibinfo{author}{\bibfnamefont{G.~L.} \bibnamefont{Moln{\' a}r}},
  \bibnamefont{and} \bibinfo{author}{\bibfnamefont{L.}~\bibnamefont{Koltay}},
  \bibinfo{journal}{Nuclear Instruments and Methods in Physics Research Section
  A: Accelerators, Spectrometers, Detectors and Associated Equipment}
  \textbf{\bibinfo{volume}{418}}, \bibinfo{pages}{374 } (\bibinfo{year}{1998}),
  ISSN \bibinfo{issn}{0168-9002},
  \urlprefix\url{http://www.sciencedirect.com/science/article/pii/S0168900298007785}.

\bibitem[{\citenamefont{Rose and Brink}(1967)}]{Rose1967}
\bibinfo{author}{\bibfnamefont{H.~J.} \bibnamefont{Rose}} \bibnamefont{and}
  \bibinfo{author}{\bibfnamefont{D.~M.} \bibnamefont{Brink}},
  \bibinfo{journal}{Rev. Mod. Phys.} \textbf{\bibinfo{volume}{39}},
  \bibinfo{pages}{306} (\bibinfo{year}{1967}), ISSN \bibinfo{issn}{0034-6861},
  \urlprefix\url{http://link.aps.org/doi/10.1103/RevModPhys.39.306}.

\bibitem[{\citenamefont{Smith et~al.}(2018)\citenamefont{Smith, MacLean,
  Ashfield, Chester, Garnsworthy, and Svensson}}]{Smith2018}
\bibinfo{author}{\bibfnamefont{J.~K.} \bibnamefont{Smith}},
  \bibinfo{author}{\bibfnamefont{A.~D.} \bibnamefont{MacLean}},
  \bibinfo{author}{\bibfnamefont{W.}~\bibnamefont{Ashfield}},
  \bibinfo{author}{\bibfnamefont{A.}~\bibnamefont{Chester}},
  \bibinfo{author}{\bibfnamefont{A.~B.} \bibnamefont{Garnsworthy}},
  \bibnamefont{and} \bibinfo{author}{\bibfnamefont{C.~E.}
  \bibnamefont{Svensson}}, \bibinfo{journal}{In Press, accepted manuscript in
  Nucl. Instruments Methods Phys. Res. Sect. A Accel. Spectrometers, Detect.
  Assoc. Equip.}  (\bibinfo{year}{2018}),
  \urlprefix\url{https://doi.org/10.1016/j.nima.2018.10.097}.

\bibitem[{\citenamefont{Fagg and Hanna}(1959)}]{Fagg1959}
\bibinfo{author}{\bibfnamefont{L.~W.} \bibnamefont{Fagg}} \bibnamefont{and}
  \bibinfo{author}{\bibfnamefont{S.~S.} \bibnamefont{Hanna}},
  \bibinfo{journal}{Rev. Mod. Phys.} \textbf{\bibinfo{volume}{31}},
  \bibinfo{pages}{711} (\bibinfo{year}{1959}),
  \urlprefix\url{https://link.aps.org/doi/10.1103/RevModPhys.31.711}.

\bibitem[{\citenamefont{Alikhani et~al.}(2012)\citenamefont{Alikhani, Givechev,
  Heinz, John, Leske, Lettmann, Moller, Pietralla, and Roder}}]{Alikhani2012}
\bibinfo{author}{\bibfnamefont{B.}~\bibnamefont{Alikhani}},
  \bibinfo{author}{\bibfnamefont{A.}~\bibnamefont{Givechev}},
  \bibinfo{author}{\bibfnamefont{A.}~\bibnamefont{Heinz}},
  \bibinfo{author}{\bibfnamefont{P.}~\bibnamefont{John}},
  \bibinfo{author}{\bibfnamefont{J.}~\bibnamefont{Leske}},
  \bibinfo{author}{\bibfnamefont{M.}~\bibnamefont{Lettmann}},
  \bibinfo{author}{\bibfnamefont{O.}~\bibnamefont{Moller}},
  \bibinfo{author}{\bibfnamefont{N.}~\bibnamefont{Pietralla}},
  \bibnamefont{and} \bibinfo{author}{\bibfnamefont{C.}~\bibnamefont{Roder}},
  \bibinfo{journal}{Nuclear Instruments and Methods in Physics Research Section
  A: Accelerators, Spectrometers, Detectors and Associated Equipment}
  \textbf{\bibinfo{volume}{675}}, \bibinfo{pages}{144 } (\bibinfo{year}{2012}),
  ISSN \bibinfo{issn}{0168-9002},
  \urlprefix\url{http://www.sciencedirect.com/science/article/pii/S016890021200174X}.

\bibitem[{\citenamefont{Schlitt et~al.}(1994)\citenamefont{Schlitt, Maier,
  Friedrichs, Albers, Bauske, von Brentano, Heil, Herzberg, Kneissl, Margraf
  et~al.}}]{Schlitt1994}
\bibinfo{author}{\bibfnamefont{B.}~\bibnamefont{Schlitt}},
  \bibinfo{author}{\bibfnamefont{U.}~\bibnamefont{Maier}},
  \bibinfo{author}{\bibfnamefont{H.}~\bibnamefont{Friedrichs}},
  \bibinfo{author}{\bibfnamefont{S.}~\bibnamefont{Albers}},
  \bibinfo{author}{\bibfnamefont{I.}~\bibnamefont{Bauske}},
  \bibinfo{author}{\bibfnamefont{P.}~\bibnamefont{von Brentano}},
  \bibinfo{author}{\bibfnamefont{R.~D.} \bibnamefont{Heil}},
  \bibinfo{author}{\bibfnamefont{R.~D.} \bibnamefont{Herzberg}},
  \bibinfo{author}{\bibfnamefont{U.}~\bibnamefont{Kneissl}},
  \bibinfo{author}{\bibfnamefont{J.}~\bibnamefont{Margraf}},
  \bibnamefont{et~al.}, \bibinfo{journal}{Nuclear Instruments and Methods in
  Physics Research Section A: Accelerators, Spectrometers, Detectors and
  Associated Equipment} \textbf{\bibinfo{volume}{337}}, \bibinfo{pages}{416 }
  (\bibinfo{year}{1994}), ISSN \bibinfo{issn}{0168-9002},
  \urlprefix\url{http://www.sciencedirect.com/science/article/pii/0168900294911118}.

\bibitem[{\citenamefont{Hutter et~al.}(2002)\citenamefont{Hutter, Babilon,
  Bayer, Galaviz, Hartmann, Mohr, M{\" o}ller, Rochow, Savran, Sonnabend
  et~al.}}]{Hutter2002}
\bibinfo{author}{\bibfnamefont{C.}~\bibnamefont{Hutter}},
  \bibinfo{author}{\bibfnamefont{M.}~\bibnamefont{Babilon}},
  \bibinfo{author}{\bibfnamefont{W.}~\bibnamefont{Bayer}},
  \bibinfo{author}{\bibfnamefont{D.}~\bibnamefont{Galaviz}},
  \bibinfo{author}{\bibfnamefont{T.}~\bibnamefont{Hartmann}},
  \bibinfo{author}{\bibfnamefont{P.}~\bibnamefont{Mohr}},
  \bibinfo{author}{\bibfnamefont{S.}~\bibnamefont{M{\" o}ller}},
  \bibinfo{author}{\bibfnamefont{W.}~\bibnamefont{Rochow}},
  \bibinfo{author}{\bibfnamefont{D.}~\bibnamefont{Savran}},
  \bibinfo{author}{\bibfnamefont{K.}~\bibnamefont{Sonnabend}},
  \bibnamefont{et~al.}, \bibinfo{journal}{Nuclear Instruments and Methods in
  Physics Research Section A: Accelerators, Spectrometers, Detectors and
  Associated Equipment} \textbf{\bibinfo{volume}{489}}, \bibinfo{pages}{247 }
  (\bibinfo{year}{2002}), ISSN \bibinfo{issn}{0168-9002},
  \urlprefix\url{http://www.sciencedirect.com/science/article/pii/S0168900202005788}.

\bibitem[{\citenamefont{Akkoyun et~al.}(2012)\citenamefont{Akkoyun, Algora,
  Alikhani, Ameil, de~Angelis, Arnold, Astier, Ata√ß, Aubert, Aufranc
  et~al.}}]{AGATA}
\bibinfo{author}{\bibfnamefont{S.}~\bibnamefont{Akkoyun}},
  \bibinfo{author}{\bibfnamefont{A.}~\bibnamefont{Algora}},
  \bibinfo{author}{\bibfnamefont{B.}~\bibnamefont{Alikhani}},
  \bibinfo{author}{\bibfnamefont{F.}~\bibnamefont{Ameil}},
  \bibinfo{author}{\bibfnamefont{G.}~\bibnamefont{de~Angelis}},
  \bibinfo{author}{\bibfnamefont{L.}~\bibnamefont{Arnold}},
  \bibinfo{author}{\bibfnamefont{A.}~\bibnamefont{Astier}},
  \bibinfo{author}{\bibfnamefont{A.}~\bibnamefont{Ata√ß}},
  \bibinfo{author}{\bibfnamefont{Y.}~\bibnamefont{Aubert}},
  \bibinfo{author}{\bibfnamefont{C.}~\bibnamefont{Aufranc}},
  \bibnamefont{et~al.}, \bibinfo{journal}{Nuclear Instruments and Methods in
  Physics Research Section A: Accelerators, Spectrometers, Detectors and
  Associated Equipment} \textbf{\bibinfo{volume}{668}}, \bibinfo{pages}{26 }
  (\bibinfo{year}{2012}), ISSN \bibinfo{issn}{0168-9002},
  \urlprefix\url{http://www.sciencedirect.com/science/article/pii/S0168900211021516}.

\bibitem[{\citenamefont{Lee et~al.}(2003)\citenamefont{Lee, Deleplanque, and
  Vetter}}]{GRETINA}
\bibinfo{author}{\bibfnamefont{I.~Y.} \bibnamefont{Lee}},
  \bibinfo{author}{\bibfnamefont{M.~A.} \bibnamefont{Deleplanque}},
  \bibnamefont{and} \bibinfo{author}{\bibfnamefont{K.}~\bibnamefont{Vetter}},
  \bibinfo{journal}{Reports on Progress in Physics}
  \textbf{\bibinfo{volume}{66}}, \bibinfo{pages}{1095} (\bibinfo{year}{2003}),
  \urlprefix\url{http://stacks.iop.org/0034-4885/66/i=7/a=201}.

\bibitem[{\citenamefont{Wood et~al.}(1999)\citenamefont{Wood, Zganjar, Coster,
  and Heyde}}]{Wood1999}
\bibinfo{author}{\bibfnamefont{J.~L.} \bibnamefont{Wood}},
  \bibinfo{author}{\bibfnamefont{E.~F.} \bibnamefont{Zganjar}},
  \bibinfo{author}{\bibfnamefont{C.~D.} \bibnamefont{Coster}},
  \bibnamefont{and} \bibinfo{author}{\bibfnamefont{K.}~\bibnamefont{Heyde}},
  \bibinfo{journal}{Nuclear Physics A} \textbf{\bibinfo{volume}{651}},
  \bibinfo{pages}{323 } (\bibinfo{year}{1999}).

\bibitem[{\citenamefont{Becker and Steffen}(1969)}]{Becker1969}
\bibinfo{author}{\bibfnamefont{A.~J.} \bibnamefont{Becker}} \bibnamefont{and}
  \bibinfo{author}{\bibfnamefont{R.~M.} \bibnamefont{Steffen}},
  \bibinfo{journal}{Phys. Rev.} \textbf{\bibinfo{volume}{180}},
  \bibinfo{pages}{1043} (\bibinfo{year}{1969}).

\bibitem[{\citenamefont{Kib{\' e}di et~al.}(2008)\citenamefont{Kib{\' e}di,
  Burrows, Trzhaskovskaya, Davidson, and Nestor}}]{Kibedi2008}
\bibinfo{author}{\bibfnamefont{T.}~\bibnamefont{Kib{\' e}di}},
  \bibinfo{author}{\bibfnamefont{T.~W.} \bibnamefont{Burrows}},
  \bibinfo{author}{\bibfnamefont{M.~B.} \bibnamefont{Trzhaskovskaya}},
  \bibinfo{author}{\bibfnamefont{P.~M.} \bibnamefont{Davidson}},
  \bibnamefont{and} \bibinfo{author}{\bibfnamefont{C.~W.}
  \bibnamefont{Nestor}}, \bibinfo{journal}{Nuclear Instruments and Methods in
  Physics Research Section A: Accelerators, Spectrometers, Detectors and
  Associated Equipment} \textbf{\bibinfo{volume}{589}}, \bibinfo{pages}{202 }
  (\bibinfo{year}{2008}), ISSN \bibinfo{issn}{0168-9002},
  \urlprefix\url{http://www.sciencedirect.com/science/article/pii/S0168900208002520}.

\bibitem[{\citenamefont{Hager and Seltzer}(1968)}]{Hager1968}
\bibinfo{author}{\bibfnamefont{R.~S.} \bibnamefont{Hager}} \bibnamefont{and}
  \bibinfo{author}{\bibfnamefont{E.~C.} \bibnamefont{Seltzer}},
  \bibinfo{journal}{Nucl. Data A} \textbf{\bibinfo{volume}{4}},
  \bibinfo{pages}{397} (\bibinfo{year}{1968}).

\bibitem[{\citenamefont{Huang and Kang}(2016)}]{NDS198}
\bibinfo{author}{\bibfnamefont{X.}~\bibnamefont{Huang}} \bibnamefont{and}
  \bibinfo{author}{\bibfnamefont{M.}~\bibnamefont{Kang}},
  \bibinfo{journal}{Nuclear Data Sheets} \textbf{\bibinfo{volume}{133}},
  \bibinfo{pages}{221 } (\bibinfo{year}{2016}), ISSN \bibinfo{issn}{0090-3752},
  \urlprefix\url{http://www.sciencedirect.com/science/article/pii/S0090375216000156}.

\bibitem[{\citenamefont{Mach et~al.}(1989)\citenamefont{Mach, Gill, and
  Moszy\'nski}}]{Mach1989}
\bibinfo{author}{\bibfnamefont{H.}~\bibnamefont{Mach}},
  \bibinfo{author}{\bibfnamefont{R.}~\bibnamefont{Gill}}, \bibnamefont{and}
  \bibinfo{author}{\bibfnamefont{M.}~\bibnamefont{Moszy\'nski}},
  \bibinfo{journal}{Nuclear Instruments and Methods in Physics Research Section
  A: Accelerators, Spectrometers, Detectors and Associated Equipment}
  \textbf{\bibinfo{volume}{280}}, \bibinfo{pages}{49 } (\bibinfo{year}{1989}),
  ISSN \bibinfo{issn}{0168-9002},
  \urlprefix\url{http://www.sciencedirect.com/science/article/pii/0168900289912722}.

\bibitem[{\citenamefont{Moszy\'nski and Mach}(1989)}]{Moszynski1989}
\bibinfo{author}{\bibfnamefont{M.}~\bibnamefont{Moszy\'nski}} \bibnamefont{and}
  \bibinfo{author}{\bibfnamefont{H.}~\bibnamefont{Mach}},
  \bibinfo{journal}{Nuclear Instruments and Methods in Physics Research Section
  A: Accelerators, Spectrometers, Detectors and Associated Equipment}
  \textbf{\bibinfo{volume}{277}}, \bibinfo{pages}{407 } (\bibinfo{year}{1989}),
  ISSN \bibinfo{issn}{0168-9002},
  \urlprefix\url{http://www.sciencedirect.com/science/article/pii/0168900289907705}.

\bibitem[{\citenamefont{Mach et~al.}(1991)\citenamefont{Mach, Wohn, Moln\'ar,
  Sistemich, Hill, Moszy\'nski, Gill, Krips, and Brenner}}]{Mach1991}
\bibinfo{author}{\bibfnamefont{H.}~\bibnamefont{Mach}},
  \bibinfo{author}{\bibfnamefont{F.}~\bibnamefont{Wohn}},
  \bibinfo{author}{\bibfnamefont{G.}~\bibnamefont{Moln\'ar}},
  \bibinfo{author}{\bibfnamefont{K.}~\bibnamefont{Sistemich}},
  \bibinfo{author}{\bibfnamefont{J.~C.} \bibnamefont{Hill}},
  \bibinfo{author}{\bibfnamefont{M.}~\bibnamefont{Moszy\'nski}},
  \bibinfo{author}{\bibfnamefont{R.}~\bibnamefont{Gill}},
  \bibinfo{author}{\bibfnamefont{W.}~\bibnamefont{Krips}}, \bibnamefont{and}
  \bibinfo{author}{\bibfnamefont{D.}~\bibnamefont{Brenner}},
  \bibinfo{journal}{Nuclear Physics A} \textbf{\bibinfo{volume}{523}},
  \bibinfo{pages}{197 } (\bibinfo{year}{1991}), ISSN \bibinfo{issn}{0375-9474},
  \urlprefix\url{http://www.sciencedirect.com/science/article/pii/037594749190001M}.

\bibitem[{\citenamefont{R{\' e}gis et~al.}(2013)\citenamefont{R{\' e}gis, Mach,
  Simpson, Jolie, Pascovici, Saed-Samii, Warr, Bruce, Degenkolb, Fraile
  et~al.}}]{Regis2013}
\bibinfo{author}{\bibfnamefont{J.-M.} \bibnamefont{R{\' e}gis}},
  \bibinfo{author}{\bibfnamefont{H.}~\bibnamefont{Mach}},
  \bibinfo{author}{\bibfnamefont{G.~S.} \bibnamefont{Simpson}},
  \bibinfo{author}{\bibfnamefont{J.}~\bibnamefont{Jolie}},
  \bibinfo{author}{\bibfnamefont{G.}~\bibnamefont{Pascovici}},
  \bibinfo{author}{\bibfnamefont{N.}~\bibnamefont{Saed-Samii}},
  \bibinfo{author}{\bibfnamefont{N.}~\bibnamefont{Warr}},
  \bibinfo{author}{\bibfnamefont{A.}~\bibnamefont{Bruce}},
  \bibinfo{author}{\bibfnamefont{J.}~\bibnamefont{Degenkolb}},
  \bibinfo{author}{\bibfnamefont{L.~M.} \bibnamefont{Fraile}},
  \bibnamefont{et~al.}, \bibinfo{journal}{Nuclear Instruments and Methods in
  Physics Research Section A: Accelerators, Spectrometers, Detectors and
  Associated Equipment} \textbf{\bibinfo{volume}{726}}, \bibinfo{pages}{191 }
  (\bibinfo{year}{2013}), ISSN \bibinfo{issn}{0168-9002},
  \urlprefix\url{http://www.sciencedirect.com/science/article/pii/S0168900213007377}.

\bibitem[{\citenamefont{Martin}(2013)}]{NDS152}
\bibinfo{author}{\bibfnamefont{M.~J.} \bibnamefont{Martin}},
  \bibinfo{journal}{Nuclear Data Sheets} \textbf{\bibinfo{volume}{114}},
  \bibinfo{pages}{1497 } (\bibinfo{year}{2013}), ISSN
  \bibinfo{issn}{0090-3752},
  \urlprefix\url{http://www.sciencedirect.com/science/article/pii/S0090375213000744}.

\bibitem[{\citenamefont{R{\' e}gis et~al.}(2010)\citenamefont{R{\' e}gis,
  Pascovici, Jolie, and Rudigier}}]{Regis2010}
\bibinfo{author}{\bibfnamefont{J.-M.} \bibnamefont{R{\' e}gis}},
  \bibinfo{author}{\bibfnamefont{G.}~\bibnamefont{Pascovici}},
  \bibinfo{author}{\bibfnamefont{J.}~\bibnamefont{Jolie}}, \bibnamefont{and}
  \bibinfo{author}{\bibfnamefont{M.}~\bibnamefont{Rudigier}},
  \bibinfo{journal}{Nuclear Instruments and Methods in Physics Research Section
  A: Accelerators, Spectrometers, Detectors and Associated Equipment}
  \textbf{\bibinfo{volume}{622}}, \bibinfo{pages}{83 } (\bibinfo{year}{2010}),
  ISSN \bibinfo{issn}{0168-9002},
  \urlprefix\url{http://www.sciencedirect.com/science/article/pii/S0168900210016578}.

\bibitem[{\citenamefont{R{\' e}gis et~al.}(2016)\citenamefont{R{\' e}gis,
  Saed-Samii, Rudigier, Ansari, Dannhoff, Esmaylzadeh, Fransen, Gerst, Jolie,
  Karayonchev et~al.}}]{Regis2016}
\bibinfo{author}{\bibfnamefont{J.-M.} \bibnamefont{R{\' e}gis}},
  \bibinfo{author}{\bibfnamefont{N.}~\bibnamefont{Saed-Samii}},
  \bibinfo{author}{\bibfnamefont{M.}~\bibnamefont{Rudigier}},
  \bibinfo{author}{\bibfnamefont{S.}~\bibnamefont{Ansari}},
  \bibinfo{author}{\bibfnamefont{M.}~\bibnamefont{Dannhoff}},
  \bibinfo{author}{\bibfnamefont{A.}~\bibnamefont{Esmaylzadeh}},
  \bibinfo{author}{\bibfnamefont{C.}~\bibnamefont{Fransen}},
  \bibinfo{author}{\bibfnamefont{R.-B.} \bibnamefont{Gerst}},
  \bibinfo{author}{\bibfnamefont{J.}~\bibnamefont{Jolie}},
  \bibinfo{author}{\bibfnamefont{V.}~\bibnamefont{Karayonchev}},
  \bibnamefont{et~al.}, \bibinfo{journal}{Nuclear Instruments and Methods in
  Physics Research Section A: Accelerators, Spectrometers, Detectors and
  Associated Equipment} \textbf{\bibinfo{volume}{823}}, \bibinfo{pages}{72 }
  (\bibinfo{year}{2016}), ISSN \bibinfo{issn}{0168-9002},
  \urlprefix\url{http://www.sciencedirect.com/science/article/pii/S016890021630170X}.

\bibitem[{\citenamefont{Pritychenko et~al.}(2016)\citenamefont{Pritychenko,
  Birch, Singh, and Horoi}}]{Pritychenko2016}
\bibinfo{author}{\bibfnamefont{B.}~\bibnamefont{Pritychenko}},
  \bibinfo{author}{\bibfnamefont{M.}~\bibnamefont{Birch}},
  \bibinfo{author}{\bibfnamefont{B.}~\bibnamefont{Singh}}, \bibnamefont{and}
  \bibinfo{author}{\bibfnamefont{M.}~\bibnamefont{Horoi}},
  \bibinfo{journal}{Atomic Data and Nuclear Data Tables}
  \textbf{\bibinfo{volume}{107}}, \bibinfo{pages}{1 } (\bibinfo{year}{2016}),
  ISSN \bibinfo{issn}{0092-640X},
  \urlprefix\url{http://www.sciencedirect.com/science/article/pii/S0092640X15000406}.

\bibitem[{\citenamefont{R{\' e}gis et~al.}(2012)\citenamefont{R{\' e}gis,
  Rudigier, Jolie, Blazhev, Fransen, Pascovici, and Warr}}]{Regis2012}
\bibinfo{author}{\bibfnamefont{J.-M.} \bibnamefont{R{\' e}gis}},
  \bibinfo{author}{\bibfnamefont{M.}~\bibnamefont{Rudigier}},
  \bibinfo{author}{\bibfnamefont{J.}~\bibnamefont{Jolie}},
  \bibinfo{author}{\bibfnamefont{A.}~\bibnamefont{Blazhev}},
  \bibinfo{author}{\bibfnamefont{C.}~\bibnamefont{Fransen}},
  \bibinfo{author}{\bibfnamefont{G.}~\bibnamefont{Pascovici}},
  \bibnamefont{and} \bibinfo{author}{\bibfnamefont{N.}~\bibnamefont{Warr}},
  \bibinfo{journal}{Nuclear Instruments and Methods in Physics Research Section
  A: Accelerators, Spectrometers, Detectors and Associated Equipment}
  \textbf{\bibinfo{volume}{684}}, \bibinfo{pages}{36 } (\bibinfo{year}{2012}),
  ISSN \bibinfo{issn}{0168-9002},
  \urlprefix\url{http://www.sciencedirect.com/science/article/pii/S0168900212004937}.

\bibitem[{\citenamefont{Bucurescu et~al.}(2016)\citenamefont{Bucurescu,
  CƒÉta-Danil, Ciocan, Costache, Deleanu, Dima, Filipescu, Florea,
  Ghi≈£ƒÉ, Glodariu et~al.}}]{Bucurescu2016}
\bibinfo{author}{\bibfnamefont{D.}~\bibnamefont{Bucurescu}},
  \bibinfo{author}{\bibfnamefont{I.}~\bibnamefont{CƒÉta-Danil}},
  \bibinfo{author}{\bibfnamefont{G.}~\bibnamefont{Ciocan}},
  \bibinfo{author}{\bibfnamefont{C.}~\bibnamefont{Costache}},
  \bibinfo{author}{\bibfnamefont{D.}~\bibnamefont{Deleanu}},
  \bibinfo{author}{\bibfnamefont{R.}~\bibnamefont{Dima}},
  \bibinfo{author}{\bibfnamefont{D.}~\bibnamefont{Filipescu}},
  \bibinfo{author}{\bibfnamefont{N.}~\bibnamefont{Florea}},
  \bibinfo{author}{\bibfnamefont{D.}~\bibnamefont{Ghi≈£ƒÉ}},
  \bibinfo{author}{\bibfnamefont{T.}~\bibnamefont{Glodariu}},
  \bibnamefont{et~al.}, \bibinfo{journal}{Nuclear Instruments and Methods in
  Physics Research Section A: Accelerators, Spectrometers, Detectors and
  Associated Equipment} \textbf{\bibinfo{volume}{837}}, \bibinfo{pages}{1 }
  (\bibinfo{year}{2016}), ISSN \bibinfo{issn}{0168-9002},
  \urlprefix\url{http://www.sciencedirect.com/science/article/pii/S0168900216308798}.

\bibitem[{\citenamefont{Rudigier et~al.}(2017)\citenamefont{Rudigier,
  Lalkovski, Gamba, Bruce, Podoly{\' a}k, Regan, Carpenter, Zhu, Ayangeakaa,
  Anderson et~al.}}]{Rudigier2017}
\bibinfo{author}{\bibfnamefont{M.}~\bibnamefont{Rudigier}},
  \bibinfo{author}{\bibfnamefont{S.}~\bibnamefont{Lalkovski}},
  \bibinfo{author}{\bibfnamefont{E.~R.} \bibnamefont{Gamba}},
  \bibinfo{author}{\bibfnamefont{A.~M.} \bibnamefont{Bruce}},
  \bibinfo{author}{\bibfnamefont{Z.}~\bibnamefont{Podoly{\' a}k}},
  \bibinfo{author}{\bibfnamefont{P.~H.} \bibnamefont{Regan}},
  \bibinfo{author}{\bibfnamefont{M.}~\bibnamefont{Carpenter}},
  \bibinfo{author}{\bibfnamefont{S.}~\bibnamefont{Zhu}},
  \bibinfo{author}{\bibfnamefont{A.~D.} \bibnamefont{Ayangeakaa}},
  \bibinfo{author}{\bibfnamefont{J.~T.} \bibnamefont{Anderson}},
  \bibnamefont{et~al.}, \bibinfo{journal}{Acta Physica Polonica B}
  \textbf{\bibinfo{volume}{48}}, \bibinfo{pages}{351} (\bibinfo{year}{2017}),
  \urlprefix\url{http://www.actaphys.uj.edu.pl/findarticle?series=Reg&vol=48&page=351}.

\bibitem[{\citenamefont{Levy et~al.}(2014)\citenamefont{Levy, Pearson, Kiefl,
  Man{\'e}, Morris, and Voss}}]{Levy2013}
\bibinfo{author}{\bibfnamefont{C.~D.~P.} \bibnamefont{Levy}},
  \bibinfo{author}{\bibfnamefont{M.~R.} \bibnamefont{Pearson}},
  \bibinfo{author}{\bibfnamefont{R.~F.} \bibnamefont{Kiefl}},
  \bibinfo{author}{\bibfnamefont{E.}~\bibnamefont{Man{\'e}}},
  \bibinfo{author}{\bibfnamefont{G.~D.} \bibnamefont{Morris}},
  \bibnamefont{and} \bibinfo{author}{\bibfnamefont{A.}~\bibnamefont{Voss}},
  \bibinfo{journal}{Hyperfine Interactions} \textbf{\bibinfo{volume}{225}},
  \bibinfo{pages}{165} (\bibinfo{year}{2014}), ISSN \bibinfo{issn}{1572-9540},
  \urlprefix\url{https://doi.org/10.1007/s10751-013-0896-4}.

\bibitem[{\citenamefont{Shimoda et~al.}(2014)\citenamefont{Shimoda, Tajiri,
  Kura, Odahara, Suga, Hirayama, Imai, Miyatake, Pearson, Levy
  et~al.}}]{Shimoda2013}
\bibinfo{author}{\bibfnamefont{T.}~\bibnamefont{Shimoda}},
  \bibinfo{author}{\bibfnamefont{K.}~\bibnamefont{Tajiri}},
  \bibinfo{author}{\bibfnamefont{K.}~\bibnamefont{Kura}},
  \bibinfo{author}{\bibfnamefont{A.}~\bibnamefont{Odahara}},
  \bibinfo{author}{\bibfnamefont{M.}~\bibnamefont{Suga}},
  \bibinfo{author}{\bibfnamefont{Y.}~\bibnamefont{Hirayama}},
  \bibinfo{author}{\bibfnamefont{N.}~\bibnamefont{Imai}},
  \bibinfo{author}{\bibfnamefont{H.}~\bibnamefont{Miyatake}},
  \bibinfo{author}{\bibfnamefont{M.}~\bibnamefont{Pearson}},
  \bibinfo{author}{\bibfnamefont{C.~D.~P.} \bibnamefont{Levy}},
  \bibnamefont{et~al.}, \bibinfo{journal}{Hyperfine Interactions}
  \textbf{\bibinfo{volume}{225}}, \bibinfo{pages}{183} (\bibinfo{year}{2014}),
  ISSN \bibinfo{issn}{1572-9540},
  \urlprefix\url{https://doi.org/10.1007/s10751-013-0895-5}.

\bibitem[{\citenamefont{Lynch}(2015)}]{Lynch2015}
\bibinfo{author}{\bibfnamefont{K.~M.} \bibnamefont{Lynch}},
  \emph{\bibinfo{title}{Laser Assisted Nuclear Decay Spectroscopy}}
  (\bibinfo{publisher}{Springer International Publishing},
  \bibinfo{address}{Cham}, \bibinfo{year}{2015}), pp. \bibinfo{pages}{53--68},
  ISBN \bibinfo{isbn}{978-3-319-07112-1},
  \urlprefix\url{https://doi.org/10.1007/978-3-319-07112-1_6}.

\bibitem[{\citenamefont{Lynch et~al.}(2012)\citenamefont{Lynch, Rajabali,
  Aghaei-Khozani, Billowes, Bissell, Blanc, Cheal, Cocolios, Schepper, Dewolf
  et~al.}}]{Lynch2012}
\bibinfo{author}{\bibfnamefont{K.~M.} \bibnamefont{Lynch}},
  \bibinfo{author}{\bibfnamefont{M.~M.} \bibnamefont{Rajabali}},
  \bibinfo{author}{\bibfnamefont{H.}~\bibnamefont{Aghaei-Khozani}},
  \bibinfo{author}{\bibfnamefont{J.}~\bibnamefont{Billowes}},
  \bibinfo{author}{\bibfnamefont{M.~L.} \bibnamefont{Bissell}},
  \bibinfo{author}{\bibfnamefont{F.~L.} \bibnamefont{Blanc}},
  \bibinfo{author}{\bibfnamefont{B.}~\bibnamefont{Cheal}},
  \bibinfo{author}{\bibfnamefont{T.~E.} \bibnamefont{Cocolios}},
  \bibinfo{author}{\bibfnamefont{S.~D.} \bibnamefont{Schepper}},
  \bibinfo{author}{\bibfnamefont{K.}~\bibnamefont{Dewolf}},
  \bibnamefont{et~al.}, \bibinfo{journal}{Journal of Physics: Conference
  Series} \textbf{\bibinfo{volume}{381}}, \bibinfo{pages}{012128}
  (\bibinfo{year}{2012}),
  \urlprefix\url{http://stacks.iop.org/1742-6596/381/i=1/a=012128}.

\bibitem[{\citenamefont{Dunlop et~al.}(2016)\citenamefont{Dunlop, Bildstein,
  Dillmann, Jungclaus, Svensson, Andreoiu, Ball, Bernier, Bidaman, Boubel
  et~al.}}]{Dunlop2016}
\bibinfo{author}{\bibfnamefont{R.}~\bibnamefont{Dunlop}},
  \bibinfo{author}{\bibfnamefont{V.}~\bibnamefont{Bildstein}},
  \bibinfo{author}{\bibfnamefont{I.}~\bibnamefont{Dillmann}},
  \bibinfo{author}{\bibfnamefont{A.}~\bibnamefont{Jungclaus}},
  \bibinfo{author}{\bibfnamefont{C.~E.} \bibnamefont{Svensson}},
  \bibinfo{author}{\bibfnamefont{C.}~\bibnamefont{Andreoiu}},
  \bibinfo{author}{\bibfnamefont{G.~C.} \bibnamefont{Ball}},
  \bibinfo{author}{\bibfnamefont{N.}~\bibnamefont{Bernier}},
  \bibinfo{author}{\bibfnamefont{H.}~\bibnamefont{Bidaman}},
  \bibinfo{author}{\bibfnamefont{P.}~\bibnamefont{Boubel}},
  \bibnamefont{et~al.}, \bibinfo{journal}{Phys. Rev. C}
  \textbf{\bibinfo{volume}{93}}, \bibinfo{pages}{062801}
  (\bibinfo{year}{2016}),
  \urlprefix\url{https://link.aps.org/doi/10.1103/PhysRevC.93.062801}.

\bibitem[{\citenamefont{Garnsworthy
  et~al.}(2017{\natexlab{b}})\citenamefont{Garnsworthy, Bowry, Olaizola, Holt,
  Stroberg, Cruz, Georges, Hackman, MacLean, Measures
  et~al.}}]{Garnsworthy2017-50Sc}
\bibinfo{author}{\bibfnamefont{A.~B.} \bibnamefont{Garnsworthy}},
  \bibinfo{author}{\bibfnamefont{M.}~\bibnamefont{Bowry}},
  \bibinfo{author}{\bibfnamefont{B.}~\bibnamefont{Olaizola}},
  \bibinfo{author}{\bibfnamefont{J.~D.} \bibnamefont{Holt}},
  \bibinfo{author}{\bibfnamefont{S.~R.} \bibnamefont{Stroberg}},
  \bibinfo{author}{\bibfnamefont{S.}~\bibnamefont{Cruz}},
  \bibinfo{author}{\bibfnamefont{S.}~\bibnamefont{Georges}},
  \bibinfo{author}{\bibfnamefont{G.}~\bibnamefont{Hackman}},
  \bibinfo{author}{\bibfnamefont{A.~D.} \bibnamefont{MacLean}},
  \bibinfo{author}{\bibfnamefont{J.}~\bibnamefont{Measures}},
  \bibnamefont{et~al.}, \bibinfo{journal}{Phys. Rev. C}
  \textbf{\bibinfo{volume}{96}}, \bibinfo{pages}{044329}
  (\bibinfo{year}{2017}{\natexlab{b}}),
  \urlprefix\url{https://link.aps.org/doi/10.1103/PhysRevC.96.044329}.

\bibitem[{\citenamefont{Dunlop et~al.}(2017)\citenamefont{Dunlop, Svensson,
  Ball, Leslie, Andreoiu, Bernier, Bidaman, Bildstein, Bowry, Burbadge
  et~al.}}]{Dunlop2017}
\bibinfo{author}{\bibfnamefont{M.~R.} \bibnamefont{Dunlop}},
  \bibinfo{author}{\bibfnamefont{C.~E.} \bibnamefont{Svensson}},
  \bibinfo{author}{\bibfnamefont{G.~C.} \bibnamefont{Ball}},
  \bibinfo{author}{\bibfnamefont{J.~R.} \bibnamefont{Leslie}},
  \bibinfo{author}{\bibfnamefont{C.}~\bibnamefont{Andreoiu}},
  \bibinfo{author}{\bibfnamefont{N.}~\bibnamefont{Bernier}},
  \bibinfo{author}{\bibfnamefont{H.}~\bibnamefont{Bidaman}},
  \bibinfo{author}{\bibfnamefont{V.}~\bibnamefont{Bildstein}},
  \bibinfo{author}{\bibfnamefont{M.}~\bibnamefont{Bowry}},
  \bibinfo{author}{\bibfnamefont{C.}~\bibnamefont{Burbadge}},
  \bibnamefont{et~al.}, \bibinfo{journal}{Phys. Rev. C}
  \textbf{\bibinfo{volume}{96}}, \bibinfo{pages}{045502}
  (\bibinfo{year}{2017}),
  \urlprefix\url{https://link.aps.org/doi/10.1103/PhysRevC.96.045502}.

\end{thebibliography}

\end{document}